\documentclass[
 aps,
 prb,
 reprint,
 amsmath,amssymb,
 superscriptaddress,
 floatfix,
 longbibliography,
 nofootinbib
]{revtex4-2}

\usepackage{graphicx}
\graphicspath{{./Figures/}}
\usepackage{dcolumn}
\usepackage{bm}
\usepackage{hyperref}
\usepackage{xcolor}
\usepackage[noabbrev]{cleveref}
\usepackage{comment}
\usepackage{dsfont}
\usepackage{slashed}
\usepackage{scalerel}
\usepackage{cancel}
\usepackage{nicefrac}
\usepackage{xfrac}
\usepackage{textcomp}
\usepackage{romannum}
\usepackage{mathtools}
\usepackage{bbold}
\usepackage{xparse}
\usepackage{tikz}
\usetikzlibrary{positioning,decorations.pathreplacing,calc,tikzmark,arrows.meta}
\usepackage{quantikz}
\usepackage{xcolor}
\usepackage{braket}
\usepackage{physics}
\usepackage{float}
\tikzset{
    tkone/.style={fill=pink!30},
    tktwo/.style={fill=purple!25},
    epsblock/.style={fill=purple!45},
    wblock/.style={fill=pink!40,draw=black},
    vblock/.style={fill=white,draw=black},
    pertblock/.style={fill=pink!40},
    su4gate/.style={fill=purple!28},
    perturb/.style={fill=pink!45},
    measgate/.style={fill=gray!20},
    pauligate/.style={fill=purple!70, draw=black, text=white}    }

\tikzset{
  pbc-first/.style={
    draw=none,
    fill=white,
    minimum width=8mm,
    minimum height=6mm,
    path picture={
      \draw
        (path picture bounding box.north west) --
        (path picture bounding box.south west) --
        (path picture bounding box.south east) --
        (path picture bounding box.north east);
    }
  },
  pbc-last/.style={
    draw=none,
    fill=white,
    minimum width=8mm,
    minimum height=6mm,
    path picture={
      \draw
        (path picture bounding box.south west) --
        (path picture bounding box.north west) --
        (path picture bounding box.north east) --
        (path picture bounding box.south east);
    }
  }
}
\begin{document}

% ----------------------------------------------------------------
% TITLE & AUTHORS (REVTeX 4-2 / PRB style)
% ----------------------------------------------------------------

\title{Non-Unitary Quantum Butterfly Effect: Theory and Experiment}

\author{Paul Argyle}
\email{pargyle1@umd.edu}
\affiliation{Department of Physics, University of Maryland, College Park, Maryland 20742, USA}
\affiliation{Joint Quantum Institute, University of Maryland and National Institute of Standards and Technology, College Park, Maryland 20742, USA}

\author{Djamil Lakhdar-Hamina}
\affiliation{Department of Physics, University of Maryland, College Park, Maryland 20742, USA}
\affiliation{Joint Quantum Institute, University of Maryland and National Institute of Standards and Technology, College Park, Maryland 20742, USA}

\author{Sarah H. Miller}
\affiliation{Applied Research Laboratory for Intelligence and Security, University of Maryland, College Park, Maryland 20742, USA}
\affiliation{Artificial Intelligence Interdisciplinary Institute at Maryland, University of Maryland, College Park, Maryland 20742, USA}

\author{Victor Galitski}
\affiliation{Department of Physics, University of Maryland, College Park, Maryland 20742, USA}
\affiliation{Joint Quantum Institute, University of Maryland and National Institute of Standards and Technology, College Park, Maryland 20742, USA}

\date{\today}

\begin{abstract}

We present a generalization of the out of time order correlator (OTOC) for non-unitary systems. For a Haar-random brickwork qudit-circuit, peppered with local, projective measurement, we calculate the infinite-temperature, ensemble-average OTOC. Utilizing transfer matrix machinery, we find that the OTOC asymptotically decays exponentially and that the decay depends on the rate of measurement. In the thermodynamic limit, with sparse measurement, the OTOC approximately exhibits a Gaussian shape. Phenomenologically, this corresponds to the development of a causal scrambling front where unlike the unitary case, the signal decays in the bulk of the causal cone. We develop a Monte Carlo framework to compute the OTOC. We then estimate the OTOC directly via quantum simulation and on Quantinuum's trapped-ion quantum computer. We resolve the unitary and non-unitary OTOC phenomenology despite the presence of hardware noise. Our analytic results characterize how local nonunitary channels affect OTOC decay, providing a quantitative benchmark for the cumulative effects of a broad class of hardware noise on many-body dynamics.

\end{abstract}

\maketitle

\section{Introduction}

The out-of-time-ordered correlator (OTOC) has emerged as a central diagnostic
of quantum chaos and information scrambling. Inspired by the work of Larkin and Ovchinnikov in the
context of superconductivity \cite{LarkinOvchinnikov1969}, the OTOC acquired renewed significance through
the work of Kitaev~\cite{Kitaev2014Hidden} and Maldacena, Shenker, and Stanford~\cite{MaldacenaShenkerStanford2016}, which connected its
early-time growth to quantum information scrambling and established a bound
on the rate of quantum chaos in certain systems~\cite{Liao2018}. More recently, developments in quantum
information and computing have made it possible to study scrambling
directly in quantum circuits \cite{yao2016interferometricapproachprobingfast, Swingle_2016,Gaerttner2017OTOC,Swingle2018Unscrambling,Nahum2018OperatorSpreading,PhysRevX.8.021013,Mi2021InformationScrambling,GoogleQuantumAI2025Constructive}. At the same time, mid-circuit measurement has become the enabling primitive for quantum error correction, adaptive and feedforward algorithms, and measurement-based architectures in which measurement is a computational resource rather than a nuisance \cite{minn}. Yet how a measurement disturbs qubits other than the one measured remains among the least characterized aspects of present-day hardware.

Furthermore, random quantum circuits have become powerful metaphors for nonequilibrium many-body physics. Such circuits provide minimal yet nontrivial models of many-body evolution: locality gives rise to a causal light-cone structure, while their statistical structure permits analytical and numerical access to otherwise complicated processes such as operator spreading, scrambling, and measurement-induced dynamics \cite{PhysRevX.7.031016,jonay2018coarsegraineddynamicsoperatorstate,PhysRevX.9.031009}.

For Haar-random unitary circuits, Nahum \textit{et al.} showed that the OTOC
exhibits a propagating scrambling front characterized by a butterfly velocity
and a front whose fluctuations and broadening encode the underlying operator
dynamics \cite{Nahum2018OperatorSpreading}. These results provide a transparent description of
scrambling in closed, unitary many-body systems. Here, we extend this framework
to \textit{non-unitary monitored dynamics} by introducing local measurements into the
random-circuit evolution. This allows us to investigate how measurement
modifies the propagation and decay of the OTOC, thereby connecting the physics
of measurement-induced dynamics with that of quantum chaos and information
scrambling. We derive a number of analytical results and numerically calculate the OTOC utilizing a Monte Carlo technique. Furthermore,  we experimentally measure the OTOC using Quantinuum's H2 trapped-ion quantum computer and find results broadly consistent with theory. The central physical result admits a simple interpretation. In the unitary circuit, the OTOC saturates behind the propagating front: the system retains a persistent memory of the local perturbation even as information
scrambles throughout the degrees of freedom.
Measurement qualitatively changes this behavior. Although the perturbation
continues to propagate causally, repeated measurements progressively erase its
signature from the OTOC. In this sense,
monitoring induces a form of dynamical ``amnesia'': the circuit progressively
forgets the perturbation. Operationally, the monitored OTOC therefore measures how far, and for how long, the backaction of an unrecorded measurement is felt elsewhere in the register.

\section{Theory}

Consider a system of $N$ qudits, which we refer to as sites and label $j = 0,1,\ldots,N-1$. Throughout this work, indices are understood modulo the size of the lattice on which they are defined. Each qudit has local Hilbert-space dimension $q$. We designate sites $x$ and $y$ as the probe and perturbation sites, respectively. The full Hilbert space is $\mathcal{H} = (\mathbb{C}^q)^{\otimes N}$. Let $\mathcal{L}(\mathcal{H})$ denote the space of linear operators on $\mathcal{H}$, and let $\varepsilon$ be a quantum channel acting on the full system; that is a completely positive, trace preserving linear map (CPTP) $\varepsilon: \mathcal{L}(\mathcal{H}) \rightarrow \mathcal{L}(\mathcal{H}) $. Let $\mathcal{W}$ be a single-qudit channel. Its action on site $y$, with the identity channel acting elsewhere, is denoted $\mathcal{W}_y = \mathcal{I}_q^{\otimes y} \otimes \mathcal{W} \otimes \mathcal{I}_q^{N-y-1}$ where $\mathcal{I}_q$ is the single-qudit identity channel. Let $V$ be a single-qudit Hermitian operator and define $V_x = I^{\otimes x}_q \otimes  V \otimes I^{\otimes N -x-1}_q $. We refer to $\varepsilon$, $\mathcal{W}_y$, and $V_x$ as the evolution channel, perturbation channel, and probe operator, respectively. 

\begin{figure}[t]
\centering
\newsavebox{\measdialbox}\sbox{\measdialbox}{\begin{tikzpicture}[x=1em,y=1em,baseline=0.6em]\path[use as bounding box] (0,0) rectangle (2,1.5); \draw[line width=0.8pt] (0.1cm,0.75em-0.06cm) to[bend left=50] (2em-0.1cm,0.75em-0.06cm); \draw[line width=0.8pt,-{Latex[scale=0.6]}] (1em,0.75em-0.1cm) -- (2em-0.1cm,0.75em+0.2cm); \node[inner sep=0,anchor=south east] at (2.15,-0.15) {$\scriptscriptstyle V$};\end{tikzpicture}}
% ============================================================
% CIRCUITS
% ============================================================
\begin{minipage}[b]{0.45\columnwidth}
\centering
\rotatebox{90}{%
\begin{quantikz}[
    row sep=0.38cm,
    column sep=0.48cm
]
\lstick{\rotatebox{-90}{\scriptsize $q_0$}}
& \qw
& \gate[4,style=epsblock]
    {\rotatebox{-90}{\scriptsize $\varepsilon$}}
& \qw
& \qw
\\
\lstick{\rotatebox{-90}{\scriptsize $q_y$}}
& \phantomgate{\rule[-2.8pt]{12.8pt}{10.6pt}}
&
& \qw
& \qw
\\
\lstick{\rotatebox{-90}{\scriptsize $\cdots$}}
& \qw
&
& \qw
& \qw
\\
\lstick{\rotatebox{-90}{\scriptsize $q_x$}}
& \qw
&
& \gate[style={vblock,inner sep=0pt}]
    {\rotatebox{-90}{\usebox{\measdialbox}}}
& \qw
\end{quantikz}%
}
\end{minipage}
\hfill
\begin{minipage}[b]{0.45\columnwidth}
\centering
\rotatebox{90}{%
\begin{quantikz}[
    row sep=0.38cm,
    column sep=0.48cm
]
\lstick{\rotatebox{-90}{\scriptsize $q_0$}}
& \qw
& \gate[4,style=epsblock]
    {\rotatebox{-90}{\scriptsize $\varepsilon$}}
& \qw
& \qw
\\
\lstick{\rotatebox{-90}{\scriptsize $q_y$}}
& \gate[style=wblock]
    {\rotatebox{-90}{\scriptsize $\mathcal W$}}
&
& \qw
& \qw
\\
\lstick{\rotatebox{-90}{\scriptsize $\cdots$}}
& \qw
&
& \qw
& \qw
\\
\lstick{\rotatebox{-90}{\scriptsize $q_x$}}
& \qw
&
& \gate[style={vblock,inner sep=0pt}]
    {\rotatebox{-90}{\usebox{\measdialbox}}}
& \qw
\end{quantikz}%
}
\end{minipage}
% ============================================================
% PANEL LABELS
% ============================================================
\vspace{0.3em}
\begin{minipage}{0.45\columnwidth}
\centering
\textbf{(a)}
\end{minipage}
\hfill
\begin{minipage}{0.45\columnwidth}
\centering
\textbf{(b)}
\end{minipage}
\caption{
Schematic representation of the two circuit protocols.
\textbf{(a)} Evolution under $\varepsilon$ followed by measurement
of $V$ at site $q_x$.
\textbf{(b)} The local perturbation $\mathcal W$ is applied at site
$q_y$ before evolution under $\varepsilon$, followed by the same
measurement of $V$ at site $q_x$.
}
\label{fig:WV_protocol}
\end{figure}
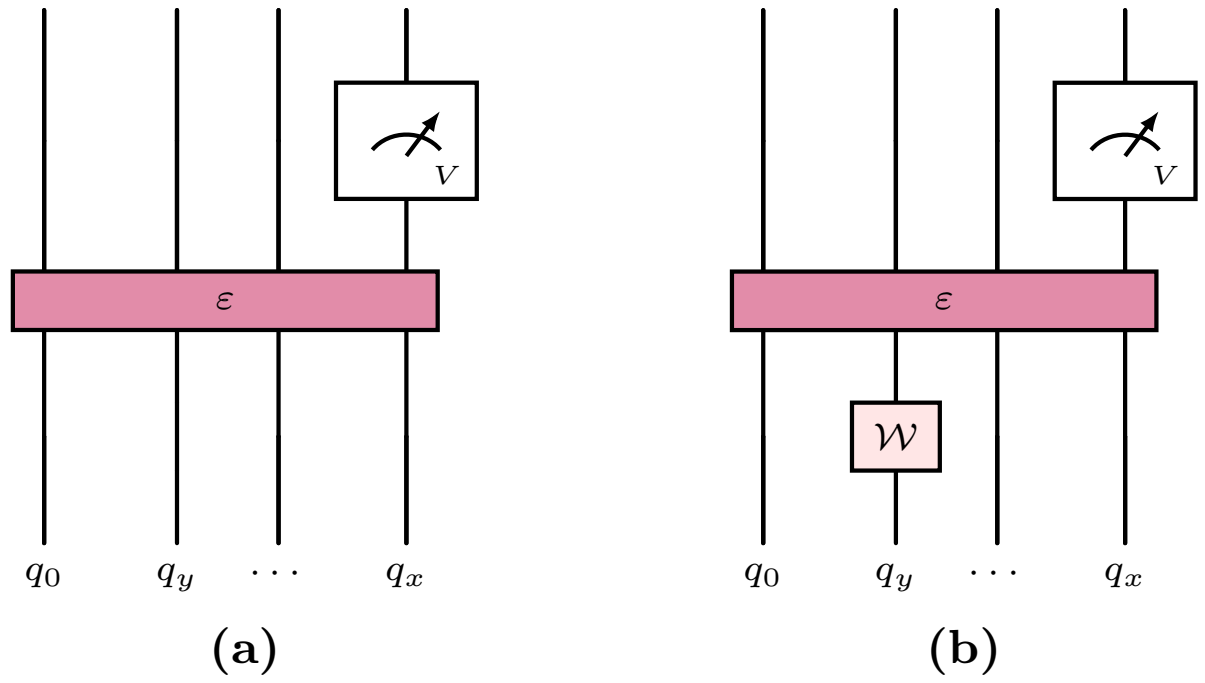

Every quantum channel $\varepsilon$ on $\mathcal{L} (\mathcal{H})$ has a unique adjoint map $\varepsilon^\dagger$, defined by 
\begin{equation}
\Tr \bigl( O \varepsilon(\rho) \bigr) = \Tr \bigl( \varepsilon^\dagger (O) \rho \bigr)
\end{equation}
for every state $\rho$ and system operator $O$. We define the response operator $\Delta_{x,y}$ by
\begin{equation}
\Delta_{x,y} =\mathcal{W}^\dagger_y  (\varepsilon^\dagger(V_x)) - \varepsilon^\dagger(V_x)
\label{eq:response_operator}.
\end{equation}
For a given density operator $\rho$, we call the expectation value of $\Delta^2_{x,y}$ the generalized out of time order correlator (OTOC), or simply the OTOC, and write
\begin{equation}
C_\rho(x,y) =  \Tr \Bigl( \rho \Delta^2_{x,y} \Bigr).
\label{generalized_otoc}
\end{equation}
When the perturbation channel is unitary, with $\mathcal{W}_y(O) = W_y O W^\dagger_y$ for a unitary operator $W_y$, the generalized OTOC reduces to the form used in Ref.~\cite{ZanardiAnand2021}
\begin{equation}
C_\rho (x,y) = \Tr \bigl( \rho [\varepsilon^\dagger(V_x),W_y]^\dagger [\varepsilon^\dagger(V_x),W_y] \bigr).
\end{equation}
When the evolution channel is also unitary, this expression reduces to the conventional squared commutator OTOC. The OTOC $C_\rho (x,y)$ bounds the change in a probe expectation value at site $x$ caused by a perturbation at site $y$. Consider two $N$-qudit quantum circuits depicted in Fig.~\ref{fig:WV_protocol}(a) and (b). For (a), an initial state $\rho$ evolves under the channel $\varepsilon$, after which the observable $V_x$ is measured. The resulting expectation value is $m_A(x) = \Tr \Bigl( \rho \varepsilon^\dagger(V_x) \Bigr)$. Circuit (b) differs from (a) only by the insertion of the channel $\mathcal{W}_y$ immediately before the evolution channel. The expectation value of this measurement is $m_B(x,y) =\Tr \Bigl( \rho \mathcal{W}^\dagger_y(\varepsilon^\dagger(V_x)) \Bigr)$. The generalized OTOC gives the upper bound
\begin{equation}
\begin{aligned}
(m_B(x,y) - m_A(x))^2  
&= |\Tr ( \rho \Delta_{x,y})|^2 \\ 
&\leq \Tr(\rho \Delta^2_{x,y})= C_\rho(x,y).
\end{aligned}
\label{eq:otoc_bound}
\end{equation}

This follows from the Cauchy-Schwarz inequality for the Hilbert-Schmidt inner product, applied to $\sqrt{\rho}$ and $\sqrt{\rho} \Delta_{x,y}$. Moreover, when the perturbation channel is unital, the average of $(m_B(x,y) -m_A(x))^2$ over Haar-random pure input states satisfies
\begin{equation}
E_\psi[ (m_B(x,y) -m_A(x))^2] = \frac{C_{\rho_{\infty}}(x,y)}{q^N+1}  ,
\label{eq:average_response}
\end{equation}
where $\rho_{\infty}$ denotes the maximally mixed (infinite temperature) density operator 
\begin{equation}
\rho_{\infty} = q^{-N}I_{\mathcal{H}}.
\label{equation:infinite-temp}
\end{equation}
The average is over states $\ket{\psi} = U \ket{\psi_0}$, where $\ket{\psi_0}$ is any fixed pure state and $U$ is Haar random. This follows from the second Haar moment for random pure states and from the fact $\Tr(\rho_{\infty} \Delta_{x,y}) =0$  when $\mathcal{W}_y$ is unital~\cite{CollinsSniady2006Haar}. Equations~\eqref{eq:otoc_bound} and~\eqref{eq:average_response} provide model-independent operational interpretations of the generalized OTOC in Eq.~\eqref{generalized_otoc}. For many physical systems, we expect the spatial and temporal dependence of the non-unitary OTOC, like that of its unitary counterpart, to serve as a useful proxy for more direct diagnostics of sensitivity to local perturbations and information spreading. 
%In Appendix~[1], we test this expectation by comparing the generalized OTOC numerically with the worst-case response %$||\Delta_{x,y}||_\infty$, the diamond distance $|| \varepsilon \circ \mathcal{W}_y -\varepsilon||_{\diamond}$, and mutual information for %the circuits studied in this work~\cite{Swingle2018Unscrambling,AharonovKitaevNisan1998Diamond,NielsenChuang2010}.

In what follows, we fix the input state for the OTOC to be the infinite-temperature density operator. For any system operator $O$, let $||O||_2 = \sqrt{\Tr(O^\dagger O)}$ denote the Hilbert-Schmidt norm. The infinite-temperature OTOC can be expressed as 
\begin{equation}
C(x,y) = C_{\rho_{\infty}}(x,y) =  \frac{1}{q^N} ||\Delta_{x,y} ||^2_2.
\end{equation}

In this work, we compute the ensemble-averaged OTOC for a monitored random circuit. The circuit consists of brickwork layers of two-qudit gates interspersed with local projective mid-circuit measurements. Each gate is drawn independently from the Haar measure. The unitary layers alternate between even and odd bonds. Even layers act on the pairs of sites $(0,1),(2,3),\ldots,(N-2,N-1)$. Under open boundary conditions ($\mathrm{OBC}$), odd layers act on sites $(1,2),(3,4), \ldots,(N-3,N-2)$; under periodic boundary conditions ($\mathrm{PBC}$), they additionally include the pair $(N-1,0)$.

Between successive unitary layers, each site is independently measured with probability $p$, so measurement events are not correlated between sites and layers. Measurements are performed in the eigenbasis of the qudit clock operator $Z$, a generalization of the Pauli $Z$ operator:
\begin{equation}
Z = \sum_{n=0}^{q-1} \omega^n \ket{n} \bra{n}, \qquad \omega = e^{2 \pi i/q}.
\end{equation}
We label the circuit depth by the integer time $t$. We take $t=0$ to denote the time immediately after the first, even unitary layer and before the first measurement layer. Subsequent values of $t$ correspond to the analogous time immediately after each successive unitary layer. Thus, even and odd values of $t$ correspond to even and odd unitary layers, respectively.

We calculate the OTOC averaged over realizations of the random circuit. Let $U$ collectively denote the realization of the Haar-random gates, $\Gamma$ the spacetime pattern of measurement locations, and $\mu_\Gamma$ the corresponding measurement outcomes. We write $\varepsilon^{(U,\Gamma,\mu_\Gamma)}_t$ for the trajectory-resolved evolution through time $t$, and $C[\varepsilon]$ for the OTOC associated with the evolution map $\varepsilon$. Because $C[\varepsilon]$ is nonlinear in $\varepsilon$, averaging the OTOC and evaluating the OTOC of the averaged evolution are generally not equivalent. Throughout the remainder of the paper, we assume that the measurement outcomes have been pre-averaged, yielding an effective dephasing channel. Appendix~\ref{tm-derivation} shows the relation, with the dependence on the measurement probability explicit:
\begin{multline}
\mathbb{E}_{U} \bigl[ C \bigl[ \mathbb{E}_{\Gamma,\mu_\Gamma}\bigl[ \varepsilon^{(U,\Gamma,\mu_\Gamma)}_t \bigr] \bigr] \bigr](p)\\
= \mathbb{E}_{U,\Gamma} \bigl[ C \bigl[ \mathbb{E}_{\mu_\Gamma}\bigl[ \varepsilon^{(U,\Gamma,\mu_\Gamma)}_t \bigr] \bigr] \bigr](p(2-p)).
\label{corr_vs_uncorr}
\end{multline}
Henceforth, $\overline{C}$ denotes 
\begin{equation}
\overline{C} = \mathbb{E}_{U,\Gamma} \bigl[ C \bigl[ \mathbb{E}_{\mu_\Gamma}\bigl[ \varepsilon^{(U,\Gamma,\mu_\Gamma)}_t \bigr] \bigr] \bigr]
\end{equation}
and we suppress the labels $(U,\Gamma,\mu_\Gamma)$ on the evolution channel.

In Appendix~\ref{app:perturbation-dependence}, we show that the dependence of the average OTOC on the perturbation channel factorizes into a channel dependent amplitude and a universal spatiotemporal profile. Let
\begin{equation}
W^{(r,s)} = X^r Z^s \text{, } \qquad (r,s)\in \mathbb{Z}^2_q \setminus  \{ (0,0)\}
\end{equation}
be any nonidentity Weyl operator, and let $\mathcal{U}^{(r,s)}(O) = W^{(r,s)} O (W^{(r,s)})^\dagger$ denote the corresponding unitary perturbation channel acting on a system operator $O$. For an arbitrary single-qudit perturbation channel $\mathcal{W}$,
\begin{equation}
\overline{C}[\mathcal{W}] (x,y,t) = \mathcal{A}(\mathcal{W}) \overline{C} [\mathcal{U}^{(r,s)}] (x,y,t),
\label{eq:pert-dependence}
\end{equation}
where the real amplitude satisfies $0\leq \mathcal{A} \leq 1$. Any single-qudit channel $\Phi$ can be expressed in the Weyl basis 
\begin{equation}
\Phi(O) = \sum_{r,s,t,u =0}^{q-1} \chi_{r,s,t,u}(\Phi) W^{(r,s)} O W^{(t,u) \dagger},
\end{equation}
where the channel dependent rank-four tensor $\chi$ is
\begin{equation}
\chi_{r,s,t,u} (\Phi) = \frac{1}{q^3} \sum_{a,b=0}^{q-1} \Tr \Bigl[
W^{(a,b)} \Phi \bigl(W^{(r,s)\dagger} W^{(a,b)\dagger} W^{(t,u)}\bigr) \Bigr].
\end{equation}
We can treat the tensor $\chi_{r,s,t,u}$ as a $q^2 \times q^2$ matrix and define its Hilbert-Schmidt norm by
\begin{equation}
|| \chi||_2 = \sqrt{\sum_{r,s,t,u \in \mathbb{Z}_q} |\chi_{r,s,t,u}|^2}.\end{equation}
The amplitude is then given by
\begin{equation}
\mathcal{A}(\mathcal{W}) = \frac{1}{2} ||\chi(\mathcal{W}) - \chi(\mathcal{I}_{q})||^2_2.
\end{equation}
Thus, the amplitude $\mathcal{A}(\mathcal{W})$ directly quantifies the deviation of the perturbation channel $\mathcal{W}$ from the identity channel $\mathcal{I}_q$. In particular, $\mathcal{W}$ may itself be the dephasing channel describing an unread projective measurement, in which case the OTOC quantifies the many-body backaction of a single mid-circuit measurement; the universal profile is unchanged and only the amplitude $\mathcal{A}(\mathcal{W})$ is modified. The average OTOC can be expressed as
\begin{equation}
\overline{C}(x,y,t) = \mathcal{A}(\mathcal{W}) \Bigl(\overline{R}(x,t) - \overline{Q}(x,y,t) \Bigr),
\label{equation:OTOC}
\end{equation}
where $R$ and $Q$ are defined by
\begin{equation}
\begin{aligned}
&R(x,t) = \frac{2}{q^N}\Tr \Bigl( \varepsilon^\dagger_t(V_x) \varepsilon^\dagger_t(V_x) \Bigr),\\
&Q(x,y,t) = \frac{2}{q^N}\Tr \Bigl( \varepsilon^\dagger_t (V_x)  W^{\dagger (r,s)}_y \varepsilon^\dagger_t(V_x) W^{(r,s)}_y \Bigr).
\end{aligned}
\end{equation}
Henceforth, we fix $\mathcal{A}=1$.

In Appendix~\ref{tm-derivation}, we show that the ensemble-averaged OTOC can be calculated using an auxiliary transfer matrix problem. Define the real vector space $\mathcal{V}_N = (\mathbb{R}^2)^{\otimes N}$, with orthonormal basis 
\begin{equation}
\{\ket{\boldsymbol{s}}:\boldsymbol{s} \in \{0,1\}^N \},\qquad \ket{\boldsymbol{s}} = \ket{s_0}\otimes \ldots \otimes \ket{s_{N-1}}.
\end{equation}
In the lexicographically ordered computational basis $\ket{00}, \ket{01},\ket{10},\ket{11}$, define the local transfer matrix
\begin{equation}
\begin{gathered}
M(p_0,p_1) = \begin{pmatrix}
1 & a(p_1) & a(p_0) & c(p_0, p_1) \\
0 & 0 & 0 & 0 \\
0 & 0 & 0 & 0 \\
0 & b(p_1) & b(p_0) & d(p_0,p_1)
\end{pmatrix},
\\[1.25ex]
\begin{aligned}
a(p) &= \frac{q(q+1) + p }{q(q+1)(q^2+1)},\\
b(p) &= \frac{q^2( q+1 -q p) } {(q+1)(q^2+1)},\\
c(p_0,p_1) &= \frac{q(p_0 + p_1) -(q-1) p_0 p_1 }{q^2(q+1)(q^2+1)},\\
d(p_0,p_1) &= 1 - \frac{q^3( p_0 + p_1) -q^2(q-1) p_0 p_1 }{(q+1)(q^2+1)}.
\end{aligned}
\end{gathered}
\label{eq:tm}
\end{equation}
When both arguments are equal, we use the shorthand $M(p)\equiv M(p,p)$, $c(p)\equiv c(p,p)$, and $d(p)\equiv d(p,p)$.

Let $S$ denote the one-site shift operator
\begin{equation}
S \ket{s_0,s_1,\ldots,s_{N-1}} = \ket{s_{N-1}, s_0, \ldots, s_{N-2}}.
\label{eq:shift_operator}
\end{equation}
For $\mathrm{PBC}$, the single-layer transfer matrix $T^{(t)}$ is defined by
\begin{equation}
T^{(t)}_{\mathrm{PBC}} = \begin{cases}
M(p)^{\otimes N/2}, & t\equiv 0 \pmod 2, \\
S M(p)^{\otimes N/2} S^{-1} & t \equiv 1 \pmod 2.
\end{cases}
\end{equation}
For $\mathrm{OBC}$, define $p_e=p(2-p)$ and abbreviate
\begin{equation}
M=M(p),\qquad M_L=M(p_e,p),\qquad M_R =M(p,p_e).
\end{equation}
The single-layer transfer matrix is
\begin{equation}
T^{(t)}_{\mathrm{OBC}} =
\begin{cases}
M_L \otimes M^{\otimes N/2-2} \otimes M_R &t \equiv 0 \pmod 2,\\
I_2 \otimes M^{\otimes N/2-1} \otimes I_2 & t \equiv 1 \pmod 2.
\end{cases}
\end{equation}
Let $l(x,t) = x -\frac{1}{2}[ 1-(-1)^{x+t}] $ and define the column vector\footnote{Under $\mathrm{OBC}$, if $x=0$ or $x=N-1$ is unpaired in the terminal odd layer, the local channel can be absorbed into an effective probe operator. The same transfer-matrix construction then applies with the effective probe inserted at time $t-1$, modifying only the overall prefactor. We do not consider the OTOC at these boundary sites and times.} 
\begin{equation}
\ket{v_0(x,t)} =S^{l(x,t)} \bigl(\ket{1}^{\otimes 2} \otimes \ket{0}^{\otimes N-2} \bigr).
\end{equation}
The ensemble-averaged OTOC is given by
\begin{equation}
\begin{aligned}
\overline{C}(x,y,t) &= C_0 \sum_{\substack{\mathbf{s} \in \{0,1\}^N \\ s_y=1}}\bra{\mathbf{s}} T^{(0)} T^{(1)} \ldots T^{(t-1)} \ket{v_0(x,t)},
\\[0.75ex]
C_0 &= \frac{2q^4}{q^4-1} \text{Var}_{I_q/q} (V),
\\[0.75ex]
\text{Var}_{I_q/q}(V) &= \Tr(V^2/q) -\Tr(V/q)^2.
\end{aligned}
\label{eq:brickwall_otoc}
\end{equation}
Henceforth, we normalize $V$ such that $\text{Var}_{I_q/q} (V) =1$. Although $p$ was introduced as a measurement probability, Appendix~\ref{tm-derivation} shows that Eq.~\eqref{eq:brickwall_otoc} remains valid for independently sampled unital single-qudit channels, with $p$ reinterpreted as their mean effective channel parameter in the range $p \in [0,\frac{q+1}{q}]$.\footnote{For general unital channels under $\mathrm{OBC}$, the boundary matrices involve the effective parameters of the corresponding composed channels, which are not generally determined by $p$ alone. The relation $p_{e}=p(2-p)$ applies to the measurement-induced circuit considered here.}

At $p=0$ the circuit is unitary and $\overline{C}$ reduces to the OTOC studied in Ref.~\cite{Nahum2018OperatorSpreading}. In this limit, every column of the local transfer matrix sums to one, so its coefficients can be interpreted as transition probabilities. Moreover, the local transfer matrix leaves the uniform two-site configurations invariant
\begin{equation}
M(0) \ket{00} = \ket{00} \qquad M(0) \ket{11} = \ket{11}.
\end{equation}
Every state evolving from $\ket{v_0(x,t)}$ can be expressed as a linear combination of basis states that are either uniform or contain a single contiguous block of ones surrounded by zeros. This evolution is therefore completely characterized by the two domain walls bounding this block. These domain walls perform discrete random walks and annihilate upon meeting. The OTOC $\overline{C}(x,y,t)$ is proportional to the probability that site $y$ lies between the two domain walls. This probability was calculated in Ref.~\cite{Nahum2018OperatorSpreading} using the method of images \cite{Fisher1984}. The resulting OTOC front propagates ballistically with butterfly velocity
\begin{equation}
v_B = \frac{q^2-1}{q^2+1}.
\end{equation}
Consequently, the OTOC vanishes outside the circuit's causal light cone and rises toward a plateau as the ballistic front passes the perturbation site.

\begin{figure}[t]
    \centering
    \includegraphics[width=1.0\linewidth]{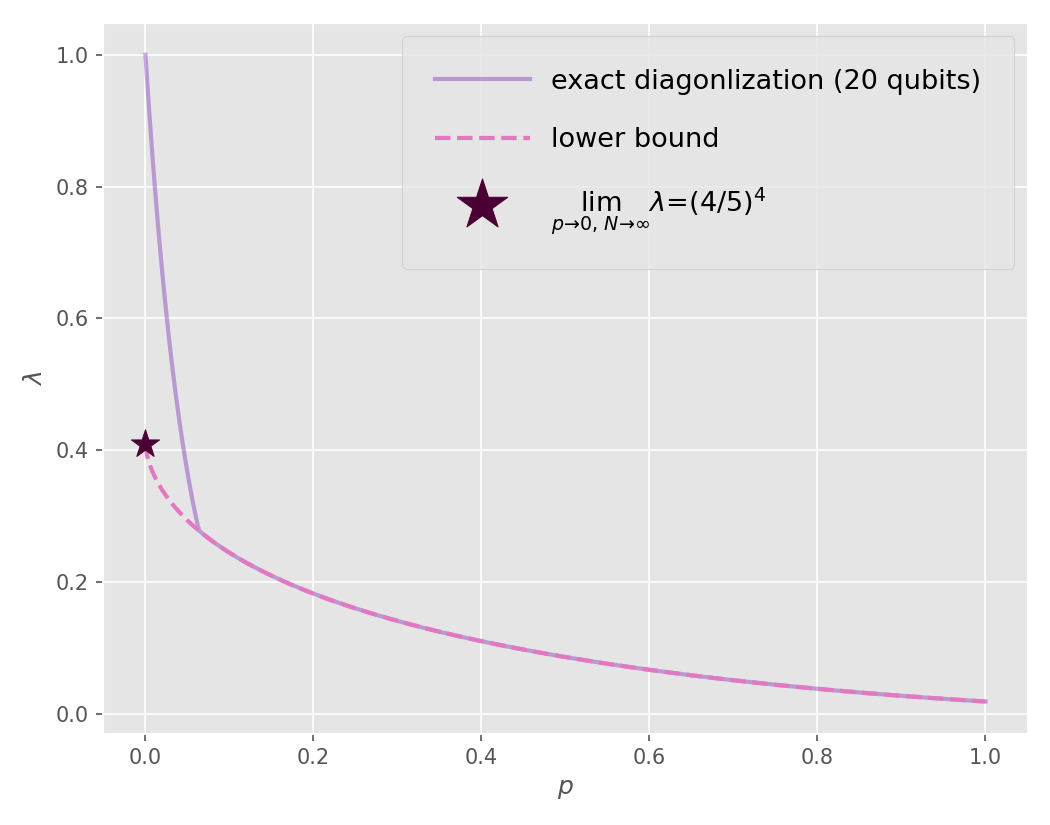}
    \caption{The transfer matrix eigenvalue $\lambda_*$  as a function of the measurement probability $p$ obtained by exact diagonalization of the 20-qubit transfer matrix, together with a lower bound on $\lambda_*$ that is valid in the thermodynamic limit. The marked point indicates the limiting value $\lim_{p \rightarrow 0^+} \lim_{N \rightarrow \infty} \lambda_*(p) = (4/5)^4$.}
    \label{fig:eigenvalue}
\end{figure}

For $p>0$, not all columns of the local transfer matrix sum to one. Instead, the coefficients satisfy 
\begin{equation}
c(p) + d(p)  = [a(p)+b(p)]^2 < 1.
\end{equation}
Moreover, $c(p)>0$ permits the nucleation of additional domain wall pairs, so the dynamics no longer preserve the single domain wall pair sector. After sufficient time has passed, the OTOC decays to zero. To characterize this decay, define the two-layer transfer matrix
\begin{equation}
T = T^{(0)} T^{(1)}.
\end{equation}
At even circuit depths,
\begin{equation}
\overline{C}(x,y,2t) = C_0 \sum_{\substack{\mathbf{s} \in \{0,1 \}^N}\\s_y=1}\bra{\mathbf{s}} T^{t} \ket{v_0(x,0)}.
\end{equation}
In Appendix~\ref{app:real_eigvals}, we show the eigenvalues of $T$ are real and lie in the interval $[0,1]$. At $p=0$, the eigenvalue $1$ is two-fold degenerate, with eigenvectors $\ket{0}^{\otimes N}$ and $\ket{1}^{\otimes N}$. The vacuum eigenvector $\ket{0}^{\otimes N}$ does not contribute to the OTOC because it is orthogonal to the row vector in Eq.~\eqref{eq:brickwall_otoc}, which has support only on states satisfying $s_y=1$. For $p>0$, the vacuum eigenvalue remains $1$, while every other eigenvalue contributing to the OTOC is strictly less than $1$. Let $\lambda_*$ denote the largest such eigenvalue. The long time behavior is then
\begin{equation}
\overline{C}(x,y,2t) \sim B(x,y,p) \lambda^t_*(p),
\end{equation}
where $B(x,y,p)$ represents the overlap between the terminal row vector in Eq.~\eqref{eq:brickwall_otoc} and the eigenvector corresponding to $\lambda_*$. In Appendix~\ref{app:perturbation-theory}, we use first-order perturbation theory to obtain
\begin{equation}
\lambda_{\text{*}} = 1- \Bigl( \frac{2 (q-1)}{q} \Bigr) \Bigl( \frac{1}{1-q^{-2N}} \Bigr) Np +O(p^2)
\end{equation}
valid when $Np \ll 1$. The limits $N\rightarrow \infty$ and $p \rightarrow 0^+$ do not commute. This is similar to spontaneous breaking of unitarity introduced in Refs.~\cite{AltlandSonner2021,Liao2022}. In Appendix~\ref{app:discontinuity}, we show that for $\mathrm{PBC}$,\footnote{Strictly, Appendix~\ref{app:discontinuity} establishes the corresponding statements for the large-$N$ limit inferior and limit superior.}
\begin{equation}
\lim_{p \rightarrow 0^+}\lim_{N\rightarrow \infty} \lambda_{*} =\Bigl( \frac{2q}{q^2+1} \Bigr)^4,
\label{eq:eigenvalue_limit}
\end{equation}
whereas, for either boundary condition,
\begin{equation}
 \lim_{N\rightarrow \infty} \lim_{p \rightarrow 0^+} \lambda_{*} = 1.
\end{equation}
For $\mathrm{OBC}$, the limiting $\mathrm{PBC}$ value in Eq.~\eqref{eq:eigenvalue_limit} also governs the bulk decay when the perturbation is located a macroscopic distance from either boundary. For every fixed $t$, its causal cone is then disjoint from the boundaries for sufficiently large $N$, so the $\mathrm{OBC}$ and $\mathrm{PBC}$ OTOCs coincide exactly; taking $N\rightarrow\infty$ before $t\rightarrow\infty$ therefore gives the same bulk decay rate. For $\mathrm{PBC}$, Fig.~\ref{fig:eigenvalue} compares $\lambda_*(p)$ obtained by exact diagonalization of the transfer matrix for $N=20$ qubits, together with a variational lower bound in the thermodynamic limit and the limiting value Eq.~\eqref{eq:eigenvalue_limit}. The construction of the variational bound is discussed in Appendix~\ref{app:lower_bound}.

Direct contraction of the transfer matrices becomes computationally prohibitive at large $N$ because the auxiliary space has dimension $2^N$. We instead develop a Monte Carlo method for estimating the OTOC.  For further details of this Monte Carlo simulation, see Appendix~\ref{app:monte-carlo}. Let 
\begin{equation}
\gamma(p) = a(p) +b(p) = 1-\frac{q-1}{q}p.
\end{equation}
Using $c(p)+d(p) = \gamma^2$, we normalize the columns of the local transfer matrix to obtain
\begin{equation}
\widehat{M}(p) = \begin{pmatrix}
1 & a/\gamma & a/\gamma & c/\gamma^2 \\
0 & 0 & 0 & 0 \\
0 & 0 & 0 & 0 \\
0 & b/\gamma & b/\gamma & d/\gamma^2
\end{pmatrix},
\end{equation}
where $a$, $b$, $c$, and $d$ are evaluated at $p$. \footnote{Under $\mathrm{OBC}$, this bulk construction applies while the geometric light cone remains disjoint from the system boundaries, as in the calculations that follow.} Every column of $\widehat{M}(p)$ sums to one, so the corresponding brickwork circuit defines a stochastic process on $N$-bit configurations. For fixed $x$ and $t$, the process begins in the configuration associated with $\ket{v_0(x,t)}$, whose only nonzero entries are at sites $l(x,t)$ and $l(x,t)+1$. A trajectory is a sequence $\boldsymbol{z} = (\boldsymbol{z^{(0)}}, \ldots, \boldsymbol{z}^{(t)})$, where $\boldsymbol{z}^{(n)} \in \{0,1\}^N$ and $z^{(n)}_j$ denotes the bit at site $j$ and time $n$. The trajectory probabilities are generated by applying $\widehat {M}(p)$ in a brickwork pattern.

The average OTOC is the weighted trajectory expectation
\begin{equation}
\overline{C} (x,y,t) = C_0 \mathbb{E} \bigl[ \gamma^{A_t(\boldsymbol{z})} z^{(t)}_y \bigr].
\end{equation}
The binary variable $z^{(t)}_{y}$ indicates whether the site $y$ is occupied at time $t$. The quantity
\begin{equation}
A_t(\boldsymbol{z}) = \sum_{n=0}^{t-1} \sum_{j=0}^{N-1} z^{(n)}_j
\end{equation}
can be interpreted as the occupied spacetime area of the trajectory.

%Appendix~[10] describes our Monte Carlo estimator $\widehat{C} (x,y,t)$ and shows that, for $N_{\mathrm{samples}}$ independently sampled %trajectories
By sampling $N_{S}$ independent trajectories we compute a Monte Carlo estimate of the OTOC $\widehat{C}(x,y,t)$. The variance of the estimate satisfies
\begin{equation}
\text{Var} \bigl[ \widehat{C} (x,y,t) \bigr] \leq \frac{C^2_0}{4 N_{S}}.
\label{eq:monte-carlo-variance}
\end{equation}

\begin{figure*}[t]
    \centering
    \includegraphics[width=0.95\textwidth]{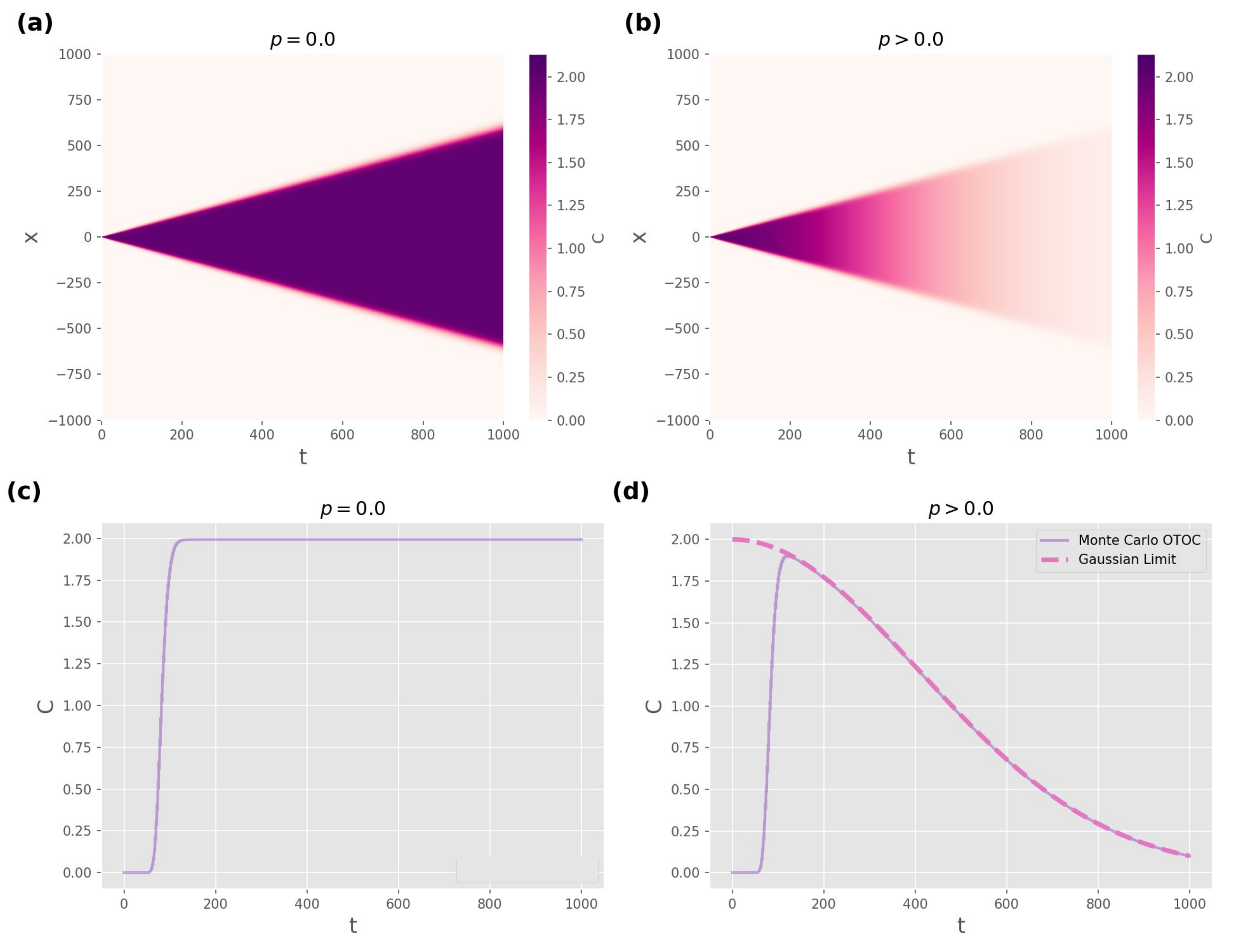}
    \caption{
    Monte Carlo estimates of the OTOC for measurement probabilities
    $p=0$ and $p=10^{-5}$, with the perturbation applied at $y=0$.
    (a,b) Full spatiotemporal profiles.
    (c,d) OTOC as a function of time at the fixed probe site $x=50$,
    along with the Gaussian limit in Eq.~\eqref{eq:gaussian}.
    The system size $N$ is chosen so that the geometric light cone
    does not reach the boundary during the time interval shown. For this figure, spatial coordinates are measured relative to the perturbation site, so that $y=0$.
    }
    \label{fig:monte_carlo}
\end{figure*}

Figure~\ref{fig:monte_carlo} shows the Monte Carlo estimates of the OTOC for $p=0$ and $p=10^{-5}$. The perturbation site is $y=0$, and $N$ is chosen sufficiently large that the geometric light cone of the circuit does not reach the system boundary within the time window shown. Panels (a,b) show the full spatiotemporal profiles, whereas panels (c,d) show the OTOC at a fixed probe site $x=50$. At each probe site, the OTOC rises to a maximum as the signal arrives and subsequently decays.

In the regime of large $N$ and small $p$ shown in Fig.~\ref{fig:monte_carlo}, the exponential decay determined by the leading transfer matrix eigenvalue describes only the tail at asymptotically late times, when the OTOC is already small at every site. It therefore does not determine the characteristic decay time of the signal. Instead, for sufficiently small $p$, the physically relevant decay is approximately Gaussian in time:
\begin{equation}
\begin{aligned}
\overline{C}(x,y,t) &\sim 2e^{- t^2/\sigma^2(p)},\\
\kappa &=\frac{(q^2-1)(q-1)}{q(q^2+1)}, \qquad \sigma^2(p) = \frac{1}{\kappa p}. 
\end{aligned}
\label{eq:gaussian}
\end{equation}

%More precisely, Appendix~[11] establishes that, for fixed $x$ and $y$ and for every $\tau >0$,
More precisely, Appendix~\ref{app:gaussian-limit} establishes that, for fixed $x$ and $y$ and for every $\tau > 0$,
\begin{equation}
\lim_{p \rightarrow 0^+} \lim_{N\rightarrow \infty} \overline{C}(x,y, \lfloor\tau p^{-1/2}\rfloor) = 2 e^{- \kappa \tau^2}.
\label{eq:gaussian_limit}
\end{equation}
The characteristic decay time scales as $p^{-1/2}$. To state this precisely, fix $\epsilon \in (0,2)$ and let 
\begin{equation}
t_*(\epsilon, p) = \sup \{ t \in \mathbb{N}: \overline{C}(x,y,t) >\epsilon \}.
\end{equation}
Then 
\begin{equation}
\lim_{p \rightarrow 0^{+}} t_*(\epsilon,p) \sqrt{p} = \sqrt{\frac{\ln(2/\epsilon)}{\kappa}}.
\end{equation}
Related Gaussian decay has been predicted for operator Loschmidt echoes in weakly dissipative many-body systems, based on operator growth~\cite{SchusterYao2023,YoshimuraSa2026}. In the limit of infinite qudit dimension, the OTOC is given by
\begin{equation}
\begin{aligned}
&\lim_{q \rightarrow \infty} \lim_{N \rightarrow \infty} \overline{C}(x,y,t)\\ 
&= \begin{cases}
2(1-p)^{t(t+1)},
& \text{if }l(y,0)-t \leq x \leq l(y,0)+1+t,  \\
0,
& \text{otherwise.}
\end{cases}
\end{aligned}
\end{equation}
The OTOC is zero outside the geometric light cone, which spreads at the maximal speed $v_B = 1$. Within it, the OTOC has no spatial dependence; measurements contribute only the spatially uniform decay factor $(1-p)^{t(t+1)}$. Because the decay is fixed analytically by $p$, a measured decay rate can be inverted to extract an effective measurement strength, which may include contributions from spectator dephasing and measurement-induced crosstalk that a nominal per-qubit readout error does not capture.

\begin{figure*}[t]
\centering
\resizebox{0.94\textwidth}{!}{%
\begin{quantikz}[row sep=0.28cm, column sep=0.28cm,
  execute at end picture={
    \coordinate (qkB) at ([yshift=-0.25cm]\tikzcdmatrixname.south);
    \draw[decorate,decoration={brace,mirror,amplitude=7pt},thick]
      (\tikzcdmatrixname-1-3.west |- qkB) -- (\tikzcdmatrixname-1-7.east |- qkB)
      node[midway,below=10pt,text height=2.2ex,text depth=0.6ex]{$\varepsilon_t^\dagger$\; (backward evolution)};
    \draw[decorate,decoration={brace,mirror,amplitude=7pt},thick]
      (\tikzcdmatrixname-1-11.west |- qkB) -- (\tikzcdmatrixname-1-15.east |- qkB)
      node[midway,below=10pt,text height=2.2ex,text depth=0.6ex]{$\varepsilon_t$\; (forward evolution)};
  }]
% q1
\lstick{$q_0$}
& \gate[style=pauligate]{I}\slice{}
& \gate[2,style=su4gate]{\scriptscriptstyle U\in SU(4)}
& \meter{}
& \qw
& \qw
& \gate[2,style=su4gate]{\scriptscriptstyle U\in SU(4)}\slice{}
& \qw
& \qw
& \qw\slice{}
& \gate[2,style=su4gate][1.3cm]{\scriptscriptstyle U^\dagger }
& \qw
& \qw
& \qw
& \gate[2,style=su4gate][1.3cm]{\scriptscriptstyle U^\dagger }
& \meter{}
\\
% q2
\lstick{$q_1$}
& \gate[style=pauligate]{X}
&
& \qw
& \gate[2,style=su4gate]{\scriptscriptstyle U\in SU(4)}
& \qw
&
& \qw
& \qw
& \qw
&
& \qw
& \gate[2,style=su4gate][1.3cm]{\scriptscriptstyle U^\dagger }
& \qw
&
& \qw
\\
% qp
\lstick{$q_y$}
& \gate[style=pauligate]{I}
& \gate[2,style=su4gate]{\scriptscriptstyle U\in SU(4)}
& \qw
&
& \qw
& \gate[2,style=su4gate]{\scriptscriptstyle U\in SU(4)}
& \qw
& \gate[style=perturb]{
    \begin{array}{c}
        X_y\\[-1pt]
        \scriptstyle t=0
    \end{array}
  }
& \qw
& \gate[2,style=su4gate][1.3cm]{\scriptscriptstyle U^\dagger }
& \qw
&
& \qw
& \gate[2,style=su4gate][1.3cm]{\scriptscriptstyle U^\dagger }
& \qw
\\
% q_{p+1}
\lstick{$q_{y+1}$}
& \gate[style=pauligate]{X}
&
& \qw
& \gate[2,style=su4gate]{\scriptscriptstyle U\in SU(4)}
& \qw
&
& \qw
& \qw
& \qw
&
& \meter{}
& \gate[2,style=su4gate][1.3cm]{\scriptscriptstyle U^\dagger }
& \qw
&
& \qw
\\
% q_{N-2}
\lstick{$q_{N-2}$}
& \gate[style=pauligate]{X}
& \gate[2,style=su4gate]{\scriptscriptstyle U\in SU(4)}
& \qw
&
& \meter{}
& \gate[2,style=su4gate]{\scriptscriptstyle U\in SU(4)}
& \qw
& \qw
& \qw
& \gate[2,style=su4gate][1.3cm]{\scriptscriptstyle U^\dagger }
& \qw
&
& \qw
& \gate[2,style=su4gate][1.3cm]{\scriptscriptstyle U^\dagger }
& \qw
\\
% qN
\lstick{$q_{N-1}$}
& \gate[style=pauligate]{I}
&
& \qw
& \qw
& \qw
&
& \qw
& \qw
& \qw
&
& \qw
& \qw
& \meter{}
&
& \qw
\end{quantikz}
}
\caption{
Example circuit realization. A randomly sampled Pauli-word
initialization is followed by alternating odd- and even-parity
brick-wall layers of Haar-random two-qubit gates
$U\in SU(4)$. The backward and forward branches,
$\varepsilon_t^\dagger$ and $\varepsilon_t$, are separated by the
local perturbation $X_y$ at $t=0$. Mid-circuit measurements $M$
occur at independently sampled spacetime locations, and the
probe qubit $q_0$ is measured at the output.
}
\label{fig:random_realization}
\end{figure*}
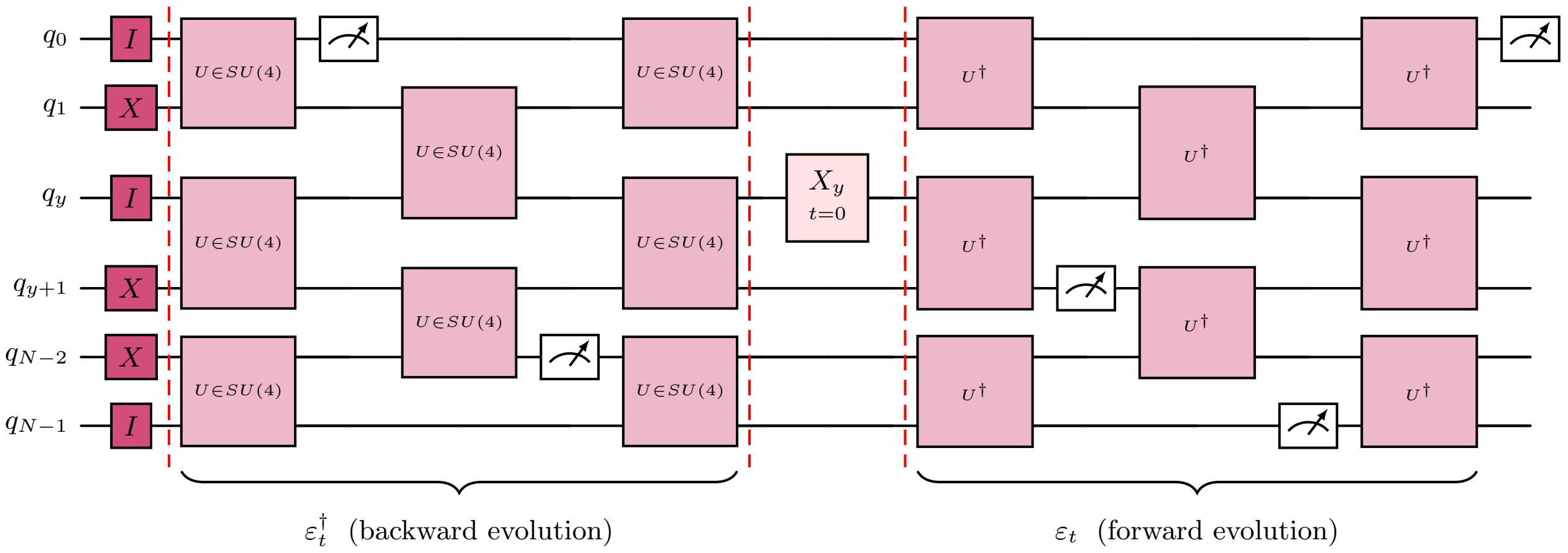

\section{Numerics}

\begin{figure}[t]
    \centering
    \begin{quantikz}[row sep=0.35cm, column sep=0.6cm]
        \lstick{\text{probe } $q_0$}
        & \gate[5,style=epsblock]{\varepsilon_t^\dagger}
        & \qw
        & \gate[5,style=epsblock]{\varepsilon_t}
        & \meter{}
        \\
        \lstick{$\vdots$}
        &
        &
        &
        &
        \\
        \lstick{$q_y$}
        &
        & \gate[style=pertblock]{\bowtie}
        &
        & \qw
        \\
        \lstick{$\vdots$}
        &
        &
        &
        &
        \\
        \lstick{$q_{N-1}$}
        &
        &
        &
        &
        \qw
    \end{quantikz}
    \caption{
        Schematic representation of the circuit protocol for a system
        of arbitrary size $N$. A local perturbation $X_y$ is inserted
        at site $y$ between the two many-body evolution channels
        $\varepsilon_t^\dagger$ and $\varepsilon_t$. The probe qubit
        $q_0$ is measured at the output.
    }
    \label{fig:butterfly_protocol}
\end{figure}
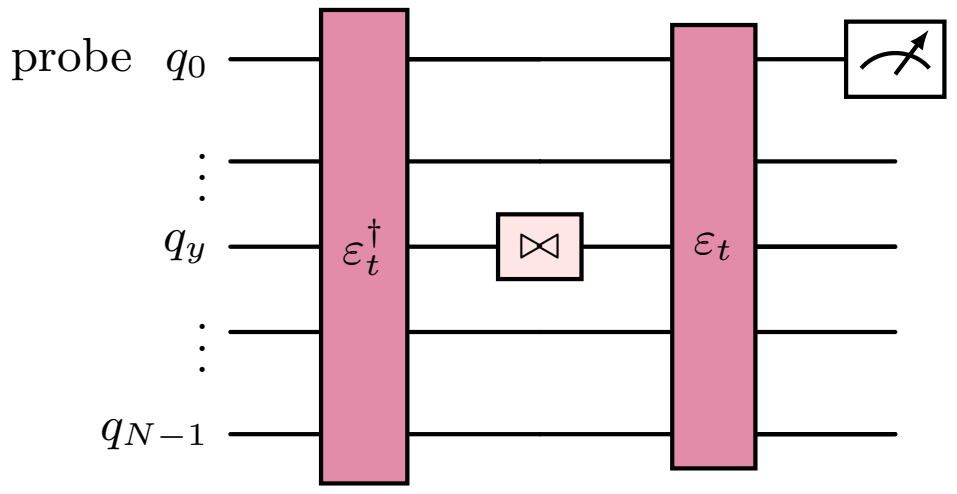

On top of Monte Carlo estimation we directly  evaluate the OTOC-quantity defined in Eq.~\eqref{equation:OTOC} using quantum circuits, both in noiseless and noisy numerical simulation and on a trapped-ion quantum processor. Hardware experiments are performed on Quantinuum’s H2 system, a quantum charge-coupled device (QCCD) architecture in which trapped ions are transported between interaction and storage regions. The device employs a linear trap with periodic boundary conditions, producing the characteristic ``race-track’’ geometry. Dynamical ion transport allows arbitrary pairs of qubits to be brought into common interaction regions and thereby realizes effective all-to-all logical connectivity without requiring a fixed nearest-neighbor interaction graph \cite{Moses_2023}. 
For local, noiseless simulations, we employ the Qiskit Aer backend. Noisy simulations are performed using a custom noise model whose parameters are summarized in Table~\ref{tab:noise_parameters}. These parameters are based on the reported specifications of the Quantinuum H2-2 emulator, supplemented by estimates where the corresponding quantities are not explicitly specified~\cite{QuantinuumH2Emulator}.  Circuits are constructed and submitted to the respective computational backends using the \texttt{pytket} API.

The quantity of interest can be decomposed according to the algebraic identity
\begin{equation}
C(t)=R(t)-Q(t),
\label{eq:C_QR}
\end{equation}
where
\begin{align}
Q(t)
&=
\frac{2}{d}
\operatorname{Tr}
\left[
\varepsilon_t^\dagger(\sigma_x^\nu)
\sigma_y^\mu
\varepsilon_t^\dagger(\sigma_x^\nu)
\sigma_y^\mu
\right],
\label{eq:Q_def}
\\
R(t)
&=
\frac{2}{d}
\operatorname{Tr}
\left[
\varepsilon_t^\dagger(\sigma_x^\nu)
\varepsilon_t^\dagger(\sigma_x^\nu)
\right].
\label{eq:R_def}
\end{align}
Here $d=2^N$ denotes the Hilbert-space dimension for an $N$-qubit system, $\sigma_y^\mu$ is a local Pauli perturbation acting on site $y$, $\sigma_x^\nu$ is the operator probed at site $x$, and $\varepsilon_t^\dagger$ denotes the Heisenberg-picture quantum channel describing evolution to time $t$. 

The normalized traces appearing in Eqs.~\eqref{eq:Q_def} and~\eqref{eq:R_def} may be written as expectation values with respect to the infinite-temperature density operator Eq.~\eqref{equation:infinite-temp}. Rather than preparing $\rho_\infty$ directly, we estimate the trace stochastically by averaging over a set of randomly prepared computational-basis product states. Starting from
\begin{equation}
|0\rangle^{\otimes N},
\end{equation}
we independently apply either $I$ or $X$ to each qubit with equal probability. A Pauli word
\begin{equation}
P_{\mathbf s}
=
\bigotimes_{j=0}^{N-1} X_j^{s_j},
\qquad
s_j\in \{0,1\},
\end{equation}
therefore prepares the computational-basis state
\begin{equation}
|\mathbf s\rangle
=
P_{\mathbf s}|0\rangle^{\otimes N}.
\end{equation}
Uniform averaging over all $2^N$ such bit strings gives exactly
\begin{equation}
\frac{1}{2^N}
\sum_{\mathbf s\in \{0,1\}^N}
|\mathbf s\rangle\langle \mathbf s|
=
\frac{\mathbb I}{2^N}
=
\rho_\infty.
\label{eq:computational_twirl}
\end{equation}
Consequently, finite random sampling of these states provides a Monte Carlo estimator of the infinite-temperature trace. Importantly, no sampling over $Y$ or $Z$ is required for this purpose: sampling $I$ and $X$ independently on each qubit is equivalent to uniformly sampling the complete computational basis.

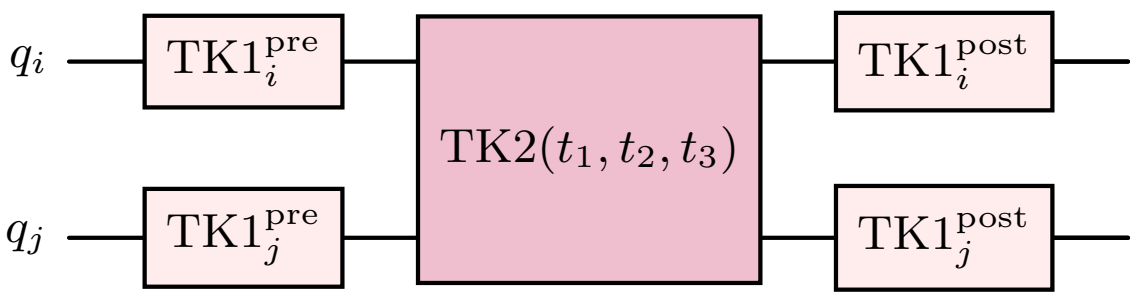
\begin{figure}[t]
    \centering
    \begin{quantikz}
        \lstick{$q_i$}
        & \gate[style=tkone]{\mathrm{TK1}_i^{\mathrm{pre}}}
        & \gate[2,style=tktwo]{\mathrm{TK2}(t_1,t_2,t_3)}
        & \gate[style=tkone]{\mathrm{TK1}_i^{\mathrm{post}}}
        & \qw
        \\
        \lstick{$q_j$}
        & \gate[style=tkone]{\mathrm{TK1}_j^{\mathrm{pre}}}
        &
        & \gate[style=tkone]{\mathrm{TK1}_j^{\mathrm{post}}}
        & \qw
    \end{quantikz}
    \caption{
        Decomposition of a Haar-random two-qubit unitary
        $U_{ij}\in SU(4)$ into local TK1 rotations and a nonlocal
        TK2 entangling gate.
    }
    \label{fig:su4_decomposition}
\end{figure}

In addition to the initial-state sampling, each circuit realization contains independently sampled two-qubit unitaries. The local two-qubit gates are drawn from the Haar measure on $SU(4)$. An arbitrary $U \in SU(4)$ admits a Cartan, or KAK, decomposition of the form
\begin{equation}
U
=
(k_1\otimes k_2)
e^{\left[
-\frac{i}{2}
\left(
\theta_x X\otimes X
+
\theta_y Y\otimes Y
\\
+
\theta_z Z\otimes Z
\right)
\right]}
(k_3\otimes k_4),
\label{eq:KAK}
\end{equation}
where $k_i\in SU(2)$. In the \texttt{pytket} convention, arbitrary one-qubit operations are represented by $\mathrm{TK1}$ gates,
\begin{equation}
\mathrm{TK1}(\alpha,\beta,\gamma)
=
R_z(\pi \alpha)R_x(\pi \beta)R_z(\pi \gamma),
\end{equation}
while the nonlocal Cartan component is represented by the $\mathrm{TK2}$ gate,
\begin{equation}
\mathrm{TK2}(a,b,c)
=
\exp\left[
-\frac{i\pi}{2}
\left(
a X\otimes X
+
b Y\otimes Y
+
c Z\otimes Z
\right)
\right].
\label{eq:TK2}
\end{equation}
Thus a Haar-distributed two-qubit unitary may be compiled exactly into single-qubit rotations surrounding a TK2 entangling operation, as shown in Fig.~\ref{fig:su4_decomposition}. The distribution of the Cartan parameters and accompanying local rotations must be chosen according to the Haar measure on $SU(4)$. Details of the sampling procedure and its implementation in the TK1–TK2 gate set are given in Appendix~\ref{app:cartan-sampling}. Finally, in the non-unitary case, where $p >0$, the measurement sites must be sampled randomly. Each site is treated as a Bernoulli random variable with probability $p$ of being measured. With respect to site-sampling, the forward and backward channels are treated \textit{independently}, although the simple algebraic relation in Eq.~\eqref{corr_vs_uncorr} relates the independent and forward-backward correlated sampling results. 

Each choice of initial state, random gates/angles, and measured sites defines a circuit realization $r$. The circuit which implements \textit{a} realization is depicted in Fig.~\ref{fig:random_realization}. For an observable $C_r(t)$, the Haar-ensemble expectation value is estimated from $N_{\mathrm r}$ independently sampled realizations as
% \begin{align}
% \overline{Q}(t)
% &=
% \frac{1}{N_{\mathrm r}}
% \sum_{r=1}^{N_{\mathrm r}}
% Q_r(t),
% \
% \overline{R}(t)=
% \frac{1}{N_{\mathrm r}}
% \sum_{r=1}^{N_{\mathrm r}}
% R_r(t),
% \end{align}
% hence
\begin{equation}
\langle C(t)\rangle_{\mathrm{Haar}}
\simeq
\frac{1}{N_{\mathrm r}}
\sum_{r=1}^{N_{\mathrm r}} C_r(t),
\end{equation}
with equality recovered in the limit $N_{\mathrm r}\rightarrow\infty$. Thus, while each two-qubit gate is sampled directly from the Haar distribution, the finite collection of circuit realizations provides a Monte Carlo approximation to the full Haar-ensemble average. Using Eq.~\ref{eq:C_QR}, for each realization we execute a pair of circuits constructed from identical random data: one containing the perturbation $X_y$, from which we estimate $Q_r(t)$, and one without the perturbation, from which we estimate $R_r(t)$. Pairing the two circuits in this way ensures that their difference is evaluated for the same realization of the random circuit. To characterize the time evolution of the operator, we construct an independent ensemble of circuit realizations at each time step. Thus, the data at time $\tau$ are not obtained by evolving the realizations sampled at $\tau-1$ forward by an additional time step. Rather, a new ensemble of realizations is independently sampled for each value of $\tau$. The initialization of computational-basis states, the sampling of Haar-random unitaries, and selection of measurement sites are controlled by a classical pseudorandom number generator (PRNG); details of the seeding procedure and reproducibility protocol are provided in Appendix~\ref{app:reproducibility-seeding}.

The probe and perturbation operators are chosen to be $\sigma^\nu_x  =Z_x$ and $\sigma^\mu_y =X_y$. We encode $|\uparrow\rangle\equiv|0\rangle$ and $|\downarrow\rangle\equiv|1\rangle$, with measurement outcomes assigned as $0\mapsto1$ and $1\mapsto-1$. If an $X$ was applied to the probe qubit during initialization, we reverse the sign assigned to its final measurement outcome. This correction is applied in both the perturbed and unperturbed circuits. The difference between the mean corrected outcomes without and with the perturbation estimates $C(t)/2$.

\begin{figure}[t]
    \centering
    \includegraphics[width=1.0\linewidth]{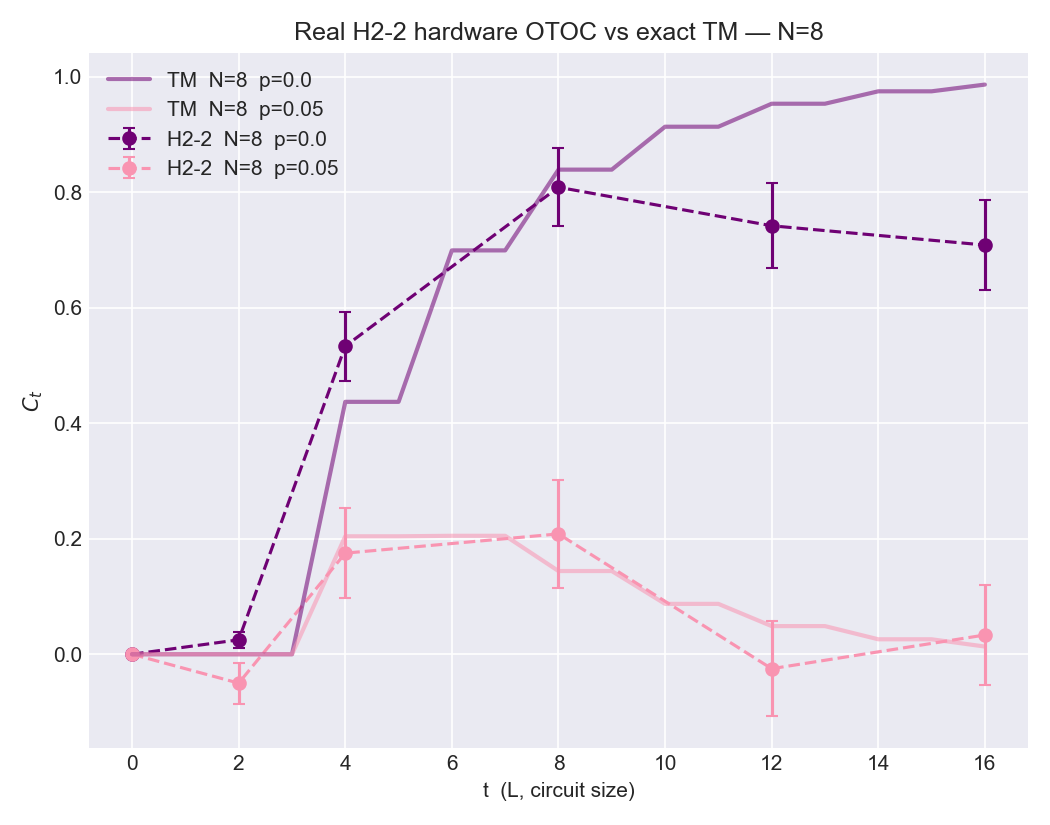}
    \caption{Averaged OTOC $\langle C(t)\rangle_{\mathrm{Haar}}/2$ as a function of time $t$ at a single site. The probe site is the uppermost qubit $x=0$, and the perturbation is applied to the fifth qubit $y=4$.  }
    \label{fig:single_site}
\end{figure}

\begin{figure*}[t]
    \centering
    % ============================================================
    % (a) SIMULATION
    % ============================================================
    \begin{minipage}{0.92\textwidth}
        \centering
        \includegraphics[width=\textwidth]
            {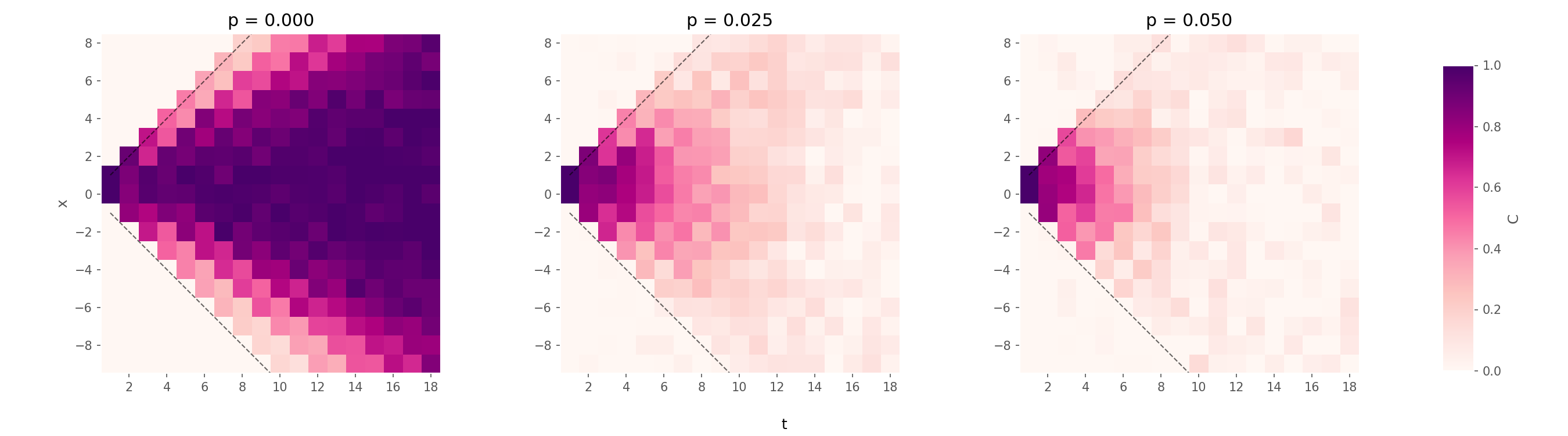}
        \vspace{0.25em}
        \raggedright
        {\Large\bfseries (a)\hspace{0.5em} SIMULATION}
    \end{minipage}
    \vspace{0.8em}

    % ============================================================
    % (b) EXPERIMENT
    % ============================================================
    \begin{minipage}{0.92\textwidth}
        \centering
        \includegraphics[width=\textwidth]
            {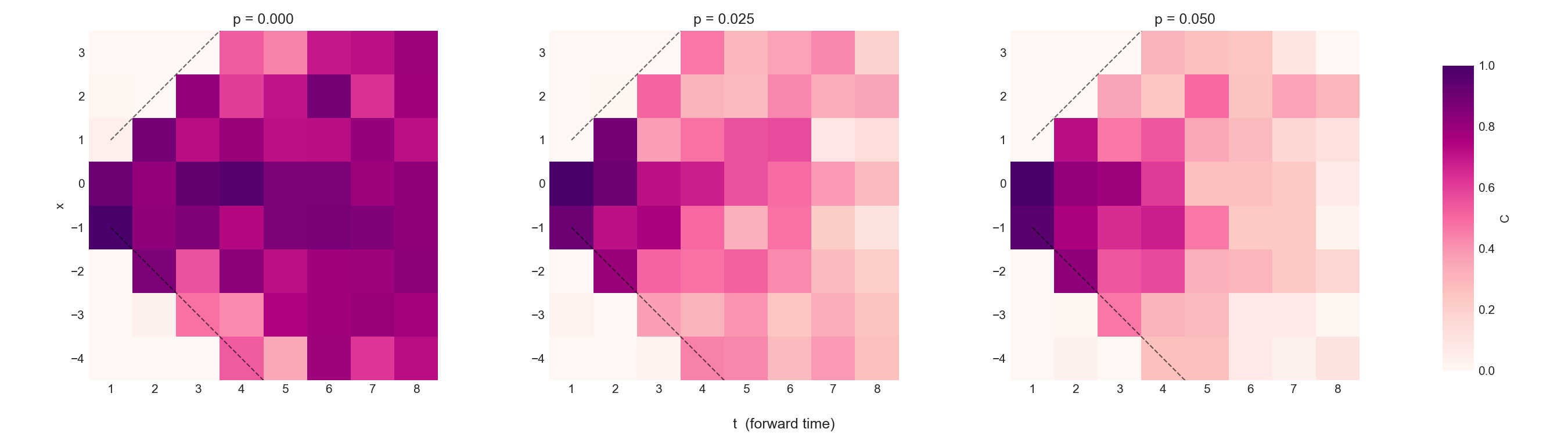}
        \vspace{0.25em}
        \raggedright
        {\Large\bfseries (b)\hspace{0.5em} EXPERIMENT}
    \end{minipage}

    \caption{
    OTOC light-cone dynamics.
    \textbf{(a)} Numerical simulation of the local circuit model.
    \textbf{(b)} Corresponding experimental results obtained on
    quantum hardware.
    }
    \label{fig:lightcones}
\end{figure*}
Owing to the computational cost and limited availability of quantum-hardware resources, hardware experiments were restricted to systems of $N=8$ qubits. This system size does not represent an intrinsic limitation of the circuit protocol. Consequently, the sampling protocol admits a natural extension to larger systems, including regimes in which direct classical simulation of the corresponding quantum dynamics becomes computationally intractable. Furthermore, since each realization is independent, the task is embarrassingly parallelizable. We made use of a tiling algorithm to execute multiple realizations as one circuit job. All the relevant code can be found here \cite{SchmetterlingEffect}. A notable advantage over previous approaches to measuring OTOCs~\cite{GoogleQuantumAI2025Constructive} is the simplicity of the present protocol, which requires no ancillary qubits or interferometric measurements and can be implemented directly using local perturbations and measurements.

% In particular, the numerical results presented in Appendix~\ref{app:self-averaging} provide evidence that the Haar-averaged OTOC exhibits self-averaging behavior, such that the number of independent circuit realizations required to obtain a fixed statistical precision need not grow exponentially with system size. 

\begin{table}[t]
    \centering
    \caption{Noise parameters used in the noisy-circuit simulations.
    The gate error parameters $p_{1\mathrm{Q}}$ and $p_{2\mathrm{Q}}$
    denote the single- and two-qubit depolarizing error probabilities,
    respectively. The readout error $p_{\mathrm{M}}$ is taken as the
    mean of the reported state-dependent measurement error probabilities.
    No explicit $T_1$ or $T_2$ relaxation times are included.}
    \label{tab:noise_parameters}
    \begin{ruledtabular}
    \begin{tabular}{lcc}
        Parameter & Description & Value \\
        \hline
        $p_{1\mathrm{Q}}$
            & Single-qubit depolarizing error
            & $7.3\times10^{-5}$ \\
        $p_{2\mathrm{Q}}$
            & Two-qubit depolarizing error
            & $1.29\times10^{-3}$ \\
        $p_{\mathrm{M}}$
            & Readout bit-flip probability
            & $1.35\times10^{-3}$ \\
        $T_1$
            & Longitudinal relaxation time
            & -- \\
        $T_2$
            & Transverse relaxation time
            & -- \\
        $t_{1\mathrm{Q}}$
            & Single-qubit gate duration
            & $50\,\mathrm{ns}$ \\
        $t_{2\mathrm{Q}}$
            & Two-qubit gate duration
            & $300\,\mathrm{ns}$ \\
    \end{tabular}
    \end{ruledtabular}
\end{table}

\begin{table}[t]
    \centering
    \caption{
        Parameters used for the numerical and hardware calculations.
        $N$ denotes the number of qubits, $L$ the circuit length, and
        $N_{\mathrm r}$ the number of random circuit realizations.
    }
    \label{tab:run_parameters}
    \begin{ruledtabular}
    \begin{tabular}{lcc}
        Parameter
        & Heat map
        & Single site
        \\
        \hline
        Number of qubits, $N$
        & $8$
        & $8$
        \\
        Circuit length, $L$
        & $8$
        & $16$
        \\
        Realizations, $N_{\mathrm r}$
        & $180$ per $L$
        & $240$ per $L$
        \\
        Perturbation
        & $X$
        & $X$
        \\
        PRNG seed
        & $1962$
        & $1962$
        \\
        Hardware system
        & H2
        & H2
    \end{tabular}
    \end{ruledtabular}
\end{table}

Results of experiments are shown in Figs.~\ref{fig:single_site} and~\ref{fig:lightcones}. Because of the analytical treatment via transfer matrix, we can compare the experimental results with exact predictions. Figure~\ref{fig:lightcones}(a) shows the OTOC dynamics for an open system of size ($N=18$). These are ``ideal'' numerics on which to benchmark our quantum hardware results. In the unitary limit, ($p=0$), the OTOC exhibits a well-defined ballistic light cone at early times. The perturbation front propagates approximately linearly, $x_{\mathrm f}(t)\sim v_B t$, demonstrating ballistic operator spreading with butterfly velocity $v_B$. This propagating front defines a sharp causal boundary: outside the light cone $C(x,t)\simeq0$, while within it the OTOC approaches its saturated value, $C(x,t)/2\to1$. Near the light-cone boundary, however, saturation is delayed over a finite spatial region, indicating broadening of the scrambling front about its ballistically propagating center. At late times the front reaches the physical boundaries of the system, where finite-size effects become important. The characteristic phenomenology of quantum scrambling is manifest: the perturbation remains causally confined within the light cone, while at late times its information becomes delocalized across the many-body system and is no longer locally accessible.

Increasing the measurement probability qualitatively modifies the dynamics lagging the scrambling front while leaving a detectable causal light-cone structure at early times. For $p=0.025$, the perturbation continues to spread quasi-ballistically, but the OTOC is retarded and no longer approaches a unitary saturation value. This effect becomes pronounced for $p=0.05$, where the perturbation initially spreads over several sites but the OTOC subsequently decays throughout the interior of the nominal light cone. Mid-circuit measurement separates the spread of the causal front from the persistence of the scrambled signal: information about the perturbation can propagate linearly while its OTOC signature is simultaneously erased by the non-unitary dynamics.

In the unitary limit ($p=0$), the most notable deviation occurs at later times, where the inter-site signal is suppressed even though the theory predicts causal saturation, $C_t/2\to 1$. This discrepancy is consistent with hardware noise, which may be modeled phenomenologically through a time-dependent effective measurement parameter, $p\to p(t)$. In the non-unitary case, the data remain broadly consistent with the analytical prediction: the OTOC reaches its maximum near $t=4$ and subsequently decays as expected. These results suggest that the non-unitary OTOC provides a nontrivial benchmark of the effects of mid-circuit measurement on many-body quantum dynamics. Two features make it useful as a hardware diagnostic: (1) the light-cone region supplies an assumption-light null test, since any signal at causally disconnected sites cannot be scrambling and may be attributed to crosstalk or correlated error; and (2) the decay within the light cone is sensitive to spectator dephasing, leakage, and imperfect reset, the failure modes that most directly degrade feedforward and syndrome-extraction cycles.

Figure~\ref{fig:lightcones}(b) shows the corresponding light-cone calculation performed on quantum hardware but for a smaller system. In the unitary limit, $p=0$, hardware noise and finite sampling do not obscure the central phenomenology: a causal light-cone structure remains clearly resolved. Notably, the measured OTOC remains strongly suppressed outside the expected light cone, indicating that hardware cross-talk and other errors do not generate an appreciable signal at causally disconnected sites. Compared with the noiseless result, however, the propagating front is substantially less sharply resolved: fluctuations obscure the precise onset of the front, while the OTOC within the light cone is systematically attenuated and fails to exhibit the same clear saturation. With the present data, these deviations cannot be cleanly separated into finite-sampling fluctuations and hardware-induced errors. Both effects become increasingly pronounced for $p>0$, where measurement-induced attenuation accentuates the degradation already present in the unitary hardware data.

\section{Conclusion}

In this work, we calculated the ensemble-averaged OTOC for Haar-random qudit circuits subjected
to interspersed local projective measurements. This calculation maps to a transfer matrix problem. We show the OTOC decays exponentially at late times and characterize how the decay rate depends on the measurement probability. The OTOC is estimated numerically using a Monte Carlo procedure. In the thermodynamic limit with sparse measurement, the OTOC decay profile is approximately Gaussian.

Using a transfer-matrix
formulation, we obtained a number of analytical results and derived explicit
forms of the OTOC in several limiting regimes.

These predictions were tested against Monte Carlo simulations and direct quantum-circuit estimation, providing
complementary numerical and experimental probes of the monitored dynamics.

We further developed a simple and transparent protocol for estimating the OTOC
directly on quantum hardware. The protocol imposes minimal requirements on the
device, requiring neither ancillary qubits nor interferometric measurements, and
our hardware results indicate that the characteristic scrambling dynamics remain
detectable in the presence of device noise and finite-sampling
fluctuations. In principle, the protocol can be extended to other echo-dynamics, as well as system sizes inaccessible to  classical simulation, providing
a path toward studying scrambling in monitored many-body systems for which
analytical and classical numerical approaches become infeasible. 

Beyond its interest for quantum chaos, the monitored OTOC addresses a practical need. Architectures that use measurement as a computational nonlinearity, such as measurement-induced neural networks \cite{minn}, depend on precisely the quantity computed here: the spatiotemporal extent over which a measurement's influence persists sets the effective receptive field and memory depth of such a network. The minimal, ancilla-free protocol demonstrated on H2 therefore provides a characterization step that can be carried directly to larger and higher-fidelity systems such as Quantinuum's Helios, where these architectures become viable at scale.

Several natural extensions remain.  The local transfer matrix derived in Appendix~\ref{tm-derivation} is valid when mid-circuit measurements are replaced by arbitrary single-qudit channels. When the local non-unitary channels are identical, the OTOC is parameterized by two real numbers, one of which is nontrivial only for non-unital channels.\footnote{Additional parameters may be needed for $\mathrm{OBC}$ to define the local transfer matrices at the boundary.} The local non-unitary channels can also vary in space and time. The method used in Appendix~\ref{tm-derivation} naturally extends to other circuit geometries composed of Haar-random gates interspersed with non-unitary channels, such as generalizations to higher dimensional spatial lattices. It will also be important to establish the relationship between the behavior of the monitored OTOC and entanglement dynamics, and to determine whether the fluctuations of the scrambling front admit universal scaling behavior or well-defined universality classes.

\begin{acknowledgements}
  The authors acknowledge sponsorship by the U.S. Air Force SEQCURE2 program at the University of Maryland's Applied Research Laboratory for Intelligence and Security. The work of V.G. is also based on research sponsored by the Air Force Office of Scientific Research under agreement number FA9550-26-1-B279; the U.S. Government is authorized to reproduce and distribute reprints for Governmental purposes notwithstanding any copyright notation thereon. We thank Quantinuum and, in particular, David Hayes for providing access to the quantum computing hardware used in this work.
\end{acknowledgements}

\bibliographystyle{apsrev4-2}
\bibliography{paper}

@article{LarkinOvchinnikov1969,
    author  = {Larkin, A. I. and Ovchinnikov, Yu. N.},
    title   = {Quasiclassical Method in the Theory of Superconductivity},
    journal = {Soviet Physics JETP},
    volume  = {28},
    number  = {6},
    pages   = {1200--1205},
    year    = {1969}
}

@article{Liao2018,
  author    = {Liao, Yunxiang and Galitski, Victor},
  title     = {Nonlinear sigma model approach to many-body quantum chaos: Regularized and unregularized out-of-time-ordered correlators},
  journal   = {Phys. Rev. B},
  volume    = {98},
  number    = {20},
  pages     = {205124},
  year      = {2018},
  month     = nov,
  publisher = {American Physical Society},
  doi       = {10.1103/PhysRevB.98.205124},
  url       = {https://doi.org/10.1103/PhysRevB.98.205124}
}

@misc{Kitaev2014Hidden,
    author       = {Kitaev, Alexei},
    title        = {Hidden Correlations in the Hawking Radiation and Thermal Noise},
    howpublished = {Talk at the Fundamental Physics Prize Symposium},
    year         = {2014},
    note         = {Stanford, November 10, 2014}
}

@article{MaldacenaShenkerStanford2016,
    author  = {Maldacena, Juan and Shenker, Stephen H. and Stanford, Douglas},
    title   = {A Bound on Chaos},
    journal = {Journal of High Energy Physics},
    volume  = {2016},
    number  = {8},
    pages   = {106},
    year    = {2016},
    doi     = {10.1007/JHEP08(2016)106},
    eprint  = {1503.01409},
    archivePrefix = {arXiv},
    primaryClass  = {hep-th}
}

@misc{yao2016interferometricapproachprobingfast,
      title={Interferometric Approach to Probing Fast Scrambling}, 
      author={Norman Y. Yao and Fabian Grusdt and Brian Swingle and Mikhail D. Lukin and Dan M. Stamper-Kurn and Joel E. Moore and Eugene A. Demler},
      year={2016},
      eprint={1607.01801},
      archivePrefix={arXiv},
      primaryClass={quant-ph},
      url={https://arxiv.org/abs/1607.01801}, 
}

@article{Swingle_2016,
   title={Measuring the scrambling of quantum information},
   volume={94},
   ISSN={2469-9934},
   url={http://dx.doi.org/10.1103/PhysRevA.94.040302},
   DOI={10.1103/physreva.94.040302},
   number={4},
   journal={Physical Review A},
   publisher={American Physical Society (APS)},
   author={Swingle, Brian and Bentsen, Gregory and Schleier-Smith, Monika and Hayden, Patrick},
   year={2016},
   month=Oct }

@article{Gaerttner2017OTOC,
    author  = {G{\"a}rttner, Martin and Bohnet, Justin G. and Safavi-Naini, Arghavan and Wall, Michael L. and Bollinger, John J. and Rey, Ana Maria},
    title   = {Measuring Out-of-Time-Order Correlations and Multiple Quantum Spectra in a Trapped-Ion Quantum Magnet},
    journal = {Nature Physics},
    volume  = {13},
    pages   = {781--786},
    year    = {2017},
    doi     = {10.1038/nphys4119}
}

@article{PhysRevX.8.021013,
  title = {Operator Hydrodynamics, OTOCs, and Entanglement Growth in Systems without Conservation Laws},
  author = {von Keyserlingk, C. W. and Rakovszky, Tibor and Pollmann, Frank and Sondhi, S. L.},
  journal = {Phys. Rev. X},
  volume = {8},
  issue = {2},
  pages = {021013},
  numpages = {19},
  year = {2018},
  month = {Apr},
  publisher = {American Physical Society},
  doi = {10.1103/PhysRevX.8.021013},
  url = {https://link.aps.org/doi/10.1103/PhysRevX.8.021013}
}

@article{Mi2021InformationScrambling,
    author  = {Mi, Xiao and Ippoliti, Matteo and Quintana, Chris

               and Greene, Amy and Chen, Zijun and Gross, Jonathan

               and Arute, Frank and Arya, Kunal and Atalaya, Juan

               and Babbush, Ryan and others},

    title   = {Information Scrambling in Quantum Circuits},
    journal = {Science},
    volume  = {374},
    number  = {6574},
    pages   = {1479--1483},
    year    = {2021},
    doi     = {10.1126/science.abg5029}

}

@article{PhysRevX.7.031016,
  title = {Quantum Entanglement Growth under Random Unitary Dynamics},
  author = {Nahum, Adam and Ruhman, Jonathan and Vijay, Sagar and Haah, Jeongwan},
  journal = {Phys. Rev. X},
  volume = {7},
  issue = {3},
  pages = {031016},
  numpages = {30},
  year = {2017},
  month = {Jul},
  publisher = {American Physical Society},
  doi = {10.1103/PhysRevX.7.031016},
  url = {https://link.aps.org/doi/10.1103/PhysRevX.7.031016}
}

@misc{jonay2018coarsegraineddynamicsoperatorstate,
      title={Coarse-grained dynamics of operator and state entanglement}, 
      author={Cheryne Jonay and David A. Huse and Adam Nahum},
      year={2018},
      eprint={1803.00089},
      archivePrefix={arXiv},
      primaryClass={cond-mat.stat-mech},
      url={https://arxiv.org/abs/1803.00089}, 
}

@article{PhysRevX.9.031009,
  title = {Measurement-Induced Phase Transitions in the Dynamics of Entanglement},
  author = {Skinner, Brian and Ruhman, Jonathan and Nahum, Adam},
  journal = {Phys. Rev. X},
  volume = {9},
  issue = {3},
  pages = {031009},
  numpages = {21},
  year = {2019},
  month = {Jul},
  publisher = {American Physical Society},
  doi = {10.1103/PhysRevX.9.031009},
  url = {https://link.aps.org/doi/10.1103/PhysRevX.9.031009}
}

@article{Moses_2023,
   title={A Race-Track Trapped-Ion Quantum Processor},
   volume={13},
   ISSN={2160-3308},
   url={http://dx.doi.org/10.1103/PhysRevX.13.041052},
   DOI={10.1103/physrevx.13.041052},
   number={4},
   journal={Physical Review X},
   publisher={American Physical Society (APS)},
   author={Moses, S. A. and Baldwin, C. H. and Allman, M. S. and Ancona, R. and Ascarrunz, L. and Barnes, C. and Bartolotta, J. and Bjork, B. and Blanchard, P. and Bohn, M. and Bohnet, J. G. and Brown, N. C. and Burdick, N. Q. and Burton, W. C. and Campbell, S. L. and Campora, J. P. and Carron, C. and Chambers, J. and Chan, J. W. and Chen, Y. H. and Chernoguzov, A. and Chertkov, E. and Colina, J. and Curtis, J. P. and Daniel, R. and DeCross, M. and Deen, D. and Delaney, C. and Dreiling, J. M. and Ertsgaard, C. T. and Esposito, J. and Estey, B. and Fabrikant, M. and Figgatt, C. and Foltz, C. and Foss-Feig, M. and Francois, D. and Gaebler, J. P. and Gatterman, T. M. and Gilbreth, C. N. and Giles, J. and Glynn, E. and Hall, A. and Hankin, A. M. and Hansen, A. and Hayes, D. and Higashi, B. and Hoffman, I. M. and Horning, B. and Hout, J. J. and Jacobs, R. and Johansen, J. and Jones, L. and Karcz, J. and Klein, T. and Lauria, P. and Lee, P. and Liefer, D. and Lu, S. T. and Lucchetti, D. and Lytle, C. and Malm, A. and Matheny, M. and Mathewson, B. and Mayer, K. and Miller, D. B. and Mills, M. and Neyenhuis, B. and Nugent, L. and Olson, S. and Parks, J. and Price, G. N. and Price, Z. and Pugh, M. and Ransford, A. and Reed, A. P. and Roman, C. and Rowe, M. and Ryan-Anderson, C. and Sanders, S. and Sedlacek, J. and Shevchuk, P. and Siegfried, P. and Skripka, T. and Spaun, B. and Sprenkle, R. T. and Stutz, R. P. and Swallows, M. and Tobey, R. I. and Tran, A. and Tran, T. and Vogt, E. and Volin, C. and Walker, J. and Zolot, A. M. and Pino, J. M.},
   year={2023},
   month=Dec }

@misc{SchmetterlingEffect,
  author       = {{Galitski Group}},
  title        = {Schmetterling Effect: Code for Measurement-Induced Butterfly Effect},
  year         = {2026},
  howpublished = {\url{https://github.com/Galitski-group/schmetterling-effect}},
  note         = {GitHub repository},
}

@misc{QuantinuumH2Emulator,
  author       = {{Quantinuum}},
  title        = {System Model H2 Emulators},
  howpublished = {Quantinuum Systems User Guide},
  year         = {2025},
  url          = {https://docs.quantinuum.com/systems/user_guide/emulator_user_guide/emulators/h2_emulators.html},
  note         = {Accessed: 2026-08-27}
}

@article{GoogleQuantumAI2025Constructive,
  author  = {{Google Quantum AI and Collaborators}},
  title   = {Observation of constructive interference at the edge of quantum ergodicity},
  journal = {Nature},
  volume  = {646},
  pages   = {825--830},
  year    = {2025},
  doi     = {10.1038/s41586-025-09526-6}
}

@article{ZanardiAnand2021,
  author  = {Zanardi, Paolo and Anand, Namit},
  title   = {Information scrambling and chaos in open quantum systems},
  journal = {Physical Review A},
  volume  = {103},
  pages   = {062214},
  year    = {2021},
  doi     = {10.1103/PhysRevA.103.062214}
}

@article{Nahum2018OperatorSpreading,
  author  = {Nahum, Adam and Vijay, Sagar and Haah, Jeongwan},
  title   = {Operator Spreading in Random Unitary Circuits},
  journal = {Physical Review X},
  volume  = {8},
  pages   = {021014},
  year    = {2018},
  doi     = {10.1103/PhysRevX.8.021014}
}

@article{CollinsSniady2006Haar,
  author  = {Collins, Beno{\^{i}}t and {\'S}niady, Piotr},
  title   = {Integration with Respect to the Haar Measure on Unitary, Orthogonal and Symplectic Group},
  journal = {Communications in Mathematical Physics},
  volume  = {264},
  pages   = {773--795},
  year    = {2006},
  doi     = {10.1007/s00220-006-1554-3}
}

@article{Swingle2018Unscrambling,
  author  = {Swingle, Brian},
  title   = {Unscrambling the Physics of Out-of-Time-Order Correlators},
  journal = {Nature Physics},
  volume  = {14},
  pages   = {988--990},
  year    = {2018},
  doi     = {10.1038/s41567-018-0295-5}
}

@article{Fisher1984,
  author  = {Fisher, Michael E.},
  title   = {Walks, Walls, Wetting, and Melting},
  journal = {Journal of Statistical Physics},
  volume  = {34},
  number  = {5--6},
  pages   = {667--729},
  year    = {1984},
  doi     = {10.1007/BF01009436}
}

@article{minn,
  title         = {Measurement-Induced Quantum Neural Network},
  author        = {Argyle, Paul and Lakhdar-Hamina, Djamil and Miller, Sarah H. and Galitski, Victor},
  journal       = {arXiv preprint arXiv:2603.19200},
  year          = {2026},
  eprint        = {2603.19200},
  archivePrefix = {arXiv},
  primaryClass  = {quant-ph},
  doi           = {10.48550/arXiv.2603.19200},
  note          = {arXiv:2603.19200 [quant-ph]}
}

@book{Watrous2018,
  author    = {Watrous, John},
  title     = {The Theory of Quantum Information},
  publisher = {Cambridge University Press},
  year      = {2018},
  doi       = {10.1017/9781316848142}
}

@article{ChuangNielsen1997,
  author  = {Chuang, Isaac L. and Nielsen, Michael A.},
  title   = {Prescription for Experimental Determination of the Dynamics of a Quantum Black Box},
  journal = {Journal of Modern Optics},
  volume  = {44},
  number  = {11--12},
  pages   = {2455--2467},
  year    = {1997},
  doi     = {10.1080/09500349708231894}
}

@book{HornJohnson2012,
  author    = {Roger A. Horn and Charles R. Johnson},
  title     = {Matrix Analysis},
  edition   = {2},
  publisher = {Cambridge University Press},
  address   = {Cambridge},
  year      = {2012},
  doi       = {10.1017/CBO9781139020411}
}

@book{GrinsteadSnell1997,
  author    = {Charles M. Grinstead and J. Laurie Snell},
  title     = {Introduction to Probability},
  edition   = {2},
  publisher = {American Mathematical Society},
  address   = {Providence, RI},
  year      = {1997},
  isbn      = {9780821807491}
}

@article{SchusterYao2023,
  author    = {Thomas Schuster and Norman Y. Yao},
  title     = {Operator Growth in Open Quantum Systems},
  journal   = {Physical Review Letters},
  volume    = {131},
  pages     = {160402},
  year      = {2023},
  doi       = {10.1103/PhysRevLett.131.160402}
}

@article{YoshimuraSa2026,
  author    = {Takato Yoshimura and Lucas S{\'a}},
  title     = {Dynamics of {Loschmidt} echoes from operator growth in noisy quantum many-body systems},
  journal   = {Physical Review B},
  volume    = {113},
  pages     = {144306},
  year      = {2026},
  doi       = {10.1103/fmr4-14vd}
}

@article{Liao2022,
  author  = {Liao, Yunxiang and Galitski, Victor},
  title   = {Emergence of many-body quantum chaos via spontaneous breaking of unitarity},
  journal = {Phys. Rev. B},
  volume  = {105},
  number  = {14},
  pages   = {L140202},
  year    = {2022},
  month   = apr,
  doi     = {10.1103/PhysRevB.105.L140202},
  eprint  = {2104.05721},
  archivePrefix = {arXiv}
}

@article{AltlandSonner2021,
  author  = {Altland, Alexander and Sonner, Julian},
  title   = {Late time physics of holographic quantum chaos},
  journal = {SciPost Phys.},
  volume  = {11},
  number  = {2},
  pages   = {034},
  year    = {2021},
  month   = aug,
  doi     = {10.21468/SciPostPhys.11.2.034},
  eprint  = {2008.02271},
  archivePrefix = {arXiv}
}

% ============================================================
% SUPPLEMENTAL MATERIAL

% ============================================================
\clearpage
\onecolumngrid
\appendix
\section{Perturbation Channel Dependence}
\label{app:perturbation-dependence}

\begin{figure}[H]
  \centering
\begin{quantikz}
\lstick{$q_0$}     &           &\gate[7]{\mathcal{F}}  &\\
\lstick{$q_1$}     &           &                       &\\
\lstick{$\vdots$}  &           &                       &\\
\lstick{$q_{y}$}   &\gate[2]{U}&                       &\\
\lstick{$q_{y+1}$} &           &                       &\\
\lstick{$\vdots$}  &           &                       &\\
\lstick{$q_{N-1}$} &           &                       &
\end{quantikz}
  \caption{ A Haar-random gate is first applied to a subset of qudit sites containing the perturbation site $y$. Irrespective of the subsequent evolution channel $\mathcal{F}$, averaging the OTOC over this initial gate removes all dependence on the perturbation $\mathcal{W}$, except for an overall multiplicative constant.  }
  \label{fig:pert-dependence}
\end{figure}
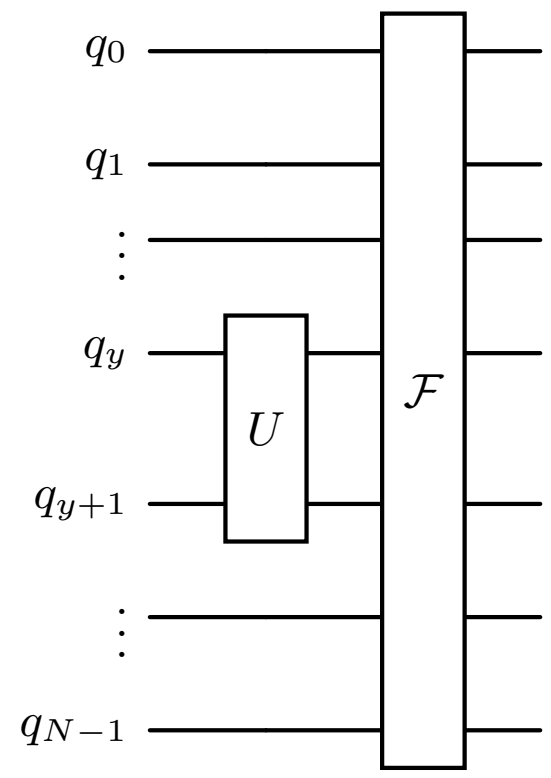
In this section, we derive Eq.~\eqref{eq:pert-dependence} of the main text, which states that the Haar-averaged OTOC depends on the perturbation channel only through an overall constant amplitude. We demonstrate this for a broader class of circuits that includes the measurement-induced circuit discussed in the main text as an example. For any positive integer $n$, let $\mathcal{H}_n = (\mathbb{C}^q)^{\otimes n}$ denote the Hilbert space of $n$ qudits with local Hilbert-space dimension $q$. For a given Hilbert space $\mathcal{H}_n$, we denote the associated space of linear operators by $\mathcal{L}(\mathcal{H}_n)$. The full system has Hilbert space $\mathcal{H}_N$, which we abbreviate to $\mathcal{H}$ when the ambient space is unambiguous. We assume $N$ is even. The qudits are labeled by sites $j =0,1,\ldots,N-1$. As in the main text, we define the single-qudit Hermitian operator $V$ and the single-qudit channel $\mathcal{W}$. Denote their embeddings at sites $x$ and $y$ by $V_x = I^{\otimes x}_q \otimes V \otimes I^{\otimes N-x-1}_q $ and $\mathcal{W}_y = \mathcal{I}_q^{\otimes y} \otimes \mathcal{W} \otimes \mathcal{I}_q^{\otimes N-y-1}$, respectively. We use $I_d$ and $\mathcal{I}_d$ to denote the $d$-dimensional identity operator and channel, respectively. 

Let $U_{d}$ denote a $d \times d$  Haar-random unitary matrix. Let $K$ be an integer satisfying $1 \leq K \leq N$. Let $\pi \in S_N$ be a permutation of the $N$ qudit sites such that the image of site $y$ lies among the first $K$ sites, i.e., $0 \leq \pi(y) < K$. Let $P_\pi$ denote the corresponding permutation operator defined by
\begin{equation}
P_{\pi} \ket{s_0,\ldots,s_{N-1}} = \ket{s_{\pi^{-1}(0)}, \ldots, s_{\pi^{-1}(N-1)}}.
\end{equation}
Define $U^{(\pi)}$ by
\begin{equation}
U^{(\pi)} = P^{\dagger}_{\pi}(U_{q^K} \otimes I^{\otimes N-K}_q) P_{\pi}.
\label{eq:embedded-unitary}
\end{equation}
Let $\mathcal{F}$ be a channel acting on the $N$-qudit Hilbert space. For a given realization of the Haar-random unitary $U_{q^K}$, define the channel $\varepsilon^{(U)}$ by
\begin{equation}
\varepsilon^{(U)} (O) = \mathcal{F} (U^{(\pi) } O U^{(\pi) \dagger}).
\label{eq:generic-channel}
\end{equation}
An example of the channel $\varepsilon^{(U)}$ is shown in Fig.~\ref{fig:pert-dependence}. The Haar-random unitary is first applied to a subset of sites that includes the perturbation site $y$. The system then evolves under the channel $\mathcal{F}$. In this section, we show that, regardless of the channel $\mathcal{F}$ governing the subsequent evolution, the Haar-averaged OTOC is independent of the details of the perturbation channel, except for an overall constant that quantifies its strength.
The Haar-averaged OTOC with respect to the infinite-temperature density operator is
\begin{equation}
 \overline{C}(x,y) = \frac{1}{q^N} \mathbb{E}_U \bigl[ || \mathcal{W}^\dagger_y (\varepsilon^{(U)\dagger}(V_x) ) -\varepsilon^{(U) \dagger}(V_x)||^2_2 \bigr],
\end{equation} 
where $|| O ||_2 =\sqrt{\Tr(O^\dagger O)}$ denotes the Hilbert-Schmidt norm and $\mathcal{W}^\dagger_y$ and $\varepsilon^{(U) \dagger}$ are the adjoint maps of $\mathcal{W}_y$ and $\varepsilon^{(U)}$, respectively. We expand this as
\begin{equation}
\begin{aligned}
\overline{C}(x,y) ={}& \frac{1}{q^N} \mathbb{E}_U \Bigl[ \Tr \bigl( \mathcal{W}^\dagger_y( \varepsilon^{(U) \dagger}(V_x)) \mathcal{W}^\dagger_y( \varepsilon^{(U) \dagger}(V_x)) \bigr) \Bigr]\\ &- \frac{2}{q^N} \mathbb{E}_U \Bigl[ \Tr \bigl( \mathcal{W}^\dagger_y( \varepsilon^{(U) \dagger}(V_x))  \varepsilon^{(U) \dagger}(V_x) \bigr) \Bigr]
+\frac{1}{q^N} \mathbb{E}_U \Bigl[ \Tr \bigl(  \varepsilon^{(U) \dagger}(V_x)  \varepsilon^{(U) \dagger}(V_x) \bigr) \Bigr].
\end{aligned}
\label{eq:expanded_OTOC}
\end{equation}

Let $X$ and $Z$ denote the $q$-dimensional qudit shift and clock operators, respectively, whose action on the computational basis is given by
\begin{equation}
X \ket{j} = \ket{j+1 \bmod q}, \qquad Z \ket{j} = \omega^j \ket{j}, \qquad \omega = e^{2 \pi i/q}.
\end{equation}
The Weyl operators are defined by 
\begin{equation}
W^{(r,s)} = X^r Z^s, \qquad r,s \in \mathbb{Z}_q,
\end{equation}
and satisfy
\begin{equation}
W^{(r,s)} W^{(t,u)} = \omega^{st} W^{(r+t,s+u)},
\label{eq:weyl-product}
\end{equation}
\begin{equation}
\Tr(W^{(r,s) \dagger} W^{(t,u)} )= q \delta_{r,t} \delta_{s,u}.
\label{eq:weyl-trace}
\end{equation}

Every channel admits a Weyl representation. We can express the perturbation channel as
\begin{equation}
\mathcal{W}(O) = \sum_{r,s,t,u =0}^{q-1} \chi_{r,s,t,u} W^{(r,s)} O W^{(t,u) \dagger}.
\end{equation}
The channel-dependent rank-four tensor $\chi$ is
\begin{equation}
\chi_{r,s,t,u} = \frac{1}{q^3} \sum_{a,b=0}^{q-1} \Tr \Bigl[
W^{(a,b)} \mathcal{W} \bigl(W^{(r,s)\dagger} W^{(a,b)\dagger} W^{(t,u)}\bigr) \Bigr].
\end{equation}
We can represent the adjoint channel acting at site $y$ as 
\begin{equation}
\mathcal{W}^\dagger_y(O) = \sum_{r,s,t,u =0}^{q-1} \chi^*_{r,s,t,u} W^{(r,s) \dagger }_y O W^{(t,u)}_y.
\label{eq:weyl-representation}
\end{equation}
Substituting Eq.~\eqref{eq:weyl-representation} into Eq.~\eqref{eq:expanded_OTOC} and applying the Weyl identity in Eq.~\eqref{eq:weyl-product} gives
\begin{equation}
\begin{aligned}
\overline{C}(x,y) &= \sum_{r,s,t,u  \in \mathbb{Z}_q} \phi_{r,s,t,u}D_{r,s,t,u}(x,y),\qquad D_{r,s,t,u}(x,y) = \frac{1}{q^N} \mathbb{E}_U \Bigl[ \Tr \Bigl( W^{(r,s) \dagger}_y  \varepsilon^{(U) \dagger}(V_x) W^{(t,u)}_y \varepsilon^{(U)\dagger} (V_x) \Bigr)  \Bigr],\\
\phi_{r,s,t,u} &= \sum_{r^\prime,s^\prime,t^\prime,u^\prime \in \mathbb{Z}_q} \chi^*_{r+t^\prime,s+u^\prime, t+r^\prime,u+s^\prime} \chi^*_{r^\prime,s^\prime,t^\prime,u^\prime} \omega^{t^\prime s-r^\prime u}  -2 \chi^*_{r,s,t,u}+\delta_{r,0} \delta_{s,0} \delta_{t,0} \delta_{u,0}.  
\end{aligned}
\end{equation}
For any positive integer $n$, define the permutation operator $F_n: \mathcal{H}_n \otimes \mathcal{H}_n \rightarrow\mathcal{H}_n \otimes \mathcal{H}_n$ by its action on computational-basis states:
\begin{equation}
\forall \boldsymbol{s}, \boldsymbol{s}^\prime \in \mathbb{Z}_q^n, \qquad  F_n (\ket{\boldsymbol{s}} \otimes \ket{\boldsymbol{s^\prime}}) = \ket{\boldsymbol{s^\prime}} \otimes \ket{\boldsymbol{s}} .
\label{eq:swap-operator}
\end{equation}
For any operators $A$, $B \in \mathcal{L}(\mathcal{H})$, 
\begin{equation}
\Tr_{\mathcal{H}} \Bigl( AB\Bigr) = \Tr_{\mathcal{H} \otimes \mathcal{H} } \Bigl( (A \otimes B) F_N \Bigr).
\label{eq:trace-expansion}
\end{equation}
The operator $F_N$ swaps each site in the first copy of the Hilbert space with the corresponding site in the second copy.
Using Eq.~\eqref{eq:trace-expansion} and substituting the assumed form for $\varepsilon^{(U) \dagger}$ from Eq.~\eqref{eq:generic-channel}, we obtain
\begin{equation}
\begin{aligned}
D_{r,s,t,u}(x,y) &= \frac{1}{q^N}\mathbb{E}_U \Bigl[ \Tr_{\mathcal{H}} \Bigl( W^{(r,s) \dagger}_y U^{(\pi) \dagger} \mathcal{F}^\dagger (V_x) U^{(\pi)} W^{(t,u)}_y U^{(\pi) \dagger} \mathcal{F}^\dagger (V_x) U^{(\pi)} \Bigr) \Bigr]\\
&=\frac{1}{q^N} \Tr_{\mathcal{H} \otimes \mathcal{H}} \Bigl( \mathbb{E}_U \Bigl[(U^{(\pi)})^{\otimes 2} (W^{(r,s) \dagger}_y \otimes W^{(t,u)}_y) (U^{(\pi) \dagger})^{\otimes 2} \Bigr] (\mathcal{F}^\dagger (V_x))^{\otimes 2} F_N \Bigr).
\end{aligned}
\end{equation}
Substituting the form of $U^{(\pi)}$ from Eq.~\eqref{eq:embedded-unitary} gives
\begin{equation}
D_{r,s,t,u}(x,y) = \frac{1}{q^N} \Tr_{\mathcal{H} \otimes \mathcal{H}} \Bigl( \mathbb{E}_U \Bigl[ (U_{q^K} \otimes I^{\otimes N-K}_q)^{\otimes 2}  (W^{(r,s) \dagger}_{\pi(y)} \otimes W^{(t,u)}_{\pi(y)}) (U^\dagger_{q^K} \otimes I^{\otimes N-K}_q)^{\otimes 2}   \Bigr]  (P_\pi \mathcal{F}^\dagger(V_x) P^{\dagger}_\pi)^{\otimes 2} F_N \Bigr).
\end{equation}
Let $\theta_K\in S_{2N}$ be the permutation defined by its action on a site index $j\in \mathbb{Z}_{2N}$:
\begin{equation}
\theta_K (j) = \begin{cases}
j,      & \text{if }0\leq j <K
            \text{ or }N+K \leq j< 2N,\\
j+K,    & \text{if }K \leq j < N,\\
j-N+K,  &\text{if }N \leq j < N+K.
\end{cases}
\label{eq:theta-k}
\end{equation}
Let $\Pi_{\theta_K} \in \mathcal{L}(\mathcal{H} \otimes \mathcal{H})$ be the corresponding permutation operator, defined by its action on computational-basis states:
\begin{equation}
\Pi_{\theta_K} \ket{s_0,s_1,\ldots,s_{2N-1}} = \ket{s_{\theta^{-1}_K(0)}, s_{\theta^{-1}_K(1)}, \ldots, s_{\theta^{-1}_K(2N-1)}}.
\end{equation}
Inserting instances of the identity $I_{\mathcal{H \otimes \mathcal{H}}} =\Pi_{\theta_K} \Pi^\dagger_{\theta_K}$ and using the fact that $\pi(y) \in \{0,1,\ldots,K-1\}$, we obtain  
\begin{equation}
\begin{aligned}
D_{r,s,t,u}(x,y) &= \frac{1}{q^N} \Tr_{\mathcal{H} \otimes \mathcal{H}} \Bigl( \Bigl\{ \mathbb{E}_{U}\Bigl[ U^{\otimes 2}_{q^K} W^{(r,s)\dagger}_{\pi(y)}W^{(t,u)}_{\pi(y)+K} (U^{\dagger}_{q^K})^{\otimes 2} \Bigr] \otimes I^{\otimes 2(N-K)}_q \Bigr \}\\  
&\quad \times\Pi_{\theta_K} (P_\pi \mathcal{F}^\dagger(V_x) P^{\dagger}_\pi)^{\otimes 2} F_N \Pi^\dagger_{\theta_K}\Bigr).
\label{eq:D-expression}
\end{aligned}
\end{equation}
In the above expression, $W^{(r,s) \dagger}_{\pi(y)}$ and $W^{(t,u)}_{\pi(y)+K}$ are understood to be elements of $ \mathcal{L} (\mathcal{H}_{2K})$. The Schur-Weyl identity gives the second-moment Haar average of an $n$-qudit unitary $U_{q^n}$ acting on an operator $O \in \mathcal{L} (\mathcal{H}_{2n})$ \cite{CollinsSniady2006Haar}:
\begin{equation}
\begin{gathered}
 \mathbb{E}_U \Bigl[ U^{\otimes 2}_{q^n} O (U^\dagger_{q^n})^{\otimes 2}\Bigr] = a_I(O)  I_{q^{2n}} + a_F(O)F_n,\\
a_I(O) = \frac{\Tr(O) -\Tr(F_nO)/q^n}{q^{2n}-1}, \qquad a_F(O) = \frac{\Tr(F_nO) - \Tr(O)/q^n}{q^{2n}-1}.
\end{gathered}
\label{eq:schur-weyl}
\end{equation}
We can apply the Schur-Weyl identity to the Haar average in Eq.~\eqref{eq:D-expression}. The coefficients are determined by
\begin{equation}
\Tr\Bigl( W^{(r,s) \dagger}_{\pi(y)} W^{(t,u)}_{\pi(y)+K} \Bigr) = q^{2K} \delta_{r,0} \delta_{s,0} \delta_{t,0} \delta_{u,0}, \qquad \Tr \Bigl(F_KW^{(r,s) \dagger}_{\pi(y)} W^{(t,u)}_{\pi(y)+K} \Bigr) =q^K \delta_{r,t} \delta_{s,u}.
\label{eq:coefficients}
\end{equation}
In Appendix~\ref{tm-derivation}, we use the Schur-Weyl identity and these coefficients to construct a transfer-matrix formalism to calculate the OTOC. However, for our present purposes, we need only the following consequences of Eq.~\eqref{eq:schur-weyl} and the coefficients in Eq.~\eqref{eq:coefficients}: $D_{r,s,t,u}(x,y)$ vanishes unless $r=t$ and $s=u$, and all nonidentity Weyl components on the diagonal are equal. We can thus express the Haar-averaged OTOC as
\begin{equation}
\begin{gathered}
\overline{C}(x,y) = \frac{1}{2}\Bigl(\sum_{(r,s) \in \mathbb{Z}^2_q\setminus \{(0,0)\}} \phi_{r,s,r,s} \Bigr) \overline{Q}(x,y) + \frac{1}{2}\phi_{0,0,0,0} \overline{R}(x),\\
\overline{R}(x) = \frac{2}{q^N}\mathbb{E}_U \Bigl[ \Tr \Bigl( \varepsilon^{(U) \dagger}(V_x) \varepsilon^{(U)\dagger} (V_x) \Bigr)  \Bigr],\\
\overline{Q}(x,y) = \frac{2}{q^N}\mathbb{E}_U \Bigl[ \Tr \Bigl( W^{(r,s) \dagger}_y  \varepsilon^{(U) \dagger}(V_x) W^{(r,s)}_y \varepsilon^{(U)\dagger} (V_x) \Bigr)  \Bigr],\qquad (r,s) \in \mathbb{Z}^2_q \setminus \{(0,0) \}.
\end{gathered}
\end{equation}
Next, we impose conditions on the tensor $\phi$ that follow from $\mathcal{W}$ being a completely positive trace-preserving (CPTP) map. The assumption that $\mathcal{W}$ is CPTP implies its adjoint map $\mathcal{W}^\dagger$ is unital, i.e., $\mathcal{W}^\dagger(I_q) =I_q$ \cite{Watrous2018}, and the matrix $\chi$ is positive semidefinite, $\chi \succeq 0$ \cite{ChuangNielsen1997}. From unitality, we obtain
\begin{equation}
\begin{aligned}
1&=\frac{1}{q} \Tr \Bigl(\mathcal{W}^\dagger(I_q)^2 \Bigr)\\
&=\frac{1}{q} \sum_{r,s,t,u \in \mathbb{Z}_q} \sum_{r^\prime, s^\prime,t^\prime, u^\prime \in \mathbb{Z}_q} \chi^*_{r,s,t,u} \chi^*_{r^\prime,s^\prime,t^\prime,u^\prime} \Tr \Bigl(W^{(r,s) \dagger} W^{(t,u)} W^{(r^\prime,s^\prime) \dagger} W^{(t^\prime,u^\prime)}  \Bigr)\\
&=  \sum_{r,s \in \mathbb{Z}_q} \sum_{r^\prime,s^\prime,t^\prime,u^\prime \in \mathbb{Z}_q} \chi^*_{r+t^\prime, s+u^\prime,r+r^\prime,s+s^\prime} \chi^*_{r^\prime,s^\prime,t^\prime,u^\prime} \omega^{s(t^\prime -r^\prime)}\\
&=\sum_{(r,s) \in \mathbb{Z}^2_q\setminus \{(0,0)\}} \Bigl( \phi_{r,s,r,s} + 2 \chi^*_{r,s,r,s}\Bigr) +\phi_{0,0,0,0} +2 \chi^*_{0,0,0,0} -1 \\
 &= \sum_{(r,s) \in \mathbb{Z}^2_q} \phi_{r,s,r,s}  + 2 \Tr \bigl(\chi^* \bigr) -1,
\end{aligned}
\label{eq:phi-identity}
\end{equation}
where we treat $\chi$ as a $q^2 \times q^2$ matrix in the Weyl representation and define $\Tr(\chi^*) = \sum_{(r,s) \in \mathbb{Z}^2_q} \chi^*_{r,s,r,s}$. Using the unitality of $\mathcal{W}^\dagger$ and the Weyl trace relation in Eq.~\eqref{eq:weyl-trace}, we obtain
\begin{equation}
1= \frac{1}{q} \Tr \Bigl(\mathcal{W}^\dagger(I_q) \Bigr) = \frac{1}{q}\sum_{r,s,t,u\in \mathbb{Z}_q} \chi^*_{r,s,t,u} \Tr\Bigl( W^{(r,s) \dagger} W^{(t,u)} \Bigr) = \Tr(\chi^*).
\label{eq:chi-unitality}
\end{equation}
Substituting Eq.~\eqref{eq:chi-unitality} into Eq.~\eqref{eq:phi-identity} gives
\begin{equation}
\sum_{r,s \in \mathbb{Z}_q} \phi_{r,s,r,s} =0.
\end{equation}
Thus, defining $\mathcal{A}(\mathcal{W}) = \phi_{0,0,0,0}/2$, we can express the OTOC associated with the perturbation channel $\mathcal{W}$ as
\begin{equation}
\overline{C}[\mathcal{W}](x,y) = \mathcal{A}(\mathcal{W}) \Bigl( \overline{R}(x) - \overline{Q}(x,y) \Bigr).
\end{equation}
We can express the amplitude in terms of $\chi(\mathcal{W})$ and the tensor $\chi(\mathcal{I}_q)$ corresponding to the identity channel, whose elements are $\chi_{r,s,t,u}(\mathcal{I}_q)=\delta_{r,0} \delta_{s,0}\delta_{t,0} \delta_{u,0}$:
\begin{equation}
\begin{aligned}
\mathcal{A}(\mathcal{W}) &= \frac{1}{2} \Bigl(1  -2 \chi^*_{0,0,0,0}(\mathcal{W}) +\sum_{r^\prime,s^\prime,t^\prime,u^\prime} \chi^*_{t^\prime ,u^\prime, r^\prime, s^\prime}(\mathcal{W}) \chi^*_{r^\prime,s^\prime,t^\prime,u^\prime}(\mathcal{W}) \Bigr)\\
&=\frac{1}{2} \Tr \Bigl( (\chi^*(\mathcal{W}) -\chi^*(\mathcal{I}_q))^2 \Bigr) = \frac{1}{2} || \chi(\mathcal{W})-\chi(\mathcal{I}_q)||^2_2.
\end{aligned}
\end{equation}
The last equality uses the fact that $\chi$ is positive semidefinite and hence Hermitian. Since $\chi$ is positive semidefinite, its eigenvalues are all non-negative, and it can be diagonalized by a unitary matrix $U^{(\chi)}$. We can write
\begin{equation}
\chi = U^{(\chi)}\Lambda U^{(\chi)\dagger} ,\qquad \Lambda = \operatorname{diag}(\lambda_i),
\end{equation}
where $\lambda_i$ are the eigenvalues of $\chi$. It follows that
\begin{equation}
\Tr (\chi^2)= \Tr(\Lambda^2) = \sum_{i} \lambda^2_i\leq (\sum_i \lambda_i)^2 =\Tr(\chi)^2 =1.
\end{equation}
Furthermore, $\chi \succeq 0$ implies $\chi_{0,0,0,0} \geq 0$. Thus the amplitude is bounded by 
\begin{equation}
\mathcal{A} \leq 1-\chi_{0,0,0,0} \leq 1.
\end{equation}
\section{Transfer Matrix Derivation}
\label{tm-derivation}
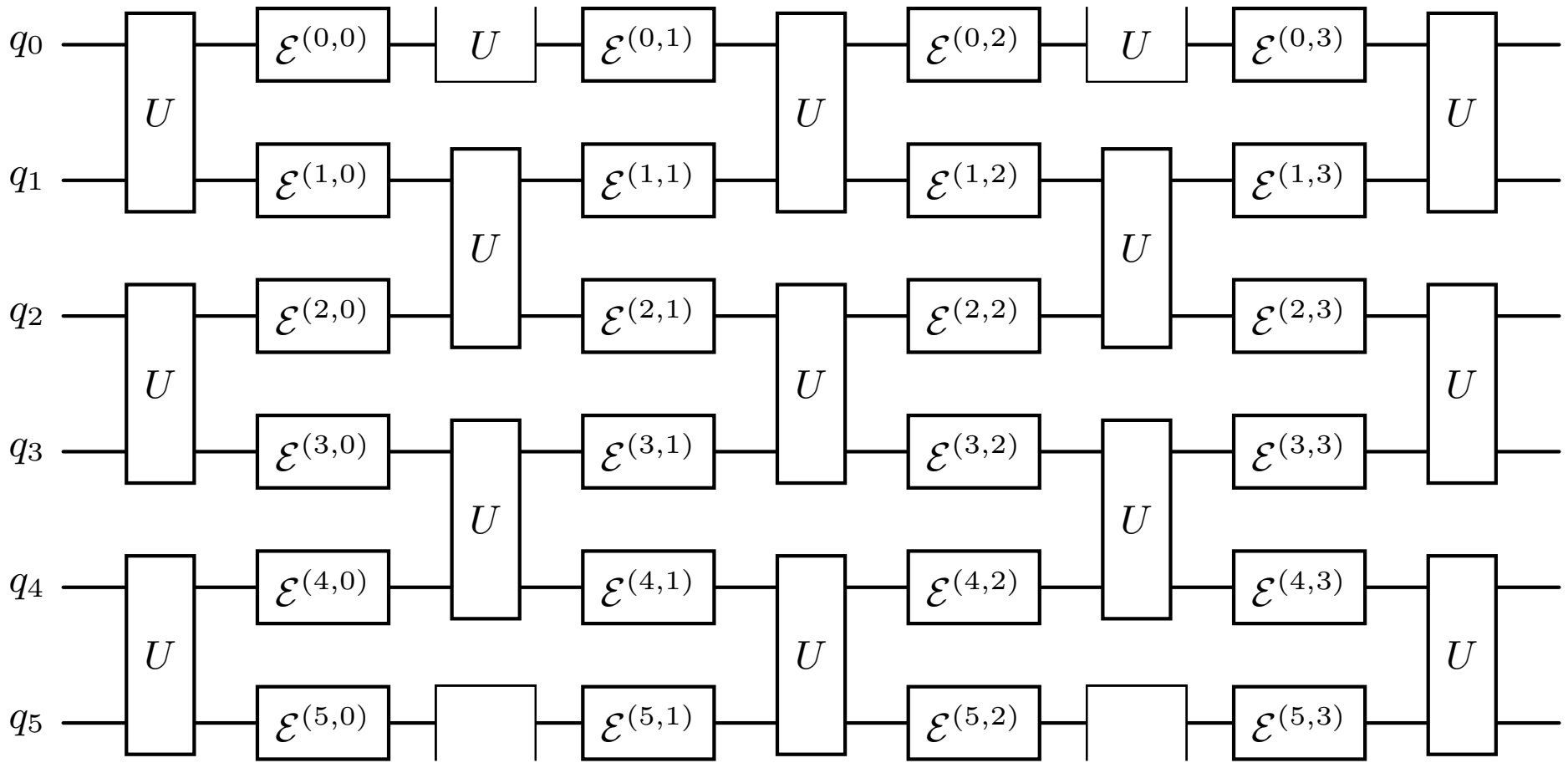
\begin{figure}[tbp]
  \centering
\begin{quantikz}
  \lstick{$q_0$}            &  \gate[2]{U}& \gate{\mathcal{E}^{(0,0)}} &\gate[style={pbc-first}]{U}&
  \gate{\mathcal{E}^{(0,1)}}&  \gate[2]{U}& \gate{\mathcal{E}^{(0,2)}} &\gate[style={pbc-first}]{U}&
  \gate{\mathcal{E}^{(0,3)}}&  \gate[2]{U}& \\
  \lstick{$q_1$}            &             & \gate{\mathcal{E}^{(1,0)}} &\gate[2]{U}&
  \gate{\mathcal{E}^{(1,1)}}&             & \gate{\mathcal{E}^{(1,2)}} &\gate[2]{U}&
  \gate{\mathcal{E}^{(1,3)}}&             & \\
  \lstick{$q_2$}            &  \gate[2]{U}& \gate{\mathcal{E}^{(2,0)}} &           &
  \gate{\mathcal{E}^{(2,1)}}&  \gate[2]{U}& \gate{\mathcal{E}^{(2,2)}} &           &
  \gate{\mathcal{E}^{(2,3)}}&  \gate[2]{U}& \\
  \lstick{$q_3$}            &             & \gate{\mathcal{E}^{(3,0)}} &\gate[2]{U}&
  \gate{\mathcal{E}^{(3,1)}}&             & \gate{\mathcal{E}^{(3,2)}} &\gate[2]{U}&
  \gate{\mathcal{E}^{(3,3)}}&             & \\
  \lstick{$q_4$}            &  \gate[2]{U}& \gate{\mathcal{E}^{(4,0)}} &           &
  \gate{\mathcal{E}^{(4,1)}}&  \gate[2]{U}& \gate{\mathcal{E}^{(4,2)}} &           &
  \gate{\mathcal{E}^{(4,3)}}&  \gate[2]{U}& \\
  \lstick{$q_5$}            &             & \gate{\mathcal{E}^{(5,0)}} &\gate[style={pbc-last}]{\phantom{U}}&
  \gate{\mathcal{E}^{(5,1)}}&             & \gate{\mathcal{E}^{(5,2)}} &\gate[style={pbc-last}]{\phantom{U}}&
  \gate{\mathcal{E}^{(5,3)}}&             & 
\end{quantikz}
  \caption{ A Haar-random brick-wall circuit with local non-unitary channels placed between the unitary layers. Each two-qudit gate is drawn independently from the Haar measure. The circuit shown has $N=6$ qudits and periodic boundary conditions. }
  \label{fig:non-unitary-brickwall}
\end{figure}

Figure~\ref{fig:non-unitary-brickwall} shows a circuit that evolves under a brick-wall arrangement of random unitaries in 1+1 dimensions, with local non-unitary channels applied between the unitary layers. Define a realization of the corresponding channel $\varepsilon_t^{(U)}$ as follows: For any permutations $f\in S_{N}$ and $g \in S_{2N}$, define the associated operators $P_{f}\in \mathcal{L}(\mathcal{H}_N)$ and $\Pi_{g} \in \mathcal{L} (\mathcal{H}_{2N})$ by their action on computational-basis states:
\begin{equation}
\begin{gathered}
P_{f} \ket{s_0,s_1,\ldots, s_{N-1}} = \ket{s_{f^{-1}(0)}, s_{f^{-1}(1)},\ldots, s_{f^{-1}(N-1)}},\\
\Pi_{g} \ket{s_{0}, s_1, \ldots, s_{2N-1}} = \ket{s_{g^{-1}(0)}, s_{g^{-1}(1)},\ldots , s_{g^{-1}(2N-1)}}.
\end{gathered}
\end{equation}
Denote the two-qudit Haar-random gate acting at sites $i$ and $j$ by
\begin{equation}
U_{i,j}  = P_{\xi_{ij}} (U_{q^2} \otimes I^{\otimes N-2}_q) P^{\dagger}_{\xi_{ij}},
\end{equation}
where $\xi_{ij}\in S_{N}$ is a permutation satisfying $\xi_{ij}(0)=i$ and $\xi_{ij}(1)=j$. Denote the corresponding channel by
\begin{equation}
u_{(i,j)}(O) =U_{i,j} O U^\dagger_{i,j}.
\end{equation}
For each time index $t \in \mathbb{Z}_{\geq 0}$, define the set of site pairs $\mathbb{S}_t$ by
\begin{equation}
\mathbb{S}_t = 
\begin{cases}
\bigl\{(0,1),(2,3),\ldots, (N-2,N-1) \bigr\},
& \text{if $t$ is even},\\
\bigl\{ (N-1,0),(1,2),\ldots,(N-3,N-2) \bigr\},
& \text{if $t$ is odd}.
\end{cases}
\end{equation}
Denote the channel for the unitary layer at time $t$ by
\begin{equation}
\mathcal{U}_t(O) = \mathop{\bigcirc}\limits_{\sigma \in \mathbb{S}_t } u_{\sigma}(O).
\label{eq:unitary-layer-channel}
\end{equation}
Since the gates act on distinct sites, the order of composition in Eq.~\eqref{eq:unitary-layer-channel} is irrelevant.
For each spatial site $j \in \mathbb{Z}_N$ and time $t \in \mathbb{Z}_{\geq 0}$, assign a single-qudit channel $\mathcal{E}^{(j,t)}$. Define the corresponding $N$-qudit channel that acts at site $j$  and acts as the identity elsewhere by $\tilde{\mathcal{E}}^{(j,t)}=\mathcal{I}^{\otimes j}_q \otimes \mathcal{E}^{(j,t)} \otimes \mathcal{I}^{\otimes N-j-1}_q$. Let $\mathcal{G}_t$ denote the channel 
\begin{equation}
\mathcal{G}_t(O) =\mathop{\bigcirc}\limits_{j \in \mathbb{Z}_N } \tilde{\mathcal{E}}^{(j,t)}(O).
\end{equation}
The full evolution channel is given by
\begin{equation}
\varepsilon^{(U)}_t(O) = \mathcal{U}_t\circ \mathcal{G}_{t-1} \circ \cdots \circ \mathcal{U}_1 \circ \mathcal{G}_0 \circ\mathcal{U}_0(O).
\end{equation}
We choose the time-indexing convention such that $\varepsilon^{(U)}_0 = \mathcal{U}_0$, $\varepsilon^{(U)}_1 = \mathcal{U}_1 \circ\mathcal{G}_{0} \circ\mathcal{U}_0$, and so forth. The Haar-random gates are sampled independently across space and time. Define a function that maps two $q^N \times q^N$ matrices to $\mathbb{C}$ by
\begin{equation}
\begin{gathered}
\overline{J}_{x,t}(A,B) = \frac{1}{q^N} \Tr_{\mathcal{H} \otimes \mathcal{H}} \Bigl( \overline{g}(x,t) (A \otimes B) F_N \Bigr),\\
\overline{g}(x,t) = \mathbb{E}_U  \bigl[ \varepsilon^{(U)\dagger}_t(V_x) \otimes \varepsilon^{(U)\dagger}_t(V_x) \bigr],
\label{eq:J}
\end{gathered}
\end{equation}
where $F_N$ is the SWAP operator defined in Eq.~\eqref{eq:swap-operator}. The Haar-averaged OTOC is
\begin{equation}
\overline{C}(x,y,t) = 2\mathcal{A}(\mathcal{W}) \Bigl( \overline{J}_{x,t}(I_{q^N} , I_{q^N}) - \overline{J}_{x,t}(W^{(r,s)\dagger}_y ,W^{(r,s)}_y) \Bigr),
\end{equation}
for any $(r,s) \in \mathbb{Z}_q^2 \setminus\{(0,0)\}$. Henceforth, we set the amplitude to $\mathcal{A}=1$. All time dependence is contained in the operator $\overline{g}(x,t)$.
Define the following maps on the duplicated Hilbert space
\begin{equation}
\omega_{(i,j)} (O) = \mathbb{E}_U \bigl[(U^\dagger_{i,j})^{\otimes 2} O U^{\otimes 2}_{i,j} \bigr],
\end{equation}
\begin{equation}
\zeta^{(t)}_{(i,j)}(O) =\omega_{(i,j)} \circ(\tilde{\mathcal{E}}^{(i,t) \dagger} \circ \tilde{\mathcal{E}}^{(j,t) \dagger})^{\otimes 2}(O),
\end{equation}
\begin{equation}
\Phi_t(O) = \mathbb{E}_U \bigl[ (\mathcal{U}^\dagger_t)^{\otimes 2} (O)  \bigr] = \mathop{\bigcirc}\limits_{\sigma \in \mathbb{S}_t } \omega_{\sigma}(O),
\end{equation}
\begin{equation}
\mathcal{T}_t(O) = \Phi_t \circ (\mathcal{G}^{\dagger}_t)^{\otimes 2} (O).
\end{equation}
In the definitions of $\Phi_t$ and $\mathcal{T}_t$, the same realization of the Haar-random gate at each pair of sites is used in both copies, while the realizations are drawn independently across spatial and temporal locations. We can express $\overline{g}(x,t)$ as
\begin{equation}
\overline{g}(x,t) = \mathcal{T}_0 \circ \mathcal{T}_{1} \circ \cdots\circ \mathcal{T}_{t-1} \circ \Phi_t(V^{\otimes 2}_x).
\end{equation}
Let $G\in \mathcal{L}(\mathcal{H} \otimes \mathcal{H})$ be an operator of the form
\begin{equation}
G =A^{(0)}_{0,N} A^{(1)}_{1,N+1} \cdots A^{(N-1)}_{N-1,2N-1}.
\end{equation}
Here, $A^{(k)}_{i,j} \in \mathcal{L}(\mathcal{H} \otimes \mathcal{H})$ acts nontrivially only at sites $i$ and $j$.
Consider the action of $\omega_{(i,j)}$  on $G$:
\begin{equation}
\begin{aligned}
\omega_{(i,j)} (G) &= \mathbb{E}_U \bigl[(U^\dagger_{i,j})^{\otimes 2} G U^{\otimes 2}_{i,j} \bigr]   \\
&= (P_{\xi_{ij}})^{\otimes 2}\mathbb{E}_U \bigl[ (U^\dagger_{q^2} \otimes I^{\otimes N-2}_{q})^{\otimes 2}(P^\dagger_{\xi_{ij}})^{\otimes 2} G (P_{\xi_{ij}})^{\otimes 2} (U_{q^2} \otimes I^{\otimes N-2}_{q})^{\otimes 2} \bigr] (P^\dagger_{\xi_{ij}})^{\otimes 2}.  
\end{aligned}
\end{equation}
Inserting factors of the identity $ I_{\mathcal{H} \otimes \mathcal{H}} = \Pi^{\dagger}_{\theta_2} \Pi_{\theta_2} $, where $\theta_2 \in S_{2N}$  is the permutation defined in Eq.~\eqref{eq:theta-k}, gives
\begin{equation}
\begin{aligned}
\omega_{(i,j)}(G) &=  (P_{\xi_{ij}})^{\otimes 2} \Pi^{\dagger}_{\theta_2} \mathbb{E}_{U} \bigl[\Pi_{\theta_2} (U^\dagger_{q^2} \otimes I^{\otimes N-2}_{q})^{\otimes 2}  \Pi^{\dagger}_{\theta_2} \Pi_{\theta_2}(P^\dagger_{\xi_{ij}})^{\otimes 2} G (P_{\xi_{ij}})^{\otimes 2} \Pi^\dagger_{\theta_2} \Pi_{\theta_2} (U_{q^2} \otimes I^{\otimes N-2}_{q})^{\otimes 2} \Pi^\dagger_{\theta_2}\bigr] \Pi_{\theta_2}(P^\dagger_{\xi_{ij}})^{\otimes 2}\\
&= P_{\xi_{ij}}^{\otimes 2} \Pi^\dagger_{\theta_2}  \mathbb{E}_U \bigl[(U^\dagger_{q^2})^{\otimes 2} \widehat{A}^{(i)}_{0,2} \widehat{A}^{(j)}_{1,3} U^{\otimes 2}_{q^2} \bigr] \prod_{k\notin \{i,j\}} A^{(k)}_{2+ \xi^{-1}_{ij}(k),N+\xi^{-1}_{ij}(k)} \Pi_{\theta_2} (P^{\dagger}_{\xi_{ij}})^{\otimes 2}.
\end{aligned}
\end{equation}
We use $\widehat{A}^{(i)}_{0,2}$ and $\widehat{A}^{(j)}_{1,3}$ to denote the four-qudit matrices defined by $A^{(i)}_{0,2} = \widehat{A}^{(i)}_{0,2} \otimes I^{\otimes 2N-4}_q$ and $A^{(j)}_{1,3} = \widehat{A}^{(j)}_{1,3} \otimes I^{\otimes 2N-4}_q$. We use the Schur-Weyl identity to express the Haar average
\begin{equation}
\mathbb{E}_U \bigl[ (U^\dagger_{q^2})^{\otimes 2} \widehat{A}^{(i)}_{0,2} \widehat{A}^{(j)}_{1,3} U^{\otimes 2}_{q^2} \bigr] = a_I(\widehat{A}^{(i)}_{0,2} \widehat{A}^{(j)}_{1,3}) I_{q^4} +a_F(\widehat{A}^{(i)}_{0,2} \widehat{A}^{(j)}_{1,3}) F_{2},
\end{equation}
where the coefficients $a_I(O)$ and $a_F(O)$, defined for any input operator $O$, are given in Eq.~\eqref{eq:schur-weyl}. For any $j \in\mathbb{Z}_N$ let $\tau_{j}\in S_{2N}$ denote the permutation that swaps $j$ and $N+j$, leaving other sites fixed. We can express $\omega_{(i,j)}$ as
\begin{equation}
\omega_{(i,j)}(G) = \sum_{s \in \{0,1\}} a^{1-s}_I(\widehat{A}^{(i)}_{0,2} \widehat{A}^{(j)}_{1,3} ) q^{2s} a^s_F(\widehat{A}^{(i)}_{0,2} \widehat{A}^{(j)}_{1,3} )  \prod_{k \in \mathbb{Z}_N \setminus \{i,j\}} A^{(k)}_{k,N+k} (\Pi_{\tau_i}/q)^s (\Pi_{\tau_j}/q)^s.
\end{equation}
Using this expression, we obtain 
\begin{equation}
\begin{aligned}
\zeta^{(t)}_{(i,j)}(G) &= \bigl[\omega_{(i,j)} \circ(\tilde{\mathcal{E}}^{(i,t) \dagger} \circ \tilde{\mathcal{E}}^{(j,t) \dagger})^{\otimes 2} \bigr] (G) \\
&= \sum_{s \in \{0,1\}} a^{1-s}_I(D^{(t)}_{i,j} ) q^{2s} a^s_F(D^{(t)}_{i,j} )  \prod_{k \in \mathbb{Z}_N \setminus \{i,j\}} A^{(k)}_{k,N+k} (\Pi_{\tau_i}/q)^s (\Pi_{\tau_j}/q)^s,\\
D^{(t)}_{i,j}&=[\mathcal{E}^{(i,t) \dagger} \otimes \mathcal{I}_q \otimes \mathcal{E}^{(i,t) \dagger} \otimes \mathcal{I}_q] (\widehat{A}^{(i)}_{0,2}) [\mathcal{I}_q \otimes \mathcal{E}^{(j,t) \dagger} \otimes \mathcal{I}_q \otimes \mathcal{E}^{(j,t)\dagger}]  (\widehat{A}^{(j)}_{1,3}).
\end{aligned}
\end{equation}
Define the quantity $r(x,t,t^\prime)$ by 
\begin{equation}
r(x,t,t^\prime) = \mathcal{T}_{t-t^\prime} \circ \mathcal{T}_{t -t^\prime +1} \circ \cdots \circ \mathcal{T}_{t-1} \circ \Phi_t(V_x^{\otimes 2}).
\end{equation}
With this definition, $\overline{g}(x,t) = r(x,t,t)$. Define $l(x,t)$ by
\begin{equation}
l(x,t) =x -\frac{1}{2}[ 1-(-1)^{x+t}].
\end{equation}
For $t^\prime = 0$, the unique two-qudit gate acting nontrivially at site $x$ is supported on sites $l(x,t)$ and $l(x,t)+1$, so we have
\begin{equation}
\begin{gathered}
\begin{aligned}
r(x,t,0) &=\Phi_t(V_x^{\otimes 2}) = \omega_{(l(x,t) , l(x,t) +1)} (V_x^{\otimes 2})\\
&= \sum_{\boldsymbol{s}  \in \{0,1\}^{N}} v_{\boldsymbol{s}}(x,t,0) \prod_{j=0}^{N-1} (\Pi_{\tau_j}/q)^{s_j},
\end{aligned}
\\
v_{\boldsymbol{s}}(x,t,0) =\Bigl(\prod_{j \in \mathbb{Z}_N \setminus \{ l(x,t),l(x,t)+1\}} \delta_{s_j,0}\Bigr) (a_I(\widehat{V}^{(x,t)}) \delta_{s_{l(x,t)},0} \delta_{s_{l(x,t)+1},0} + q^2 a_F(\widehat{V}^{(x,t)}) \delta_{s_{l(x,t)},1} \delta_{s_{l(x,t)+1},1} ),\\
\widehat{V}^{(x,t)}  = \bigl( I^{\otimes x-l(x,t)}_q \otimes V \otimes I^{\otimes 1-x+l(x,t)}_q \bigr)^{\otimes 2}.
\end{gathered}
\end{equation}
Starting from $t^\prime =0$, evolve $r(x,t,t^\prime)$ forward in $t^\prime$ by successively applying the brickwork layers formed from the maps $\zeta^{(n)}_{(i,j)}$, for $n=t-1,t-2, \ldots, t-t^\prime$. Since each $\zeta^{(n)}_{(i,j)}$ maps any input into the span of the identity and SWAP operators, $r(x,t,t^\prime)$ takes the form
\begin{equation}
r(x,t,t^\prime) = \sum_{\boldsymbol{s} \in \{0,1\}^N} v_{\boldsymbol{s}}(x,t,t^\prime) \prod_{j=0}^{N-1} (\Pi_{\tau_j}/q)^{s_j}.
\end{equation}
We can regard the coefficients $v_{\boldsymbol{s}}$ as the components of a vector in $(\mathbb{R}^{2})^{\otimes N}$. For any $x,t,t^\prime$, define the vector corresponding to $r(x,t,t^\prime)$ by 
\begin{equation}
\begin{gathered}
\begin{aligned}
\ket{v(x,t,t^\prime)} &= \sum_{\boldsymbol{s} \in \{0,1\}^N}v_{\boldsymbol{s}}(x,t,t^\prime) \ket{\boldsymbol{s}}\\
&=\frac{C_0}{2} T^{(t-t^\prime)} \cdots T^{(t-1)} (\ket{v_0(x,t)} + \frac{a_I(\widehat{V}^{(x,t)})}{q^2 a_F(\widehat{V}^{(x,t)})} \ket{0}^{\otimes N}),
\end{aligned}
\\
\ket{v_0(x,t)} = \bigotimes_{j \in \mathbb{Z}_N} \ket{\delta_{j,l(x,t) }+\delta_{j,l(x,t)+1}}.
\end{gathered}
\end{equation}
The single-layer transfer matrix $T^{(t)}$ is defined by
\begin{equation}
T^{(t)} = \prod_{(i,j) \in \mathbb{S}_t} M_{i,j}^{(t)}.
\end{equation}
The local transfer matrix $M^{(t)}_{i,j}$, expressed in the lexicographically ordered basis $\ket{00}, \ket{01},\ket{10}, \ket{11}$, is given by
\begin{equation}
\begin{gathered}
M^{(t)}_{i,j} = \begin{pmatrix}
1 & a(\tilde{p}^{(j,t)}, \Upsilon^{(j,t)}) & a(\tilde{p}^{(i,t) } ,\Upsilon^{(i,t)}) & c(\tilde{p}^{(i,t)}, \Upsilon^{(i,t)},\tilde{p}^{(j,t)},\Upsilon^{(j,t)}) \\
0 & 0 & 0 & 0 \\
0 & 0 & 0 & 0 \\
0 & b(\tilde{p}^{(j,t)},\Upsilon^{(j,t)}) & b(\tilde{p}^{(i,t)}, \Upsilon^{(i,t)}) & d(\tilde{p}^{(i,t)}, \Upsilon^{(i,t)},\tilde{p}^{(j,t)},\Upsilon^{(j,t)})
\end{pmatrix},
\\[1.25ex]
\begin{aligned}
a(\tilde{p}, \Upsilon) &= \frac{q(q+1) + \tilde{p} +q^3 \Upsilon}{q(q+1)(q^2+1)},\\
b(\tilde{p}, \Upsilon) &= \frac{q^2( q+1 -q \tilde{p} -\Upsilon) } {(q+1)(q^2+1)},\\
c(\tilde{p}_0, \Upsilon_0, \tilde{p}_1, \Upsilon_1) &= \frac{q(\tilde{p}_0 + \tilde{p}_1) -(q-1) \tilde{p}_0 \tilde{p}_1 +q^2(\Upsilon_0 +\Upsilon_1) +q^2(q-1) \Upsilon_0 \Upsilon_1}{q^2(q+1)(q^2+1)},\\
d(\tilde{p}_0, \Upsilon_0, \tilde{p}_1, \Upsilon_1) &= 1 - \frac{q^3( \tilde{p}_0 + \tilde{p}_1) -q^2(q-1) \tilde{p}_0 \tilde{p}_1 +\Upsilon_0 +\Upsilon_1 +(q-1) \Upsilon_0 \Upsilon_1 }{(q+1)(q^2+1)},
\end{aligned}
\\
\begin{aligned}
\tilde{p}^{(j,t)} &=\frac{q^2-\alpha^{(j,t)}}{q(q-1)},\\
\Upsilon^{(j,t)} &= \frac{\beta^{(j,t) }-q}{q(q-1)},\\
\alpha^{(j,t)} &= \Tr \bigl( \bigl[ (\mathcal{E}^{(j,t)})^{\otimes 2} (F_1 ) \bigr] F_1\bigr),\\
\beta^{(j,t)} &= \Tr \bigl( \mathcal{E}^{(j,t)}(I_q)^2 \bigr).
\end{aligned}
\end{gathered}
\label{eq:generalized-tm-matrix}
\end{equation}

The coefficients $a,b,c,d$ follow from Eq.~\eqref{eq:schur-weyl}. Substituting $\overline{g}(x,t) = r(x,t,t)$ into Eq.~\eqref{eq:J} gives
\begin{equation}
\begin{aligned}
\overline{J}_{x,t}(I_{q^N},I_{q^N} ) &=\frac{1}{q^N} \sum_{\boldsymbol{s} \in \{0,1\}^{N}} v_{\boldsymbol{s}}(x,t,t)  \Tr_{\mathcal{H}\otimes \mathcal{H}} \Bigl( \prod_{j}(\Pi_{\tau_j}/q)^{s_j} F_N \Bigr)\\
&=\frac{1}{q^N} \sum_{\boldsymbol{s} \in \{0,1\}^{N}} v_{\boldsymbol{s}}(x,t,t)  \Tr_{\mathcal{H}\otimes \mathcal{H}} \Bigl( \prod_{j}(\Pi_{\tau_j}/q)^{s_j} \prod_{k} \Pi_{\tau_k} \Bigr)\\
&=\sum_{\boldsymbol{s} \in \{0,1 \}^N} v_{\boldsymbol{s}}(x,t,t).
\end{aligned}
\end{equation}
Similarly, for the second term, we obtain
\begin{equation}
\begin{aligned}
\overline{J}_{x,t} (W^{(r,s) \dagger}_y, W^{(r,s)}_y \Bigr) &= \frac{1}{q^N} \sum_{\boldsymbol{s} \in \{0,1\}^N} v_{\boldsymbol{s}}(x,t,t) \Tr_{\mathcal{H} \otimes \mathcal{H}} \Bigl( \prod_{j} (\Pi_{\tau_j}/q)^{s_j} (W^{(r,s) \dagger}_y\otimes W^{(r,s)}_y )F_N \Bigr)\\
&=\frac{1}{q^N} \sum_{\boldsymbol{s} \in \{0,1\}^N} v_{\boldsymbol{s}}(x,t,t) \Tr_{\mathcal{H} \otimes \mathcal{H}} \Bigl( \prod_{j} (\Pi_{\tau_j}/q)^{s_j} (W^{(r,s) \dagger}_y W^{(r,s)}_{N+y} ) \prod_{k} \Pi_{\tau_k} \Bigr)\\
&=\sum_{\boldsymbol{s} \in \{0,1\}^N ,s_y=0} v_{\boldsymbol{s}} (x,t,t).
\end{aligned}
\end{equation}
The Haar-averaged OTOC is
\begin{equation}
\begin{gathered}
\overline{C}(x,y,t) = C_0 \sum_{\boldsymbol{s} \in \{0,1\}^{N}, s_y=1} \bra{\boldsymbol{s}}  T^{(0)} \cdots T^{(t-1)} \ket{v_0(x,t)},\\
C_0 = \frac{2q^4}{q^4-1} \operatorname{Var}_{I_q/q} (V), \qquad \text{Var}_{I_q/q}(V) = \Tr(V^2/q) -\Tr(V/q)^2.
\label{eq:appendix-otoc}
\end{gathered}
\end{equation}
Henceforth, we normalize $V$ so that $\text{Var}_{I_q/q} (V)=1$. Equation~\eqref{eq:appendix-otoc}  is valid for arbitrary local non-unitary gates. The effect of each channel $\mathcal{E}^{(j,t)}$ on the OTOC is described by two parameters $\tilde{p}^{(j,t)}$ and $\Upsilon^{(j,t)}$. Unitary channels correspond to the point $\tilde{p}^{(j,t)}=0$, $\Upsilon^{(j,t)} =0$ in this parameter space. 
Unital channels form the one-parameter family with $\Upsilon =0$, $0 \leq \tilde{p}^{(k,t)} \leq \frac{q+1}{q}$. The parameters $\tilde{p}^{(j,t)}$ and $\Upsilon^{(j,t)}$ cannot be chosen independently. A pair  $(\tilde{p}, \Upsilon)$ is attainable by a CPTP single-qudit channel if and only if they satisfy inequalities of the form
\begin{equation}
0 \leq \tilde{p} \leq \frac{q+1}{q}, \qquad 0 \leq  \Upsilon \leq \Upsilon^{(q)}_{\max} (\tilde{p}).
\label{eq:alpha-beta-regime}
\end{equation}
For $q=2$, the upper bound on $\Upsilon$ is
\begin{equation}
\Upsilon^{(2)}_{\max}(\tilde{p})=
\begin{cases}
(1- \sqrt{1-\tilde{p}})^2,
& 0 \leq \tilde{p} \leq 1,\\
3-2 \tilde{p},
& 1 \leq \tilde{p} \leq \frac{3}{2}.
\end{cases}
\end{equation}
So far, we have considered calculating the OTOC for a fixed realization of non-unitary channels. Next, we consider calculating the average OTOC over realizations of these local non-unitary channels. If the local non-unitary channels at each spacetime location are drawn independently, the average takes the form
\begin{equation}
\mathbb{E}_{\mathcal{E}} [\mathbb{E}_U [C] \bigl(\tilde{p}(\mathcal{E}), \Upsilon(\mathcal{E})\bigr) ] = \mathbb{E}_{U} [ C] \bigl( \mathbb{E}_\mathcal{E} [\tilde{p}(\mathcal{E})] , \mathbb{E}_{\mathcal{E}} [\Upsilon(\mathcal{E})]
\bigr).
\end{equation}
That is to say, we can simply replace $\tilde{p}(\mathcal{E})$ with $\mathbb{E}_{\mathcal{E}}[\tilde{p}(\mathcal{E})]$
and $\Upsilon (\mathcal{E})$ with $\mathbb{E}_{\mathcal{E}} [\Upsilon(\mathcal{E})]$. This follows because the local transfer matrix is separately affine in $\tilde{p}^{(j,t)},\Upsilon^{(j,t)}$ for each independent local channel. Averaging can effectively extend the parameter regime beyond that described by Eq.~\eqref{eq:alpha-beta-regime}.

Next, we restrict ourselves to the measurement-induced circuits discussed in the main text. At each spacetime location defined by the channel locations above, a measurement in the eigenbasis of the clock operator $Z$ is performed with probability $p$. The measurement events are independent Bernoulli trials. Let $\Gamma$ denote the spacetime pattern of measurement locations and $\mu_\Gamma$ the corresponding measurement outcomes. We write $\varepsilon^{(U,\Gamma,\mu_\Gamma)}_t$ for the trajectory-resolved evolution through time $t$, and $C[\varepsilon]$ for the OTOC associated with the evolution map $\varepsilon$. As discussed in the main text, since the OTOC is nonlinear in the evolution channel, the order of averaging is generally significant. We assume the measurement results $\mu_\Gamma$ are averaged over prior to inserting the evolution map into the OTOC.
%We defer discussion of the average post-selected quantity $\mathbb{E}_{U,\Gamma,\mu_\Gamma} [C[\varepsilon^{(U, \Gamma, \mu_\Gamma)}]]$ to Appendix~X. 
First, consider the quantity $\mathbb{E}_U[C[ \mathbb{E}_{\Gamma, \mu_{\Gamma}} [\varepsilon^{(U,\Gamma,\mu_{\Gamma})}]]]$. In this case, each local non-unitary is effectively the dephasing channel 
 \begin{equation}
 \forall j\in \mathbb{Z}_N,t \in \mathbb{Z}_{\geq 0},\qquad \mathcal{E}^{(j,t)}(O)  =  \mathcal{D}_p(O) = (1-p)O + p\Delta(O), \qquad \Delta(O)= \sum_{n=0}^{q-1} \ket{n} \bra{n} O \ket{n} \bra{n}.
 \end{equation}
 Substituting this expression into Eq.~\eqref{eq:generalized-tm-matrix} gives
 \begin{equation}
 \tilde{p} = p(2-p), \qquad\Upsilon = 0.
 \end{equation}
To evaluate $\mathbb{E}_{U,\Gamma} [C[\mathbb{E}_{\mu_{\Gamma}} [\varepsilon^{(U,\Gamma,\mu_{\Gamma})}]]]$, we average $\alpha$ and $\beta$ over the local channel realizations, obtaining
\begin{equation}
\mathbb{E}_{\mathcal{E}} [\tilde{p}(\mathcal{E})] = p\tilde{p}(1)+(1-p) \tilde{p}(0) = p  ,\qquad \mathbb{E}_{\mathcal{E}} [\Upsilon(\mathcal{E})] = 0. 
\label{eq:alpha-beta-of-p}
\end{equation}
It follows that we can relate the following averages through a change of variables in $p$:
\begin{equation}
\mathbb{E}_{U} [ C[ \mathbb{E}_{\Gamma, \mu_\Gamma}[ \varepsilon^{(U,\Gamma, \mu_{\Gamma})]}] ]](p) =  \mathbb{E}_{U,\Gamma} [ C[ \mathbb{E}_{ \mu_\Gamma}[ \varepsilon^{(U,\Gamma, \mu_{\Gamma})]}] ]](p(2-p)).
\end{equation}
Henceforth, we use $\overline{C}$ to denote $\mathbb{E}_{U,\Gamma} [C[\mathbb{E}_{\mu_{\Gamma}} [\varepsilon^{(U,\Gamma, \mu_{\Gamma})}]]]$.
Setting all $\tilde{p}^{(j,t)}  = p$ and $\Upsilon^{(j,t)}=0$ in Eq.~\eqref{eq:generalized-tm-matrix} yields the transfer matrix given in Eq.~\eqref{eq:tm} of the main text. Under $\mathrm{OBC}$, the channels acting on the unpaired boundary sites commute through the corresponding unitary layer and are composed with the channels acting on the same sites in the adjacent layer. The boundary transfer matrices are therefore obtained using the parameters of these composed channels. For the measurement-induced circuit considered here, this gives the effective boundary parameter $p_e=p(2-p)$.

\section{Eigenvalues of the Transfer Matrix are Real and Lie in the Interval $[0,1]$}
\label{app:real_eigvals}
For both $\mathrm{PBC}$ and $\mathrm{OBC}$, we show the eigenvalues of the transfer matrix $T=T^{(0)} T^{(1)}$ are real and lie in the interval $[0,1]$. We assume throughout that all local nonunitary channels are unital, so that $\Upsilon^{(j,t)}=0$ at each spacetime site. We allow these channels to vary spatially but require them to be time independent. We write $p_j = \tilde{p}^{(j,0)}\in [0,\frac{q+1}{q}]$ for the effective parameter discussed in Appendix~\ref{tm-derivation}. For either boundary condition, $T^{(0)}$ takes the form
\begin{equation}
T^{(0)} = \bigotimes_{j=0}^{N/2-1} M^{(2j,2j+1)},
\end{equation}
where the local transfer matrix is
\begin{equation}
\begin{gathered}
M^{(i,j)} = \begin{pmatrix}
1 & a(p_j) & a(p_i) & c(p_i, p_j) \\
0 & 0 & 0 & 0 \\
0 & 0 & 0 & 0 \\
0 & b(p_j) & b(p_i) & d(p_i,p_j)
\end{pmatrix},
\\[1.25ex]
\begin{aligned}
a(p) &= \frac{q(q+1) + p }{q(q+1)(q^2+1)},\\
b(p) &= \frac{q^2( q+1 -q p) } {(q+1)(q^2+1)},\\
c(p_0,p_1) &= \frac{q(p_0 + p_1) -(q-1) p_0 p_1 }{q^2(q+1)(q^2+1)},\\
d(p_0,p_1) &= 1 - \frac{q^3( p_0 + p_1) -q^2(q-1) p_0 p_1 }{(q+1)(q^2+1)}.
\end{aligned}
\end{gathered}
\label{proof_tm_matrix}
\end{equation}
The odd layer matrix $T^{(1)}$ is
\begin{equation}
T^{(1)} = \begin{cases}
\bigotimes_{j=0}^{N/2-1} M^{(2j+1,2j+2)},
& \mathrm{PBC},\\
 I_2 \otimes \bigl(\bigotimes_{j=0}^{N/2-2} M^{(2j+1,2j+2)}\bigr) \otimes I_2,
& \mathrm{OBC}.
\end{cases}
\end{equation}
For any vector $\boldsymbol{v}\in \mathbb{R}^{2^{N}}$, $T \boldsymbol{v} = T^{(0)} (T^{(1)} \boldsymbol{v})$ necessarily belongs to the image of $T^{(0)}$. Thus 
\begin{equation}
\begin{aligned}
\text{im}(T) &\subseteq \text{im}(T^{(0)}),\\
\text{rank}(T) &\leq \text{rank}(T^{(0)}) \leq 2^{N/2}.
\end{aligned}
\end{equation}
At most $2^{N/2}$ eigenvalues of $T$ are nonzero. This suggests that its nonzero spectrum can be represented by a $2^{N/2} \times 2^{N/2}$ matrix. We can make this reduction explicit by defining the matrix $K$ as follows:
\begin{equation}
\begin{gathered}
K = \bigl(\bigotimes_{j=0}^{N/2-1}V^{(2j,2j+1)} \bigr) T^{(1)} U^{\otimes N/2},
\\[1.25ex]
U = \begin{pmatrix}
1 & 0\\
0 & 0\\
0 & 0\\
0 & 1
\end{pmatrix}, \qquad
V^{(i,j)} = \begin{pmatrix}
1 & a(p_j) & a(p_i) & c(p_i,p_j)\\
0 & b(p_j) & b(p_i) & d(p_i,p_j)
\end{pmatrix},
\end{gathered}
\label{eq:Kmatrix}
\end{equation}
where $a$, $b$, $c$, and $d$ are the polynomials defined in Eq.~\eqref{proof_tm_matrix}. Let $X = U^{\otimes N/2}$ and $Y =\bigl(\bigotimes_{j=0}^{N/2-1}V^{(2j,2j+1)} \bigr) T^{(1)}$, so that $T=XY$ and $K=YX$. The matrices $XY$ and $YX$ must have the same nonzero eigenvalues~\cite{HornJohnson2012} as follows from the identity
\begin{equation}
\det(\lambda I_{2^N} -XY) =\lambda^{2^{N}-2^{N/2}} \det (\lambda I_{2^{N/2}} -YX).
\end{equation}
Thus, $K$ has the same nonzero eigenvalues as $T$. Define the matrix
\begin{equation}
\eta = W^{\otimes N/2}, \qquad W= \begin{pmatrix}
q^4 & 1 \\
1 & 1
\end{pmatrix}.
\label{eq:metric}
\end{equation}
We refer to $\eta$ as the metric. The eigenvalues of $W$ are
\begin{equation}
\omega_{\pm} = \frac{q^4+1 \pm \sqrt{(q^4-1)^2+4}}{2}.
\end{equation}
For any qudit dimension $q \geq 2$, these eigenvalues are real and positive. The eigenvalues of $\eta$ are of the form $\omega^{N/2-n}_{+}\omega^{n}_{-}$, where $n$ is an integer satisfying $0 \leq n \leq N/2$. Hence, the eigenvalues of $\eta$ are real and positive. Since $\eta$ is also real and symmetric, it follows that it is positive definite. For any vectors $\boldsymbol{v},\boldsymbol{w} \in \mathbb{C}^{2^{N/2}}$, we define
\begin{equation}
\langle\boldsymbol{v},\boldsymbol{w} \rangle_{\eta} = \boldsymbol{v}^{\dagger} \eta \boldsymbol{w}.
\label{inner_product}
\end{equation}
Since $\eta$ is positive definite, Eq.~\eqref{inner_product} defines an inner product on $\mathbb{C}^{2^{N/2}}$, which we call the $\eta$-inner product. Two vectors $\boldsymbol{v},\boldsymbol{w} $ are said to be $\eta$-orthogonal if $\langle \boldsymbol{v},\boldsymbol{w} \rangle_\eta =0$. Let $\boldsymbol{\psi} \neq \boldsymbol{0}$ be an eigenvector of $K$ with eigenvalue $\lambda$. Since $\langle \boldsymbol{\psi} , \boldsymbol{\psi}\rangle_\eta>0$, the eigenvalue can be written as
\begin{equation}
\lambda = \frac{\langle \boldsymbol{\psi},K \boldsymbol{\psi} \rangle_\eta}{\langle\boldsymbol{\psi}, \boldsymbol{\psi} \rangle_\eta}= \frac{\boldsymbol{\psi}^\dagger \eta K \boldsymbol{\psi}}{\boldsymbol{\psi}^\dagger \eta \boldsymbol{\psi}}.
\end{equation}
To prove the eigenvalues of $T$ are real and nonnegative, it suffices to show that $\eta K \succeq 0$.
The product $WV^{(i,j)}$ is
\begin{equation}
\begin{aligned}
WV^{(i,j)} &= \begin{pmatrix}
q^4 & q^4 a(p_j)+b(p_j) & q^4 a(p_i) + b(p_i) & q^4 c(p_i,p_j)+d(p_i,p_j)\\
1 & a(p_j)+b(p_j) & a(p_i)+b(p_i) & c(p_i,p_j)+d(p_i,p_j)
\end{pmatrix}
\\[1.25ex]
&= \begin{pmatrix}
q^4 & q^2 & q^2  & 1\\
1 & \gamma(p_j) & \gamma(p_i) & \gamma(p_i) \gamma(p_j)
\end{pmatrix},
\end{aligned}
\label{factorization}
\end{equation}
where $\gamma(p) =a(p)+b(p) = 1- \frac{q-1}{q}p$. The second line of Eq.~\eqref{factorization} uses the coefficients in Eq.~\eqref{proof_tm_matrix} and represents a nontrivial factorization of $WV$. In the computational basis, the matrix elements of $WV$ are
\begin{equation}
(WV^{(i,j)})_{l,mn} = F_{lm}(p_i) F_{ln}(p_j),\qquad
F(p) = \begin{pmatrix}
q^2 & 1\\
1 & \gamma(p)
\end{pmatrix}.
\end{equation}
Let $\mathcal{M}^{(i,j)}(k,l;n,m) =\bra{k,l} M^{(i,j)} \ket{n,m}$ denote the matrix elements of $M^{(i,j)}$ in the computational basis. With this notation, the matrix elements of $T^{(1)}$ are 
\begin{equation}
T^{(1)}_{\boldsymbol{x},\boldsymbol{y}} = (1-\sigma +\sigma\delta_{x_0,y_0}\delta_{x_{N-1},y_{N-1}})\prod_{k=0}^{N/2-1-\sigma} \mathcal{M}^{(2k+1,2k+2)}(x_{2k+1}, x_{2k+2}; y_{2k+1}, y_{2k+2}),
\end{equation}
where $\sigma=0$ for $\mathrm{PBC}$ and $\sigma=1$ for $\mathrm{OBC}$. The matrix elements of $\eta K$ in the computational basis are
\begin{equation}
\begin{aligned}
(\eta K)_{\boldsymbol{x},\boldsymbol{y}} &=\sum_{\boldsymbol{\alpha},\boldsymbol{\beta} \in \{0,1\}^N} \bigl(\eta \bigotimes_{j=0}^{N/2-1} V^{(2j,2j+1)}  \bigr)_{\boldsymbol{x},\boldsymbol{\alpha}}  T^{(1)}_{\boldsymbol{\alpha},\boldsymbol{\beta}} (U^{\otimes N/2})_{\boldsymbol{\beta},\boldsymbol{y}}\\
&= \sum_{\boldsymbol{\alpha},\boldsymbol{\beta} \in \{0,1\}^N}\Bigl[
\prod_{j=0}^{N/2-1} F_{x_j \alpha_{2j}}(p_{2j}) F_{x_j \alpha_{2j+1}}(p_{2j+1})\\
&\times(1-\sigma +\sigma\delta_{\alpha_0,\beta_0}\delta_{\alpha_{N-1},\beta_{N-1}}) \prod_{k=0}^{N/2-1-\sigma} \mathcal{M}^{(2k+1,2k+2)}(\alpha_{2k+1}, \alpha_{2k+2}; \beta_{2k+1}, \beta_{2k+2}) \prod_{l=0}^{N/2-1} \delta_{\beta_{2l},y_l} \delta_{\beta_{2l+1},y_l}\Bigr]\\
& =\Bigl( F_{x_0,y_0 }(p_0)  F_{x_{N/2-1},y_{N/2-1}}(p_{N-1}) \Bigr)^{\sigma} \prod_{j=0}^{N/2-1-\sigma} B^{(2j+1,2j+2)}(x_j, x_{j+1}; y_j,y_{j+1}).
\end{aligned}
\end{equation}
In the last line, we defined
\begin{equation}
B^{(i,j)}(k,l;m,n) = \sum_{s_0,s_1 \in \{0,1\} }F_{k s_0}(p_i) F_{ls_1}(p_j)  \mathcal{M}^{(i,j)}(s_0,s_1;m,n).
\end{equation}
The matrix $B^{(i,j)}$ can be written as
\begin{equation}
\begin{gathered}
B^{(i,j)} = \boldsymbol{r} \boldsymbol{r}^T  +\boldsymbol{u}^{(i,j)} \boldsymbol{u}^{(i,j)T},\\
\boldsymbol{r} = \begin{pmatrix}
q^2 \\ 1 \\ 1 \\ q^{-2}
\end{pmatrix}, \qquad
\boldsymbol{u}^{(i,j)} =
\frac{1}{\sqrt{q^4-1}}
\begin{pmatrix}
0\\ q^2 \gamma(p_j)-1\\ q^2 \gamma(p_i)-1\\ \frac{q^4 \gamma(p_i) \gamma(p_j) -1}{q^2}
\end{pmatrix}.
\end{gathered}
\end{equation}
The matrix $B^{(i,j)}$ is positive semidefinite because the following inequality holds for every $\boldsymbol{z} \in \mathbb{C}^{4}$:
\begin{equation}
\boldsymbol{z}^\dagger B^{(i,j)} \boldsymbol{z} = |\boldsymbol{r}^T \boldsymbol{z}|^2 + |\boldsymbol{u}^{(i,j) T} \boldsymbol{z}|^2 \geq 0.
\end{equation}
Let $J_2$ denote
\begin{equation}
J_2 = \begin{pmatrix}
1 & 1 \\
1 & 1
\end{pmatrix}.
\end{equation}
We have $J_2 \succeq0$ since its eigenvalues are $0$ and $2$. For each $j \in \mathbb{Z}_{N/2}$, let $\phi_j \in S_{N/2}$ be a permutation satisfying $\phi_j(j)=0$ and $\phi_j(j+1)=1$. Let $P_{\phi_j} \in \mathbb{C}^{2^{N/2} \times 2^{N/2}}$ be the associated permutation matrix that acts on a computational basis state as 
\begin{equation}
P_{\phi_j} \ket{x_0,x_1,\ldots, x_{N/2-1}} = \ket{x_{\phi^{-1}_j(0)},x_{\phi^{-1}_j(1)},\ldots, x_{\phi^{-1}_j(N/2-1)}}.
\end{equation}
Define the matrix $\tilde{B}_j$ by
\begin{equation}
\tilde{B}_j = P^\dagger_{\phi_j} \bigl( B^{(2j+1,2j+2)} \otimes J_2^{\otimes N/2-2}\bigr) P_{\phi_j}.
\end{equation}
Positive semidefiniteness is preserved under tensor products and congruence transformations: if $B,C \succeq0$, then $B\otimes C \succeq 0$, and if $B\succeq0$, then $A^\dagger B A \succeq0$ for any matrix $A$ of compatible dimensions. Since $B^{(2j+1,2j+2)} \succeq0$ and $J_2 \succeq 0$, it follows that $\tilde{B}_j \succeq0$. Let $\tilde{F}$ denote
\begin{equation}
\tilde{F} = F(p_0) \otimes J_2^{\otimes N/2-2} \otimes F(p_{N-1}).
\end{equation}
The eigenvalues of $F(p)$ are
\begin{equation}
f_{\pm} = \frac{q^2+\gamma(p) \pm \sqrt{(q^2-\gamma(p))^2+4}}{2} \geq0,
\end{equation}
so $F(p)\succeq 0$ and hence $\tilde{F} \succeq 0$. We can express $\eta K$ as
\begin{equation}
\eta K = \Bigl( \bigodot_{j=0}^{N/2-1-\sigma} \tilde{B}_j \Bigr)\odot \bigl(\sigma\tilde{F}+(1-\sigma) J^{\otimes N/2}_{2}\bigr),
\end{equation}
where $\odot$ denotes the Hadamard (elementwise) product. By the Schur-Hadamard product theorem, the Hadamard product of two positive semidefinite matrices is also positive semidefinite~\cite{HornJohnson2012}. Hence $\eta K \succeq0$.

Finally, we show the eigenvalues of $T$ have magnitude at most one. A real matrix is column-substochastic if its entries are nonnegative and every column sums to at most one. The four columns of $M^{(i,j)}$ sum to 
\begin{equation}
1,\qquad\gamma(p_j), \qquad \gamma(p_i), \qquad \gamma(p_i) \gamma(p_j).
\end{equation}
Since $p_i,p_j \in [0,\frac{q+1}{q}]$, it follows that
\begin{equation}
0 \leq \gamma(p) \leq  1.
\end{equation}
The entries of $M^{(i,j)}$ are all nonnegative. Thus, $M^{(i,j)}$ is column-substochastic. Given two column-substochastic matrices $A$ and $B$, their Kronecker product $A\otimes B$ is also column-substochastic since
\begin{equation}
\sum_{i,k} (A \otimes B)_{(i,k),(j,l)} = \bigl(\sum_i A_{ij} \bigr) \bigl( \sum_k B_{kl}\bigr) \leq 1.
\label{tensor_product_substochastic}
\end{equation}
Similarly, the product $AB$, where it is defined, is also column-substochastic since
\begin{equation}
\sum_i (AB)_{ij} = \sum_k \bigl(\sum_i A_{ik} \bigr) B_{kj} \leq \sum_k B_{kj} \leq 1.
\label{product_substochastic}
\end{equation}
By Eq.~\eqref{tensor_product_substochastic}, the layer matrices $T^{(0)}$ and $T^{(1)}$ are column-substochastic. Since the layer transfer matrices are column-substochastic, it follows from Eq.~\eqref{product_substochastic} that the transfer matrix $T=T^{(0)} T^{(1)}$ is column-substochastic. Every eigenvalue of a column-substochastic matrix has magnitude at most one. To see this, let $\boldsymbol{v} \neq \boldsymbol{0}$ denote an eigenvector of $T$ with eigenvalue $\lambda$. We have
\begin{equation}
\begin{aligned}
|\lambda|\sum_i |v_i| &= \sum_i |\sum_j T_{ij} v_j|  \\
&\leq  \sum_j(\sum_{i} T_{ij}) |v_j|\\
&\leq \sum_j |v_j|.
\end{aligned}
\end{equation}
It follows that $|\lambda| \leq 1$. Combining nonnegativity of the spectrum with the bound $|\lambda|\leq 1$, we conclude that every eigenvalue of $T$ is real and lies in the interval $[0,1]$ for both $\mathrm{PBC}$ and $\mathrm{OBC}$.

\section{Variational Lower Bound on $\lambda_*$}
\label{app:lower_bound}
\begin{figure}[t]
\centering
\includegraphics[width=1.0\linewidth]{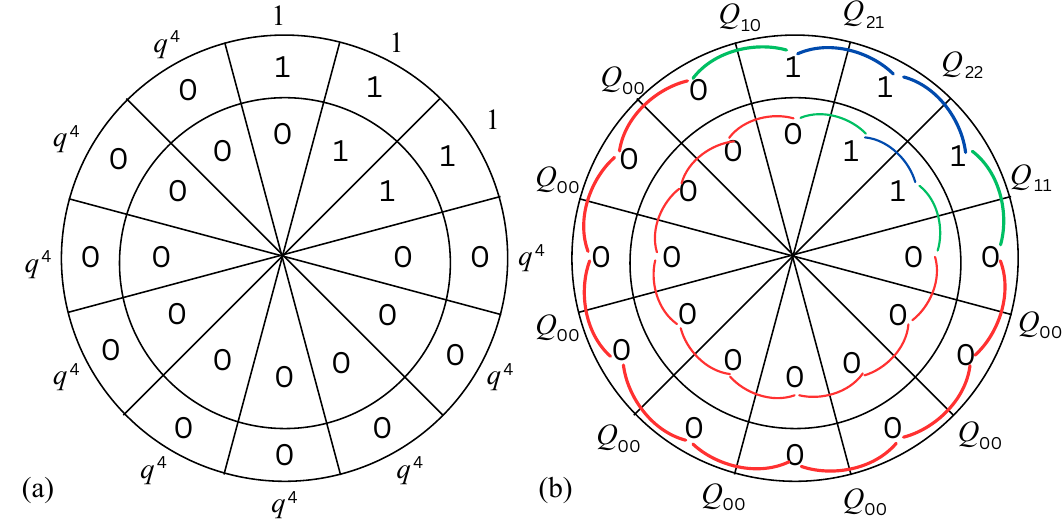}
\caption{Illustration of the configurations used in the variational calculation. (a) A pair of island states contributing to the inner product $\langle\psi_{l^\prime},\psi_l\rangle_{\eta}$. The associated weight is the product of the factors shown outside the circle, which in this example is $q^{4\times 9}$. (b) The corresponding configuration contributing to the calculation of $\langle\psi_{l^\prime},K\psi_l \rangle_{\eta}$. The weight is $Q_{00}^{8} Q_{10} Q_{21}Q_{22}Q_{11}$, as determined by the alignment of the bond variables. }
\label{fig:lower_bound_calculation}
\end{figure}

In this and all subsequent appendices, we consider the spatially uniform unital circuit described by Eq.~\eqref{eq:tm}, with $p \in [0,\frac{q+1}{q}]$ interpreted as the generalized parameter of Appendix~\ref{tm-derivation}. In the present appendix, we additionally impose periodic boundary conditions. In Appendix~\ref{app:real_eigvals}, we showed that the transfer matrix $T \in \mathbb{R}^{2^N \times 2^N}$ has the same nonzero eigenvalues as a reduced matrix $K  \in \mathbb{R}^{2^{N/2} \times 2^{N/2}}$. Within this reduced vector space, we defined an inner product induced by the positive-definite matrix $\eta$ and showed that $K$ is Hermitian with respect to this inner product. It follows that, for any nonzero vector $\boldsymbol{\psi} \in \mathbb{R}^{2^{N/2}}$ that is $\eta$-orthogonal to the vacuum, the eigenvalue $\lambda_*$ is bounded from below by the Rayleigh quotient
\begin{equation}
\lambda_* \geq \frac{\langle \boldsymbol{\psi}, K \boldsymbol{\psi} \rangle_\eta}{\langle \boldsymbol{\psi}, \boldsymbol{\psi} \rangle_{\eta} }.
\label{rayleigh_quotient}
\end{equation}
For each positive integer $m$, let $S^{(m)}\in\mathbb{R}^{2^m \times 2^m}$ be the cyclic shift operator defined by its action on computational basis states:
\begin{equation}
S^{(m)} \ket{s_0, s_1,\ldots ,s_{m-1}} = \ket{s_{m-1},s_0,s_1,\ldots,s_{m-2}}.
\end{equation}
When the dimension of the space is clear from context, we suppress the superscript and write $S$ in place of $S^{(m)}$. For $1\leq l<N/2$, define $\ket{\psi_l}$ to be a superposition of all computational-basis states that consist of an island of $l$ consecutive ones surrounded by zeros:
\begin{equation}
\ket{\psi_l} = \sum_{n=0}^{N/2-1} \ket*{\boldsymbol{s}^{(n,l)}},\qquad \ket*{\boldsymbol{s}^{(n,l)}} = S^n \bigl(\ket{1}^{\otimes l} \otimes \ket{0}^{\otimes N/2-l}\bigr).
\end{equation}
We use $\boldsymbol{s}^{(n,l)}\in \{0,1\}^{N/2}$ to denote the bit string associated with the computational-basis state $\ket*{\boldsymbol{s}^{(n,l)}}$. For real coefficients $c_l$, let $\ket*{\psi^{(L)}}$ denote a truncated superposition of island states of length at most $L$: 
\begin{equation}
\ket*{\psi^{(L)}} = \sum_{l=1}^L c_l \ket{\psi_l}.
\end{equation}
Our variational ansatz is
\begin{equation}
\ket{\psi} = \ket*{\psi^{(L)}} - \frac{\langle \boldsymbol{0},\psi^{(L)} \rangle_\eta}{\langle \boldsymbol{0},\boldsymbol{0} \rangle_\eta} \ket{\boldsymbol{0}},
\end{equation}
where $\ket{\boldsymbol{0}}=\ket{0}^{\otimes N/2}$ denotes the vacuum. Throughout this appendix, we assume $L < \frac{N}{4}$. Using the metric in Eq.~\eqref{eq:metric}, the ansatz takes the form
\begin{equation}
\ket{\psi} = \sum_{l=1}^L c_l \ket{\psi_l} - \frac{N}{2} \bigl( \sum_{l=1}^Lq^{-4l} c_l \bigr) \ket{\boldsymbol{0}}.
\label{eq:trial_state}
\end{equation}
The denominator of the Rayleigh quotient is
\begin{equation}
\langle\psi,\psi \rangle_\eta  =\sum_{l,l^\prime =1}^L c_l c_{l^\prime} \langle \psi_{l^\prime},\psi_l \rangle_{\eta} -\frac{N^2}{4} \sum_{l,l^\prime=1}^L c_l c_{l^\prime} q^{2N-4(l+l^\prime)}.
\label{eq:rquotient_denominator}
\end{equation}
We expand the inner product between island states as
\begin{equation}
\langle \psi_{l^\prime}, \psi_{l}\rangle_{\eta} = \sum_{n,m=0}^{N/2-1} \bigl( \bra{1}^{\otimes l^\prime} \otimes \bra{0}^{\otimes N/2-l^\prime} \bigr) S^{\dagger n}     S^m \bigl((\ket{0}+\ket{1})^{\otimes l} \otimes (q^4 \ket{0}+\ket{1})^{\otimes N/2-l} \bigr).
\end{equation}
This inner product can be viewed as a weighted sum over the $(N/2)^2$ configurations of two islands of lengths $l$ and $l^\prime$, with each configuration specified by the pair of positions $(n,m) \in \mathbb{Z}_{N/2}^2$. An example is depicted in Fig.~\ref{fig:lower_bound_calculation}(a). Explicitly,
\begin{equation}
\langle \psi_{l^\prime}, \psi_{l}\rangle_{\eta} = \sum_{n,m=0}^{N/2-1} \prod_{j=0}^{N/2-1} f(s^{(n,l^\prime)}_j,s^{( m,l)}_j).
\end{equation}
Here $f:\{0,1\}^2 \rightarrow \mathbb{R}$ is the function given by
\begin{equation}
f(0,0) =q^4,\qquad f(0,1)=f(1,0)=f(1,1)=1.
\end{equation}
Under the assumption $L<N/4$, the inner product evaluates to
\begin{equation}
\begin{gathered}
\langle \psi_{l^\prime},\psi_{l} \rangle_{\eta} =\frac{N}{2} q^{2N}\Bigl[ q^{-4 \max(l,l^\prime)} \bigl(|l-l^\prime| +\alpha\bigr) -q^{-4(l+l^\prime)} \bigl( l+l^\prime +\alpha-N/2 \bigr)\Bigr],\\
\alpha = \frac{q^4+1}{q^4-1}.
\end{gathered}
\end{equation}
Substituting this expression into Eq.~\eqref{eq:rquotient_denominator} yields
\begin{equation}
\langle \psi,\psi\rangle_\eta = \frac{N}{2} q^{2N} \boldsymbol{c}^T G \boldsymbol{c},
\end{equation}
where $\boldsymbol{c}=(c_1,\ldots,c_L)^T$, and $G\in \mathbb{R}^{L \times L}$ has elements
\begin{equation}
G_{l,l^\prime} =q^{-4 \max(l,l^\prime)} \bigl(|l-l^\prime| +\alpha\bigr) -q^{-4(l+l^\prime)} \bigl(l+l^\prime +\alpha\bigr),\qquad l,l^\prime =1,\ldots,L.\label{eq:G_matrix}
\end{equation}
We next evaluate the numerator of the Rayleigh quotient. Since $K \ket{\boldsymbol{0}} = \ket{\boldsymbol{0}}$ and $(\eta K)^T =\eta K$,
\begin{equation}
 \langle \boldsymbol{0}, K \psi_l \rangle_\eta = \langle \psi_l, K \boldsymbol{0} \rangle_\eta = \langle \psi_l,\boldsymbol{0}\rangle_\eta.
\end{equation}
Hence,
\begin{equation}
\langle\psi,K \psi \rangle_\eta  =\sum_{l,l^\prime =1}^L c_l c_{l^\prime} \langle \psi_{l^\prime},K\psi_l \rangle_{\eta} -\frac{N^2}{4} \sum_{l,l^\prime=1}^L c_l c_{l^\prime} q^{2N-4(l+l^\prime)}.
\end{equation}
As discussed in Appendix~\ref{app:real_eigvals}, the matrix $\eta K$ can be written as
\begin{equation}
\eta K = (WV)^{\otimes N/2} S^{(N)\dagger} M^{\otimes N/2} S^{(N)} U^{\otimes N/2},
\end{equation}
where $M$ is the local transfer matrix given in Eq.~\eqref{eq:tm} and $U$ is given in Eq.~\eqref{eq:Kmatrix}. We have the relations
\begin{equation}
\bra{0}WV = (q^2 \bra{0} +\bra{1})^{\otimes 2} ,\qquad \bra{1} WV= (\bra{0}+\gamma\bra{1})^{\otimes 2}.
\end{equation}
For convenience, define
\begin{equation}
\mu^{(n,l)}_j =\delta_{s^{(n,l)}_j, 0} q^2 + \delta_{s^{(n,l)}_j,1},\qquad \nu^{(n,l)}_j =  \delta_{s^{(n,l)}_j,0} +\delta_{s^{(n,l)}_j,1} \gamma.
\end{equation}
Using these definitions,
\begin{equation}
\begin{aligned}
\bra*{\boldsymbol{s}^{(n,l)}}\eta K &= \bra*{\boldsymbol{s}^{(n,l)}} (WV)^{\otimes N/2}S^{(N) \dagger} M^{\otimes N/2} S^{(N)} U^{\otimes N/2} \\
&= \bigotimes_{j=0}^{N/2-1} \bigl[ (\mu^{(n,l)}_j \bra{0} + \nu^{(n,l)}_j \bra{1})^{\otimes 2} \bigr] S^{(N) \dagger} M^{\otimes N/2} S^{(N)} U^{\otimes N/2}\\
&= \bigotimes_{j=0}^{N/2-1} \bigl[ (\mu^{(n,l)}_{j-1} \bra{0} + \nu^{(n,l)}_{j-1} \bra{1}) \otimes(\mu^{(n,l)}_j \bra{0} + \nu^{(n,l)}_j \bra{1})  \bigr] M^{\otimes N/2} S^{(N)} U^{\otimes N/2}\\
&= \bigotimes_{j=0}^{N/2-1} \bigl[ \mu^{(n,l)}_{j-1} \mu^{(n,l)}_{j} \bra{0,0} M +\nu^{(n,l)}_{j-1} \nu^{(n,l)}_j \bra{1,1} M\bigr] S^{(N)} U^{\otimes N/2}\\
&= \bigotimes_{j=0}^{N/2-1} \bigl[  \mu^{(n,l)}_{j-1} \mu^{(n,l)}_j \bra{0,0} + (a \mu^{(n,l)}_{j-1} \mu^{(n,l)}_{j} + b\nu^{(n,l)}_{j-1} \nu^{(n,l)}_{j}) (\bra{0,1} +\bra{1,0})\\ &\quad + (c \mu^{(n,l)}_{j-1} \mu^{(n,l)}_{j} + d\nu^{(n,l)}_{j-1} \nu^{(n,l)}_{j}) \bra{1,1} \bigr] S^{(N)} U^{\otimes N/2}.
\end{aligned}
\end{equation}
It follows that
\begin{equation}
\bra{\boldsymbol{s}^{(n,l)}} \eta K \ket{\boldsymbol{s}^{(m,l^\prime)}} = \prod_{j=0}^{N/2-1} g \Bigl((s^{(n,l)}_{j-1} ,s^{(n,l)}_j ),(s^{(m,l^\prime)}_{j-1} ,s^{(m,l^\prime)}_j ) \Bigr).
\end{equation}
Here $g:\{0,1\}^2 \times \{0,1\}^2 \rightarrow \mathbb{R}$ is a function of two bond variables. Each bond variable belongs to one of three classes: a bulk bond within a region of ones, $(1,1)$; a bulk bond within a region of zeros, $(0,0)$; or an edge bond, $(0,1)$ or $(1,0)$. We can represent the function $g$ using a symmetric matrix $Q$: 
\begin{equation}
g((s_0,s_1),(s_2,s_3)) = Q_{s_0+s_1, s_2 +s_3},\qquad Q =\begin{pmatrix}
Q_{00} & Q_{01} & Q_{02}\\
Q_{10} & Q_{11} & Q_{12}\\
Q_{20} & Q_{21} & Q_{22}
\end{pmatrix}
 =\begin{pmatrix}
q^4 & q^2 & 1\\
q^2 & a q^2 +b \gamma & a+\gamma^2 b\\
1 & a+\gamma^2 b & c+\gamma^2 d
\end{pmatrix}.
\end{equation}
We can view $\langle \psi_{l^\prime} , K \psi_l \rangle_{\eta}$ as a weighted sum over configurations indexed by $(n,m) \in \mathbb{Z}_{N/2}^2$. One such configuration is depicted in Fig.~\ref{fig:lower_bound_calculation}(b). The weight of each configuration is the product of local factors, one for each pair of bond values, with each factor given by the corresponding entry of $Q$. Under the assumption $L<N/4$, we obtain
\begin{equation}
\begin{gathered}
\begin{aligned}
\langle \psi_{l^\prime},K \psi_l \rangle_{\eta} &=
\frac{N}{2} q^{2N} q^{-4(l+l^\prime)} \Bigl[  w^{\min(l,l^\prime)-1} \bigl(2uv+(|l-l^\prime|-1)v^2 +\frac{2v^2}{w-1}+\delta_{l,l^\prime}(u-v )^2\bigr)\\
&\quad+\frac{N}{2}+2u-l-l^\prime-1-\frac{2v^2}{w-1}\Bigr],
\end{aligned}\\
u=aq^2 +b\gamma,\qquad v =q^2\bigl(a + \gamma^2 b\bigr), \qquad w=q^4\bigl(c +\gamma^2 d\bigr).
\end{gathered}
\end{equation}
At $w=1$, both the preceding expression and the definition of $R$ below are understood by continuity. It follows that
\begin{equation}
\begin{gathered}
\langle \psi ,K \psi \rangle_{\eta} =\frac{N}{2} q^{2N} \boldsymbol{c}^T R \boldsymbol{c},\\
R_{l,l^\prime} = q^{-4(l+l^\prime)} \Bigl[  w^{\min(l,l^\prime)-1} \bigl(2uv+(|l-l^\prime|-1)v^2 +\frac{2v^2}{w-1}+\delta_{l,l^\prime}(u-v )^2\bigr)\\
\quad+2u-l-l^\prime-1-\frac{2v^2}{w-1}\Bigr].
\end{gathered}
\label{eq:R_matrix}
\end{equation}
Thus, for any nonzero coefficient vector $\boldsymbol{c}\in\mathbb{R}^L $, the Rayleigh quotient gives the variational lower bound
\begin{equation}
\lambda_* \geq \frac{\boldsymbol{c}^T R \boldsymbol{c}}{\boldsymbol{c}^T G \boldsymbol{c}}.
\end{equation}
For fixed $L$, the Rayleigh quotient is independent of $N$, subject only to the requirement $L<N/4$. To optimize the bound, we set the gradient of the Rayleigh quotient with respect to the coefficients to zero. This yields the generalized eigenvalue equation
\begin{equation}
R \boldsymbol{c} =\zeta G \boldsymbol{c}.
\label{eq:generalized_eigval_eqn}
\end{equation}
Since $R$ and $G$ are real symmetric, with $G$ positive definite, Eq.~\eqref{eq:generalized_eigval_eqn} is a symmetric-definite generalized eigenvalue problem. Its eigenvalues are real, and its eigenvectors may be chosen real. The largest generalized eigenvalue $\zeta$ is the maximum value of the Rayleigh quotient over our variational family. For large $L$, the generalized eigenvalue problem as written becomes numerically ill-conditioned. For numerical stability, we apply the diagonal congruence transformations $R \rightarrow DRD$ and $G\rightarrow DGD$, where $D\in \mathbb{R}^{L\times L}$ has elements $D_{l, l^\prime} = q^{2l} \delta_{l ,l^\prime}$. These transformations leave the generalized eigenvalues unchanged. We solve the transformed problem numerically for $L=10^3$ at different values of $p$ to obtain the lower bound shown in Fig.~\ref{fig:eigenvalue}.

\section{Two-Domain-Wall Spectrum at $p=0$ and the Discontinuity of $\lambda_*$}
\label{app:discontinuity}
\subsection{Two-domain-wall spectrum at $p=0$}
Throughout this appendix, we assume periodic boundary conditions. For each computational basis state $\ket{\boldsymbol{s}}$, labeled by $\boldsymbol{s}\in \{0,1\}^{N}$, we define the domain-wall number $n_{\text{dw}} (\boldsymbol{s})$ as the number of nearest-neighbor bonds across which the neighboring binary values differ. Equivalently,
\begin{equation}
n_{\text{dw}}(\boldsymbol{s}) = \sum_{j=0}^{N-1} (1- \delta_{s_j,s_{j+1}}).
\end{equation}
Define the $r$-domain-wall subspace by
\begin{equation}
\mathcal{W}_r = \text{span} \bigl\{ \ket{\boldsymbol{s}}:n_{\text{dw}}(\boldsymbol{s}) =r \bigr\}.
\end{equation}
The OTOC is obtained by evolving an initial state in $\mathcal{W}_2$. At $p=0$, domain walls can move or annihilate but cannot be created. Consequently, $\mathcal{W}_0 \oplus \mathcal{W}_2$ is invariant under the transfer matrix. Thus, at $p=0$, only basis vectors in $\mathcal{W}_0 \oplus \mathcal{W}_{2}$ contribute to the OTOC. Let $\Pi_{\mathcal{W}_r}$ denote the projector onto $\mathcal{W}_r$. For $i,j \in \{0,2,\ldots,N\}$, define
\begin{equation}
T_{\mathcal{W}_i \rightarrow \mathcal{W}_j} = \Pi_{\mathcal{W}_j} T \bigr|_{\mathcal{W}_i}:\mathcal{W}_i \rightarrow \mathcal{W}_j.
\end{equation}
Ordering the sectors by increasing domain-wall number, the transfer matrix can be written in block-triangular form
\begin{equation}
T = \begin{pmatrix}
T_{\mathcal{W}_0 \rightarrow \mathcal{W}_0} & T_{\mathcal{W}_2 \rightarrow \mathcal{W}_0} & \cdots & T_{\mathcal{W}_{N-2} \rightarrow \mathcal{W}_0} & T_{\mathcal{W}_{N} \rightarrow\mathcal{W}_0}\\
0 & T_{\mathcal{W}_2 \rightarrow \mathcal{W}_2} & \cdots & T_{\mathcal{W}_{N-2} \rightarrow \mathcal{W}_2}& T_{\mathcal{W}_N \rightarrow \mathcal{W}_2 }\\
\vdots & \ddots & \ddots & \vdots & \vdots \\
0 & \cdots & 0 &T_{\mathcal{W}_{N-2} \rightarrow \mathcal{W}_{N-2}} & T_{\mathcal{W}_N \rightarrow \mathcal{W}_{N-2}}\\
0  & 0 & \cdots & 0 & T_{\mathcal{W}_N \rightarrow \mathcal{W}_N}
\end{pmatrix}.
\end{equation}
It follows that the spectrum of $T$ is the union of the spectra of its diagonal blocks:
\begin{equation}
\text{spec}\bigl(T\bigr) = \bigcup_{r=0}^{N/2} \text{spec} \bigl(T_{\mathcal{W}_{2r} \rightarrow \mathcal{W}_{2r}} \bigr).
\end{equation}
We now determine the eigenvalues of $T$ associated with the two-domain-wall sector. In the following subsection, we show that the limiting value in Eq.~\eqref{eq:eigenvalue_limit} is the thermodynamic limit of the largest eigenvalue in the two-domain-wall band.

 Using the cyclic shift operator $S$ defined in Eq.~\eqref{eq:shift_operator} of the main text, for $j \in \mathbb{Z}_N$ and $1 \leq l \leq N-1$, we denote by $\ket{j,l}$ the computational basis state containing a single island of $l$ consecutive ones whose leftmost site is $j$:
\begin{equation}
\ket{j,l} = S^{j} \bigl( \ket{1}^{\otimes l} \otimes \ket{0}^{\otimes N-l}\bigr).
\end{equation}
For $\sigma \in \{0,1\}$, define the projected layer transfer matrices by
\begin{equation}
T^{(\sigma)}_{\mathcal{W}_2 \rightarrow \mathcal{W}_2} = \Pi_{\mathcal{W}_2} T^{(\sigma)} \bigr|_{\mathcal{W}_2}: \mathcal{W}_2 \rightarrow \mathcal{W}_2.
\end{equation}
Since components that leave $\mathcal{W}_2$ cannot return, we have
\begin{equation}
T_{\mathcal{W}_2 \rightarrow \mathcal{W}_2} =T^{(0)}_{\mathcal{W}_2 \rightarrow \mathcal{W}_2} T^{(1)}_{\mathcal{W}_2 \rightarrow \mathcal{W}_2}.
\end{equation}
For convenience, define
\begin{equation}
a_0 = a(0) =\frac{1}{q^2+1},\qquad b_0 = b(0) =\frac{q^2}{q^2+1}.
\end{equation}
The projected layer matrices act as
\begin{equation}
\begin{aligned}
T^{(1)}_{\mathcal{W}_2 \rightarrow\mathcal{W}_2} \ket{2j,2l} &= a_0^2 (1-\delta_{l,1}) \ket{2j+1,2(l-1)} +b_0^2 (1-\delta_{l,N/2-1}) \ket{2j-1, 2(l+1)}\\
& \quad +a_0 b_0 \bigl(\ket{2j-1,2l}+ \ket{2j+1,2l} \bigr),\\
T^{(0)}_{\mathcal{W}_2 \rightarrow \mathcal{W}_2}\ket{2j+1,2l} &= a_0^2(1-\delta_{l,1})  \ket{2j+2,2(l-1)} +b_0^2 (1-\delta_{l,N/2-1}) \ket{2j,2(l+1)}\\ 
& \quad+ a_0 b_0 \bigl( \ket{2j,2l} +\ket{2j+2,2l} \bigr).
\end{aligned}
\end{equation}
It follows that
\begin{equation}
T_{\mathcal{W}_2 \rightarrow \mathcal{W}_2} \ket{2j,2l} = H^2 \ket{2j,2l},
\end{equation}
where the action of $H$ on $\ket{j,2l}$ is
\begin{equation}
H \ket{j,2l} = a_0^2 (1-\delta_{l,1})\ket{j+1,2(l-1)} +b_0^2 (1-\delta_{l,N/2-1}) \ket{j-1,2(l+1)} +a_0b_0 \bigl( \ket{j-1,2l} +\ket{j+1,2l}\bigr).
\end{equation}
We introduce two subspaces $\mathcal{W}_2^{(0)}, \mathcal{W}_2^{(1)} \subset \mathcal{W}_2$, distinguished by whether the island's leftmost site is even or odd:
\begin{equation}
\begin{aligned}
\mathcal{W}_2^{(0)}  &= \text{span} \{ \ket{2j,2l} : 0 \leq j \leq \frac{N}{2}-1, 1 \leq l \leq \frac{N}{2}-1 \},\\
\mathcal{W}_2^{(1)}  &= \text{span} \{ \ket{2j+1,2l} : 0 \leq j \leq \frac{N}{2}-1, 1 \leq l \leq  \frac{N}{2}-1 \}.
\end{aligned}
\end{equation}
The projected layer matrices satisfy
\begin{equation}
 T^{(0)}_{\mathcal{W}_2 \rightarrow \mathcal{W}_2}(\mathcal{W}_2)  \subseteq \mathcal{W}_2^{(0)},\qquad T^{(1)}_{\mathcal{W}_2 \rightarrow \mathcal{W}_2} (\mathcal{W}_2)  \subseteq \mathcal{W}_2^{(1)}.
\end{equation}
Let $\Pi_0:\mathcal{W}_2 \rightarrow \mathcal{W}_2^{(0)}$ denote the projector onto $\mathcal{W}^{(0)}_2$. Define
\begin{equation}
A =   T^{(1)}_{\mathcal{W}_2 \rightarrow \mathcal{W}_2} \big|_{\mathcal{W}^{(0)}_2}, \qquad B =T^{(0)}_{\mathcal{W}_2 \rightarrow \mathcal{W}_2} \big|_{\mathcal{W}^{(1)}_2},\qquad C =\Pi_0 T_{\mathcal{W}_2 \rightarrow \mathcal{W}_2} \big|_{\mathcal{W}_2^{(1)}}.
\end{equation}
With respect to the ordered decomposition $\mathcal{W}^{(0)}_2 \oplus \mathcal{W}_2^{(1)}$, the restriction of the two-domain-wall block takes the form
\begin{equation}
T_{\mathcal{W}_2 \rightarrow\mathcal{W}_2} \bigr|_{\mathcal{W}^{(0)}_2 \oplus \mathcal{W}_2^{(1)}}=
\begin{pmatrix}
BA & C\\
0 & 0
\end{pmatrix}.
\end{equation}
On the same space,
\begin{equation}
H^2 = \begin{pmatrix}
BA & 0 \\
0 & AB
\end{pmatrix}.
\end{equation}
The nonzero eigenvalues of $T_{\mathcal{W}_2 \rightarrow\mathcal{W}_2}$ are those of $BA$. Since $AB$ and $BA$ have the same spectrum, $H^2$ and $T_{\mathcal{W}_2 \rightarrow\mathcal{W}_2}$ have the same distinct nonzero eigenvalues, although their multiplicities differ. It follows that every nonzero eigenvalue $\lambda$ of $T$ in the two-domain-wall sector can be written as $\lambda = \mu^2$ for some eigenvalue $\mu$ of $H$. For each $n\in \mathbb{Z}_N$, consider a momentum state of the form
\begin{equation}
\ket{\phi_n} = \sum_{l=1}^{N/2-1} c^{(n)}_{l} \ket{\phi_{n,2l}}, \qquad \ket{\phi_{n,l} } =\sum_{j=0}^{N-1} e^{-\frac{i2 \pi nj }{N}} \ket{j,l}.
\end{equation}
Applying $H$ to $\ket{\phi_{n,2l}}$ gives
\begin{equation}
H \ket{\phi_{n,2l}} = e^{\frac{i 2 \pi n}{N}} a_0^2 \ket{\phi_{n,2(l-1)}} + e^{-\frac{i 2\pi n}{N}} b_0^2 \ket{\phi_{n,2(l+1)}} +2a_0 b_0 \cos(\frac{2 \pi n}{N}) \ket{\phi_{n,2l}}.
\end{equation}
The eigenvalue equation $H\ket{\phi_n} = \mu \ket{\phi_n}$ holds if the coefficients $c^{(n)}_l$ satisfy, for $1\leq l \leq \frac{N}{2}-1$,
\begin{equation}
e^{\frac{i 2\pi n}{N}}a_0^2 c^{(n)}_{l+1} + e^{-\frac{i 2 \pi n}{N}} b_0^2 c^{(n)}_{l-1} +2 a_0 b_0 \cos\Bigl( \frac{2 \pi n}{N} \Bigr) c^{(n)}_l =\mu c^{(n)}_l,
\label{eq:c_update}
\end{equation}
together with the boundary conditions
\begin{equation}
c^{(n)}_0 = c^{(n)}_{N/2} = 0.
\end{equation}
Substituting 
\begin{equation}
c^{(n)}_l = \bigl(\frac{b_0}{a_0}\bigr)^l e^{- \frac{i 2 \pi nl}{N}} d^{(n)}_l
\end{equation}
into Eq.~\eqref{eq:c_update} gives
\begin{equation}
\begin{gathered}
d^{(n)}_{l+1} + d^{(n)}_{l-1} + 2 \cos \bigl(\frac{2 \pi n}{N} \bigr) d^{(n)}_l = \frac{\mu}{a_0 b_0}d^{(n)}_l,\\
d^{(n)}_0 = d^{(n)}_{N/2} =0.
\end{gathered}
\end{equation}
Up to an overall normalization of the eigenvectors, the solutions are
\begin{equation}
\begin{gathered}
d_l^{(n,m)} = \sin \Bigl( \frac{2 \pi m l}{N} \bigr),\\
\mu(n,m) = 2a_0b_0 \Bigl( \cos \frac{2 \pi n}{N} + \cos \frac{2 \pi m}{N} \Bigr),
\end{gathered}
\end{equation}
where $n \in \mathbb{Z}_N$ and $m \in \mathbb{Z}_{N/2}\setminus \{0\}$. The corresponding eigenvalues of $T$ are $\lambda(n,m) = \mu^2(n,m)$. The largest eigenvalue of $T$ in the two-domain-wall band is $\mu^2(0,1)$, whose thermodynamic limit is
\begin{equation}
\lim_{N\rightarrow \infty} \mu^2(0,1)  =  \Bigl(\frac{2q}{q^2+1}\Bigr)^4.
\end{equation}
In the next subsection, we show that this is the limiting value in Eq.~\eqref{eq:eigenvalue_limit}.

\subsection{Discontinuity in $\lambda_*$}
Let $\widetilde{T}$ denote the vacuum-removed transfer matrix
\begin{equation}
\widetilde{T} = \bigl(I-\ket{0}^{\otimes N} \bra{0}^{\otimes N} \bigr) T \bigl( I-\ket{0}^{\otimes N} \bra{0}^{\otimes N} \bigr)= \bigl(I-\ket{0}^{\otimes N} \bra{0}^{\otimes N} \bigr) T.
\end{equation}
The second equality follows because
\begin{equation}
T \ket{0}^{\otimes N} = \ket{0}^{\otimes N}.
\end{equation}
This construction replaces the vacuum eigenvalue with zero while leaving the remaining eigenvalues unchanged. It therefore follows from Appendix~\ref{app:real_eigvals} that all eigenvalues of $\widetilde{T}$ are real and lie in $[0,1]$. Define $\lambda_*$ to be the largest eigenvalue of $\widetilde{T}$. At $p=0$, for any $N$, $\ket{1}^{\otimes N}$ is an eigenvector of $\widetilde{T}$ with eigenvalue $1$. Since $\widetilde{T}$ depends continuously on $p$, for any $N$,
\begin{equation}
\lim_{p \rightarrow 0^+} \lambda_*(N,p) =1.
\end{equation}
Consequently,
\begin{equation}
\lim_{N \rightarrow \infty} \lim_{p \rightarrow 0^+} \lambda_*(N,p) =1.
\end{equation}
We now show that
\begin{equation}
\lim_{p \rightarrow 0^+} \limsup_{N \rightarrow \infty} \lambda_*(N,p) =\lim_{p \rightarrow 0^+} \liminf_{N \rightarrow \infty} \lambda_*(N,p)= \Bigl(\frac{2q}{q^2+1}\Bigr)^4.
\label{eq:app_lambda_limit}
\end{equation}
We establish this by a squeeze argument: entrywise nonnegativity of the transfer matrix gives an upper bound through a suitable positive left test vector, while a Rayleigh quotient gives a lower bound. The two bounds coincide as $p\rightarrow 0^+$ after taking the large-$N$ limit superior and limit inferior, respectively.

Let $\boldsymbol{v}\in\mathbb{R}^{2^N}$ have strictly positive components $v_i>0$. Suppose that, for some nonnegative scalar $\theta$, the following componentwise inequality holds:
\begin{equation}
\boldsymbol{v}^T \widetilde{T} \leq \theta \boldsymbol{v}^T.
\label{eq:theta_inequality}
\end{equation}
Then every eigenvalue $\lambda$ of $\widetilde{T}$ satisfies
\begin{equation}
\lambda \leq \theta.
\label{eq:theta_bound}
\end{equation}
The scalar $\theta$ is a Collatz-Wielandt upper bound on the spectral radius $\rho(\widetilde{T})$~\cite{HornJohnson2012}.
Let $\boldsymbol{u} \neq\boldsymbol{0}$ be a right eigenvector of $\widetilde{T}$ with eigenvalue $\lambda$.
For each component $u_i$,
\begin{equation}
\begin{aligned}
|\lambda||u_i| &= \bigl|\sum_j \widetilde{T}_{ij} u_j \bigr|\leq \sum_j \bigl| \widetilde{T}_{ij} u_j \bigr|= \sum_j\widetilde{T}_{ij} |u_j|.
\end{aligned}
\end{equation}
The last equality follows from the entrywise nonnegativity of $\widetilde{T}$. Combining this with Eq.~\eqref{eq:theta_inequality}, we obtain
\begin{equation}
|\lambda| \sum_i v_i |u_i| \leq  \sum_{ij} v_i \widetilde{T}_{ij} |u_j| \leq \theta \sum_j v_j |u_j|.
\end{equation}
Because $\boldsymbol{v}$ is strictly positive and $\boldsymbol{u} \neq \boldsymbol{0}$, the sum $\sum_i v_i|u_i|$ is positive. Dividing by this sum and using the nonnegativity of the eigenvalues of $\widetilde{T}$ establishes Eq.~\eqref{eq:theta_bound}. Next, fix the test row vector to be
\begin{equation}
\bra{\boldsymbol{v}} = \sum_{\boldsymbol{s} \in \{0,1\}^N} q^{-h(\boldsymbol{s})} \bra{\boldsymbol{s}},\qquad h(\boldsymbol{s}) = \sum_{j=0}^{N-1} s_j.
\end{equation}
Here $h(\boldsymbol{s})$ is the Hamming weight of the bit string $\boldsymbol{s}$. We can express the vacuum-removed transfer matrix $\widetilde{T}$ in terms of vacuum-removed layer transfer matrices as
\begin{equation}
\widetilde{T} = \widetilde{T}^{(0)} \widetilde{T}^{(1)},\qquad \sigma \in \{0,1\}, \quad \widetilde{T}^{(\sigma)} =  (I -\ket{0}^{\otimes N} \bra{0}^{\otimes N})T^{(\sigma)}.
\end{equation}
For any $\boldsymbol{s} \in \{ 0,1\}^{N}$,
\begin{equation}
\begin{aligned}
\bra{\boldsymbol{v}} T^{(\sigma)} \ket{\boldsymbol{s}} &= \prod_{j=0}^{N/2-1} \Bigl( \bra{00} M \ket{s_{2j+\sigma}s_{2j+1+\sigma}}+q^{-2} \bra{11} M \ket{s_{2j+\sigma} s_{2j+1+\sigma}} \bigr)\\
&=A^{n^{(\sigma)}_{01,10}(\boldsymbol{s})} B^{n^{(\sigma)}_{11} (\boldsymbol{s})} \bra{\boldsymbol{v}} \boldsymbol{s} \rangle.
\end{aligned}
\end{equation}
The coefficients $A$ and $B$ are
\begin{equation}
A(p) = qa(p) +q^{-1} b(p),\qquad B(p)=q^{2}c(p) +d(p).
\end{equation}
For $p\in [0, \frac{q+1}{q}]$, both $A(p)$ and $B(p)$ are nonnegative and satisfy
\begin{equation}
\frac{1}{q} \leq A(p) \leq \frac{2q}{q^2+1} <1, \qquad \frac{1}{q^2} \leq B(p) \leq 1. 
\end{equation}
The exponents $n^{(\sigma)}_{01,10} (\boldsymbol{s})$ and $n^{(\sigma)}_{11} (\boldsymbol{s})$ are
\begin{equation}
\begin{aligned}
n^{(\sigma)}_{01,10}(\boldsymbol{s}) &= \sum_{j=0}^{N/2-1} \bigl| s_{2j+\sigma+1} - s_{2j +\sigma} \bigr|, \\
n^{(\sigma)}_{11}(\boldsymbol{s}) &= \sum_{j=0}^{N/2-1} s_{2j + \sigma} s_{2j+\sigma +1}.
\end{aligned}
\end{equation}
It follows that
\begin{equation}
\begin{aligned}
\bra{\boldsymbol{v}} \widetilde{T}^{(\sigma)} \ket{\boldsymbol{s}} &= A^{n^{(\sigma)}_{01,10}(\boldsymbol{s})} B^{n^{(\sigma)}_{11} (\boldsymbol{s})} \bra{\boldsymbol{v}} \boldsymbol{s} \rangle -  \bra{\boldsymbol{0}}T^{(\sigma)} \ket{\boldsymbol{s}} \\
&\leq A^{n^{(\sigma)}_{01,10}(\boldsymbol{s})} B^{n^{(\sigma)}_{11} (\boldsymbol{s})} \bra{\boldsymbol{v}} \boldsymbol{s} \rangle.
\end{aligned}
\end{equation}
Define $\mathcal{R}^{(\sigma)}$ by
\begin{equation}
\mathcal{R}^{(\sigma)} = \{ \boldsymbol{s} \in \{0,1\}^{N} \setminus\{\boldsymbol{0}\}: \forall j \in \mathbb{Z}_{N/2} , s_{2j+\sigma}=s_{2j+\sigma+1} \}.
\end{equation}
We start by showing
\begin{equation}
\max_{\boldsymbol{s}^\prime \in \mathcal{R}^{(1)}} A^{n^{(0)}_{01,10} (\boldsymbol{s}^\prime)} B^{n^{(0)}_{11} (\boldsymbol{s}^\prime)} =\max_{\boldsymbol{s}^\prime \in \mathcal{R}^{(0)}} A^{n^{(1)}_{01,10} (\boldsymbol{s}^\prime)} B^{n^{(1)}_{11} (\boldsymbol{s}^\prime)} =\max \bigl(A^2,B^{N/2} \bigr).
\label{eq:maxAB}
\end{equation}
To evaluate this maximum, consider $\boldsymbol{s} \in \mathcal{R}^{(\sigma)}$. For every $ j \in \mathbb{Z}_{N/2}$, we have $s_{2j+\sigma} =s_{2j+1 +\sigma}$. For the all-ones string,
\begin{equation}
n^{(1-\sigma)}_{01,10} (\boldsymbol{1}) =  0,\qquad n^{(1-\sigma)}_{11}(\boldsymbol{1}) = \frac{N}{2}.
\end{equation}
Because the vacuum is excluded, every string other than the all-ones string contains at least two domain walls, and therefore
\begin{equation}
n^{(1-\sigma)}_{01,10} (\boldsymbol{s}) \geq 2.
\end{equation}
Its weight is consequently at most $A^2$. This value is attained by a string containing a single paired block of ones surrounded by zeros, for which $n^{(1-\sigma)}_{01,10}=2$ and $n^{(1-\sigma)}_{11}=0$. Since the all-ones string has weight $B^{N/2}$, Eq.~\ref{eq:maxAB} follows. Let
\begin{equation}
\mathcal{V}^{(\sigma)} =\text{span}\{ \ket{\boldsymbol{s}}: \boldsymbol{s} \in \mathcal{R}^{(\sigma)}\}.
\end{equation}
Since
\begin{equation}
\text{im}(\widetilde{T}^{(\sigma)}) \subseteq \mathcal{V}^{(\sigma)}, \qquad \text{im}(\widetilde{T}) \subseteq \mathcal{V}^{(0)},
\end{equation}
the subspace $\mathcal{V}^{(0)}$ is invariant under $\widetilde{T}$, and every eigenvector with a nonzero eigenvalue lies in $\mathcal{V}^{(0)}$. It therefore suffices to apply the positive-vector bound to $\widetilde{T}|_{\mathcal{V}^{(0)}}$. For any $\boldsymbol{s} \in \mathcal{R}^{(0)}$,
\begin{equation}
\begin{aligned}
\bra{\boldsymbol{v}} \widetilde{T} \ket{\boldsymbol{s}} &= \sum_{\boldsymbol{s}^\prime \in \{0,1\}^N} \bra{\boldsymbol{v}} \widetilde{T}^{(0)} \ket{\boldsymbol{s}^\prime}\bra{\boldsymbol{s}^\prime}  \widetilde{T}^{(1)} \ket{\boldsymbol{s}}\\
&= \sum_{\boldsymbol{s}^\prime \in \mathcal{R}^{(1)} } \bra{\boldsymbol{v}} \widetilde{T}^{(0)} \ket{\boldsymbol{s}^\prime}\bra{\boldsymbol{s}^\prime}  \widetilde{T}^{(1)} \ket{\boldsymbol{s}}\\
&\leq \sum_{\boldsymbol{s}^\prime \in \mathcal{R}^{(1)} }  A^{n^{(0)}_{01,10} (\boldsymbol{s}^\prime)} B^{n^{(0)}_{11} (\boldsymbol{s}^\prime)} \bra{\boldsymbol{v}} \boldsymbol{s}^\prime \rangle \bra{\boldsymbol{s}^\prime} \widetilde{T}^{(1)} \ket{\boldsymbol{s}}\\
&\leq \Bigl( \max_{\boldsymbol{s}^\prime \in \mathcal{R}^{(1)}} A^{n^{(0)}_{01,10} (\boldsymbol{s}^\prime)} B^{n^{(0)}_{11} (\boldsymbol{s}^\prime)} \Bigr) \bra{\boldsymbol{v}} \widetilde{T}^{(1)} \ket{\boldsymbol{s}}\\
& \leq\Bigl( \max_{\boldsymbol{s}^\prime \in \mathcal{R}^{(1)}} A^{n^{(0)}_{01,10} (\boldsymbol{s}^\prime)} B^{n^{(0)}_{11} (\boldsymbol{s}^\prime)} \Bigr) \Bigl( A^{n^{(1)}_{01,10} (\boldsymbol{s})} B^{n^{(1)}_{11} (\boldsymbol{s})} \Bigr)\bra{\boldsymbol{v}} \boldsymbol{s} \rangle\\
& \leq \Bigl( \max_{\boldsymbol{s}^\prime \in \mathcal{R}^{(1)}} A^{n^{(0)}_{01,10} (\boldsymbol{s}^\prime)} B^{n^{(0)}_{11} (\boldsymbol{s}^\prime)} \Bigr) \Bigl( \max_{\boldsymbol{s}^{\prime \prime} \in \mathcal{R}^{(0)}} A^{n^{(1)}_{01,10} (\boldsymbol{s}^{\prime \prime})} B^{n^{(1)}_{11} (\boldsymbol{s}^{\prime \prime})} \Bigr) \bra{\boldsymbol{v}} \boldsymbol{s} \rangle\\
&= \max \bigl(A^2, B^{N/2} \bigr)^2 \bra{\boldsymbol{v}} \boldsymbol{s} \rangle.
\end{aligned}
\end{equation}
Applying Eq.~\eqref{eq:theta_bound} with $\theta=\max \bigl( A^2(p), B^{N/2}(p) \bigr)^2$ gives
\begin{equation}
\lambda_*(N,p) \leq \max \bigl( A^2(p), B^{N/2}(p) \bigr)^2.
\end{equation}
For any $p>0$, $B(p)<1$, so the preceding bound gives
\begin{equation} 
 \limsup_{N\rightarrow \infty} \lambda_*(N,p) \leq A^4(p).
\end{equation}
Taking $p\rightarrow 0^+$ gives
\begin{equation}
 \limsup_{p\rightarrow 0^+}\limsup_{N\rightarrow \infty} \lambda_*(N,p) \leq A^4(0) =\Bigl(\frac{2q}{q^2+1}\Bigr)^4.
\end{equation}
This gives the upper bound for the squeeze argument. We next derive the lower bound.

In Appendix~\ref{app:lower_bound}, we derive a variational lower bound on $\lambda_*$ by calculating the Rayleigh quotient in Eq.~\eqref{rayleigh_quotient} for a trial state. The trial state $\ket{\psi} \in \mathbb{R}^{2^{N/2}}$, given in Eq.~\eqref{eq:trial_state}, is constructed from two-domain-wall states with lengths up to $L\in \mathbb{N}$. The variational ansatz is parameterized by real coefficients $c_l$, with $l=1,2,\ldots,L$. The variational bound on $\lambda_*$ is
\begin{equation}
\lambda_*(N,p) \geq \frac{\langle \psi, K \psi \rangle_\eta}{\langle \psi,\psi \rangle_\eta} = \frac{1}{Z_L}\sum_{l,l^\prime=1}^L c_l c_{l^\prime} R_{l l^\prime} (p),\qquad Z_L =\sum_{m,m^\prime=1}^L c_m c_{m^\prime} G_{m m^\prime},
\end{equation}
where $G$ and $R(p)$ are defined in Eqs.~\eqref{eq:G_matrix} and \eqref{eq:R_matrix}. Apart from the restriction that $N > 4L$, this bound is $N$ independent. Thus, it is a valid bound in the thermodynamic limit for any $p\in [0,\frac{q+1}{q}]$ and any $L \in \mathbb{N}$. Taking $\liminf_{N\rightarrow\infty}$, followed by $\liminf_{p\rightarrow0^+}$, and using continuity of $R(p)$, gives
\begin{equation}
\liminf_{p\rightarrow 0^+} \liminf_{N \rightarrow \infty}  \lambda_*(N,p) \geq  \frac{\sum_{l,l^\prime=1}^L c_l c_{l^\prime} R_{l l^\prime} (0)}{\sum_{m,m^\prime=1}^L c_m c_{m^\prime} G_{m m^\prime} }.
\label{eq:rayleigh_quotient_limit}
\end{equation}
Let $X$ be the $(L+2) \times (L+2)$ matrix with the following nonzero elements:
\begin{equation}
\begin{gathered}
\qquad X_{l l} =2 a_0 b_0,\qquad X_{l+1l} = b_0^2 \qquad (1\leq l \leq L+1),\\
X_{l-1 l} = a_0^2, \qquad(2\leq l \leq L+1),\\
a_0 = \frac{1}{q^2+1},\qquad b_0=\frac{q^2}{q^2+1}.
\end{gathered}
\end{equation}
The remaining matrix elements of $X$ are zero. For each $\kappa \in \mathbb{R}$ and $l,m \in \mathbb{Z}$, define
\begin{equation}
\Phi^{(\kappa)}(l,m) =q^{-4 \max(l,m)} (|l-m|+ \kappa) - (l+m+\kappa) q^{-4(l+m)}.
\end{equation}
We can express the entries of $R(0)$ and $G$ as
\begin{equation}
\begin{gathered}
G_{lm} =\Phi^{(\alpha)} (l,m),\qquad R_{lm}(0) =\Phi^{(\alpha -2 \beta)}(l,m) + \beta^2 q^{-4l} \delta_{lm},\\
\alpha = \frac{q^4+1}{q^4-1},\qquad \beta =b_0^2-a_0^2.
\end{gathered}
\end{equation}
First, we show that the $R$ matrix at $p=0$ can be expressed as
\begin{equation}
R_{lm} (0) =\sum_{k=1}^{L+2} \Phi^{(\alpha)}(l,k) (X^2)_{km}.
\label{eq:R_X_relation}
\end{equation}
For $1\leq l \leq L$ and $1\leq m \leq L+1$, a direct calculation gives
\begin{equation}
\sum_{k=1}^{L+2} \Phi^{(\kappa)}(l,k) X_{km} = \Phi^{(\kappa-\beta)}(l,m) + \xi(\kappa)q^{-4l} \delta_{l,m},\qquad \xi(\kappa ) =a_0^2 +b_0^2 - \kappa \beta.
\label{eq:phi_X_update}
\end{equation}
This expression is obtained by separately checking the three cases $l \leq m-1$, $l=m$, and $l \geq m+1$. Since $\xi(\alpha)=0$ and $\xi(\alpha-\beta)=\beta^2$, applying Eq.~\eqref{eq:phi_X_update} twice yields Eq.~\eqref{eq:R_X_relation}.

Next, we fix the coefficients to be
\begin{equation}
c_l =q^{2l} \sin \Bigl(\frac{ \pi l}{L}\Bigr),\qquad l=1,2,\ldots,L.
\end{equation}
Substituting this ansatz into the right-hand side of Eq.~\eqref{eq:rayleigh_quotient_limit} and using Eq.~\eqref{eq:R_X_relation}, we obtain the $p=0$ Rayleigh quotient:
\begin{equation}
\begin{gathered}
\frac{\langle \psi,K \psi \rangle_\eta}{\langle \psi,\psi \rangle_{\eta}} = \nu_L^2 + \frac{\mathcal{B}_L}{Z_L},\\
\nu_L=\Bigl(\frac{2q}{q^2+1}\Bigr)^2 \cos^2 \bigl(\frac{\pi}{2L}\bigr),\\
\mathcal{B}_L =\sum_{l=1}^{L}c_l c_{L-1}\Bigl[a_0^2 b_0^2  \Phi^{(\alpha)}(l,L-1) +(2a_0 b_0^3 +b_0^2 \nu_L) \Phi^{(\alpha)}(l,L) +b_0^4  \Phi^{(\alpha)}(l,L+1) \Bigr].
\end{gathered}
\end{equation}
Since $\mathcal{B}_L/Z_L$ is manifestly nonnegative, we have
\begin{equation}
\liminf_{p\rightarrow 0^+} \liminf_{N \rightarrow \infty} \lambda_*(N,p) \geq \nu_L^2 +\frac{\mathcal{B}_L}{Z_L} \geq \nu_L^2.
\end{equation}
Taking $L\rightarrow \infty$ gives
\begin{equation}
\liminf_{p\rightarrow 0^+} \liminf_{N \rightarrow \infty} \lambda_*(N,p) \geq \Bigl(\frac{2q}{q^2+1}\Bigr)^4.
\end{equation}
Together with the upper bound, this establishes the result in Eq.~\eqref{eq:app_lambda_limit}.

\section{Decay Rate from First-Order Perturbation Theory}
\label{app:perturbation-theory}
At $p=0$, $\lambda=1$ is a twofold degenerate eigenvalue of the transfer matrix, with right eigenspace
\begin{equation}
\text{span}\{\ket{R_0},\ket{R_1}\},\qquad \ket{R_0}=\ket{0}^{\otimes N},\qquad \ket{R_1} = \ket{1}^{\otimes N}.
\label{eq:right_eigvecs}
\end{equation}
The vacuum state $\ket{R_0}$ remains a right eigenvector with eigenvalue $1$ for all $p$, while the degeneracy is lifted for $p>0$. The other eigenvalue that splits from $1$ is $\lambda_*$. We use degenerate perturbation theory to calculate $\lambda_*$ to first order in $p$. The resulting expansion is valid in the regime $Np\ll 1$.

At $p=0$, the left eigenspace associated with $\lambda =1$ is spanned by
\begin{equation}
\bra{L_0} = \sum_{\boldsymbol{s} \in \{0,1\}^N} \Bigl( \frac{q^{-2h(\boldsymbol{s})} - q^{-2N}} {1-q^{-2N}} \Bigr) \bra{\boldsymbol{s}},\qquad \bra{L_1} = \sum_{\boldsymbol{s}\in\{0,1\}^N} \Bigl( \frac{1-q^{-2h(\boldsymbol{s})}}{1-q^{-2N}} \Bigr) \bra{\boldsymbol{s}},
\end{equation}
where $h(\boldsymbol{s}) =\sum_{j=0}^{N-1} \delta_{s_j,1}$ is the Hamming weight of the bit string $\boldsymbol{s}$. These left eigenvectors are biorthonormal to the right eigenvectors in Eq.~\eqref{eq:right_eigvecs}:
\begin{equation}
\forall n,m \in\{0,1\},\qquad \bra{L_n} R_m\rangle = \delta_{n,m}.
\end{equation}
The row vectors $\bra{L_0}$ and $\bra{L_1}$ are linear combinations of the row vectors
\begin{equation}
\begin{aligned}
\bra{u}&=\sum_{\boldsymbol{s} \in \{0,1\}^N} \bra{\boldsymbol{s}} =(\bra{0}+\bra{1})^{\otimes N},\\
\bra{v}&= \sum_{\boldsymbol{s} \in \{0,1\}^N} q^{-2 h(\boldsymbol{s})} \bra{\boldsymbol{s}} = (\bra{0}+q^{-2}\bra{1})^{\otimes N}.
\end{aligned}
\end{equation}
The local transfer matrix satisfies
\begin{equation}
\sum_{\boldsymbol{s}\in \{0,1\}^2} \bra{\boldsymbol{s}} M(0) =\sum_{\boldsymbol{s}\in \{0,1\}^2} \bra{\boldsymbol{s}},\qquad \sum_{\boldsymbol{s}\in \{0,1\}^2} q^{-2h(\boldsymbol{s})}\bra{\boldsymbol{s}} M(p) =\sum_{\boldsymbol{s}\in \{0,1\}^2} q^{-2h(\boldsymbol{s})}\bra{\boldsymbol{s}}.
\label{eq:local_tm_left_eigvecs}
\end{equation}
Since $\bra{u}$ and $\bra{v}$ factorize as tensor products, it follows from Eq.~\eqref{eq:local_tm_left_eigvecs} that, for both $\mathrm{PBC}$ and $\mathrm{OBC}$, $\bra{u}$ and $\bra{v}$ are left eigenvectors of $T$ at $p=0$ with eigenvalue $1$. The weighted row vector $\bra{v}$ remains a left eigenvector with eigenvalue $1$ for all $p$.

Expand the transfer matrix to first order in $p$:
\begin{equation}
T(p) = T(0)+pT^\prime(0)+O(p^2).
\end{equation}
The first-order corrections to the degenerate eigenvalue are the eigenvalues of the $2 \times 2$ matrix $E$ with elements
\begin{equation}
E_{nm} = \bra{L_n} T^\prime(0) \ket{R_m}.
\end{equation}
Since $T(p)\ket{R_0} =\ket{R_0}$ for all $p$, differentiating at $p=0$ gives $T^\prime(0) \ket{R_0} =\boldsymbol{0}$. It follows that the zeroth column of $E$ vanishes:
\begin{equation}
E_{00} =E_{10}=0.
\end{equation}
Using the definitions above and the fact that $\bra{v} T(p) =\bra{v}$ for all $p$,
\begin{equation}
\bra{v} T^\prime (0 )\ket{R_1} =0 =E_{01} +q^{-2N} E_{11}.
\end{equation}
We also have
\begin{equation}
\bra{u} T^\prime(0) \ket{R_1} =E_{01} +E_{11}  = \frac{-2N (q-1)}{q}.
\end{equation}
The eigenvalues of $E$ are therefore $0$ and 
\begin{equation}
E_{11} = -\frac{2N(q-1)}{q(1-q^{-2N})}.
\end{equation}
Thus, for both $\mathrm{PBC}$ and $\mathrm{OBC}$,\footnote{\mbox{For $\mathrm{OBC}$, this assumes the measurement-induced relation $p_e=p(2-p)$}.} the first-order expansion is
\begin{equation}
\lambda_*(N,p)= 1-\Bigl(\frac{2(q-1)}{q}\Bigr) \Bigl(\frac{1}{1-q^{-2N}}\Bigr) Np +O(p^2).
\end{equation}

\section{Monte Carlo Simulation}
\label{app:monte-carlo}
For simplicity, we impose periodic boundary conditions and take the local nonunitary channels to be unital and uniform in space and time.
The general case discussed in Appendix~\ref{tm-derivation} admits a similar Monte Carlo simulation. For $r \in\mathbb{Z}_N$ and $t\in\mathbb{N}$, define
\begin{equation}
\Omega(r,t) = \sum_{\boldsymbol{s}\in \{0,1\}^N, s_r=1} \bra{\boldsymbol{s}}T^{(t)} \ldots T^{(2)}T^{(1)} \ket*{\boldsymbol{z}^{(0)}}, \quad \ket*{\boldsymbol{z}^{(0)}}= \bigl(\ket{1}^{\otimes 2} \otimes \ket{0}^{\otimes N-2} \bigr).
\label{eq:omega}
\end{equation}
The average OTOC can be expressed as\footnote{\parbox[t]{0.95\linewidth}{\raggedright This step assumes $\mathrm{PBC}$ and that the nonunitary channels are uniform in space and time. Under these assumptions, the OTOC at all $x$ and $y$ can be evaluated by evolving a single initial computational-basis state.}}
\begin{equation}
\begin{gathered}
\overline{C}(x,y,t) = C_0\Omega(y-l(x,t),t).
\end{gathered}
\end{equation}
Thus, at fixed $t$, the average OTOC for all $x$ and $y$ can be obtained by computing $\Omega(r,t)$ for all $r \in \mathbb{Z}_N$.
We use $\boldsymbol{z}= (\boldsymbol{z}^{(0)},\boldsymbol{z}^{(1)}, \ldots , \boldsymbol{z}^{(t)})$ to denote a trajectory of $N$-bit configurations, where $\boldsymbol{z}^{(n)} \in \{0,1\}^N$. Define the set of trajectories $\mathcal{Z}_t$ by
\begin{equation}
\mathcal{Z}_t = \bigl\{\boldsymbol{z}=(\boldsymbol{z}^{(0)},\ldots,\boldsymbol{z}^{(t)}) \in (\{0,1\}^N)^{t+1}: \boldsymbol{z}^{(0)}=(1,1,0,\ldots,0) \bigr\}.
\end{equation}
By inserting resolutions of the identity, we can write $\Omega(r,t)$ as a weighted sum over trajectories:
\begin{equation}
\Omega(r,t) =  \sum_{\boldsymbol{z} \in \mathcal{Z}_t, z_r^{(t)} =1} \mathcal{K}(\boldsymbol{z}),\qquad 
\mathcal{K}(\boldsymbol{z}) = \prod_{n=0}^{t-1} \prod_{j=0}^{N/2-1} \bra{z^{(n+1)}_{2j+n+1},z^{(n+1)}_{2j+n+2}} M \ket{z^{(n)}_{2j+n+1},z^{(n)}_{2j+n+2}}.
\label{eq:sum_histories}
\end{equation}
The column sums of the local transfer matrix $M(p)$ are
\begin{equation}
1,\qquad \gamma,\qquad \gamma,\qquad \gamma^2,
\end{equation}
where $\gamma(p)=a(p)+b(p)=1-\frac{q-1}{q}p$. Define 
\begin{equation}
D = \text{diag} (1,\gamma,\gamma,\gamma^2),\qquad \widehat{M} = MD^{-1}.
\end{equation}
It follows that we can write Eq.~\eqref{eq:sum_histories} as
\begin{equation}
\begin{gathered}
\Omega(r,t) =  \sum_{\boldsymbol{z} \in \mathcal{Z}_t, z_r^{(t)} =1} P_t(\boldsymbol{z})\gamma^{A_t(\boldsymbol{z})},\qquad A_t(\boldsymbol{z}) = \sum_{n=0}^{t-1} \sum_{j=0}^{N-1} z^{(n)}_j ,\\
P_t(\boldsymbol{z}) = \prod_{n=0}^{t-1} \prod_{j=0}^{N/2-1} \bra{z^{(n+1)}_{2j+n+1},z^{(n+1)}_{2j+n+2}} \widehat{M} \ket{z^{(n)}_{2j+n+1},z^{(n)}_{2j+n+2}}.
\end{gathered}
\end{equation}
Because $\widehat{M}$ is nonnegative and column-stochastic, $P_t(\boldsymbol{z})$ satisfies
\begin{equation}
0\leq P_t(\boldsymbol{z}) \leq1,\qquad \sum_{\boldsymbol{z}\in \mathcal{Z}_t} P_t(\boldsymbol{z})=1.
\end{equation}
We can therefore interpret $P_t(\boldsymbol{z})$ as the probability of the trajectory $\boldsymbol{z}$, with local transition probabilities given by the matrix elements of $\widehat{M}$. Thus,
\begin{equation}
\Omega(r,t) =\mathbb{E}[\gamma^{A_t(\boldsymbol{z})} z^{(t)}_r].
\end{equation}
By sampling $N_{S}$ independent trajectories, we compute a Monte Carlo estimate of $\Omega(r,t)$, which we denote $\widehat{\Omega}^{(N_S)}(r,t)$. 

Next, we derive a bound on the variance of this estimate. For any site $r$, time $t$, and trajectory $\boldsymbol{z}\in\mathcal{Z}_t$,
\begin{equation}
0 \leq \gamma^{A_t(\boldsymbol{z})}z^{(t)}_r \leq 1.
\end{equation}
It follows that
\begin{equation}
(\gamma^{A_t(\boldsymbol{z})}z^{(t)}_r)^2 \leq \gamma^{A_t(\boldsymbol{z})} z^{(t)}_r.
\end{equation}
Hence,
\begin{equation}
\begin{aligned}
\text{Var}(\gamma^{A_t(\boldsymbol{z})}z^{(t)}_r ) &= \mathbb{E}[(\gamma^{A_t(\boldsymbol{z})}z^{(t)}_r)^2]-\mathbb{E}[\gamma^{A_t(\boldsymbol{z})}z^{(t)}_r]^2\\
&\leq  \mathbb{E}[\gamma^{A_t(\boldsymbol{z})}z^{(t)}_r]-\mathbb{E}[\gamma^{A_t(\boldsymbol{z})}z^{(t)}_r]^2\\
&=\Omega(r,t) -\Omega^2(r,t) \\
&= \frac{1}{4} - \bigl(\Omega(r,t)-\frac{1}{2} \bigr)^2\\
&\leq \frac{1}{4}.
\end{aligned}
\end{equation}
This is the $[0,1]$ case of Popoviciu's inequality on variances. It establishes the bound in Eq.~\eqref{eq:monte-carlo-variance} of the main text.

\section{Gaussian Limit}
\label{app:gaussian-limit}
To establish the limit in Eq.~\eqref{eq:gaussian_limit}, it is sufficient to show that, for fixed $r$ and every $\tau >0$,
\begin{equation}
\lim_{p\rightarrow0^+} \lim_{N\rightarrow \infty} \Omega(r,\lfloor \tau p^{-1/2} \rfloor) =(1-q^{-4}) e^{- \kappa \tau^2}, \qquad \kappa = \frac{(q^2-1)(q-1)}{q(q^2+1)},
\label{eq:omega_gaussian_limit}
\end{equation}
where $\Omega(r,t)$ is defined in Eq.~\eqref{eq:omega}. When we wish to make the dependence of $\Omega(r,t)$ on $p$ explicit, we write $\Omega_p(r,t)$. Taking the thermodynamic limit of $\Omega(r,t)$ is equivalent to specifying that for all times considered, $N$ is large enough that the geometric light cone is unaffected by the finite system size. In the remainder of this appendix, we assume the thermodynamic limit has been taken, so that spatial sites $j\in\mathbb{Z}$ are defined on an infinite one-dimensional lattice.
\begin{figure}[h]
\centering
\includegraphics[width=0.55\linewidth]{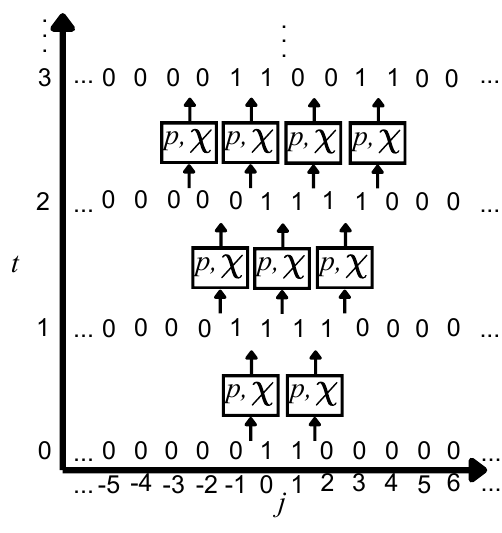}
\caption{ Depiction of the Monte Carlo process. For each time $t\in\mathbb{N}$ and pair of bond sites $e \in\mathbb{S}_t$, an independent random real number is drawn from the uniform distribution $\chi^{(t)}_{e} \sim \mathcal{U}([0,1])$.  The initial state is given in Eq.~\eqref{eq:initial_state}. States at later times are determined by the realization of random numbers and the $p$-dependent local transition probabilities.}
\label{fig:mc_process}
\end{figure}
To start, we give an explicit definition of the Monte Carlo process discussed in Appendix~\ref{app:monte-carlo} in the thermodynamic limit. An illustration of this Monte Carlo process is shown in Fig.~\ref{fig:mc_process}. We call the function $\boldsymbol{z}^{(t)}: \mathbb{Z} \rightarrow\{0,1\}$ the state at time $t$. We use $z^{(t)}_j$ to denote this function evaluated at lattice site $j$. The initial state is specified by
\begin{equation}
z^{(0)}_j = \delta_{j0} +\delta_{j1}.
\label{eq:initial_state}
\end{equation}
At every time $t\in \mathbb{N}$, define the set of bond sites
\begin{equation}
\mathbb{S}_t = \{ (2j-t,2j+1-t): j=0,\ldots,t \}.
\end{equation}
For each time $t$ and each pair of sites $e \in\mathbb{S}_t$, we independently draw a random real number $\chi^{(t)}_e \sim \mathcal{U} ([0,1])$, where $\mathcal{U}([0,1])$ denotes the uniform distribution on $[0,1]$. We denote a joint realization of these random variables over all times and bonds by $\boldsymbol{\chi}$ and the set of all such realizations by $\mathcal{X}$. We use $\boldsymbol{\chi}^{(t)}$ to denote the realization of all random numbers used in the transition from $\boldsymbol{z}^{(t-1)}$ to $\boldsymbol{z}^{(t)}$. For any $p \in[0,\frac{q+1}{q}]$ and $\chi \in[0,1]$, let $G_{p,\chi}: \{0,1\}^2 \rightarrow \{0,1\}^2$ be the function defined by its action on any $x=(x_0,x_1) \in \{0,1\}^2$:
\begin{equation}
\begin{gathered}
G_{p,\chi} (x) =
\begin{cases}
(0,0) , & \text{if } h(x) =0,\\
(0,0), & \text{if } h(x) =1 \text{ and } \chi \geq \widehat{b}(p),\\
(1,1), & \text{if } h(x) = 1 \text{ and } \chi < \widehat{b}(p),\\
(0,0), & \text{if } h(x) = 2 \text{ and } \chi \geq \widehat{d}(p),\\
(1,1), & \text{if } h(x) =2 \text{ and } \chi < \widehat{d}(p),
\end{cases}\\
\widehat{b}(p) =\frac{b(p)}{\gamma(p)},\qquad \widehat{d}(p)= \frac{d(p)}{\gamma^2(p)},\qquad h(x) = \delta_{x_0 1} +\delta_{x_{1}1}.
\end{gathered}
\label{eq:G_transition}
\end{equation}
The state at every time is defined by the initial state in Eq.~\eqref{eq:initial_state} and the recursion relation
\begin{equation}
\forall t \in \mathbb{N},\forall j \in \mathbb{Z}_{t+1}: \qquad(z^{(t)}_{2j-t}(p,\boldsymbol{\chi}) ,z^{(t)}_{2j-t+1} (p,\boldsymbol{\chi})) = G_{p, \chi^{(t)}_{(2j-t,2j-t+1)}}\bigl( (z^{(t-1)}_{2j-t}(p,\boldsymbol{\chi}) ,z^{(t-1)}_{2j-t+1} (p,\boldsymbol{\chi})) \bigr).
\end{equation}
Sites outside the geometric light cone remain zero. For given values of $p$ and $t$ and a given realization $\boldsymbol{\chi}$, the Monte Carlo process defines a unique trajectory given by the tuple
\begin{equation}
\mathcal{Z}(t,p,\boldsymbol{\chi}) = \bigl(\boldsymbol{z}^{(0)}, \boldsymbol{z}^{(1)}(p,\boldsymbol{\chi}),\ldots , \boldsymbol{z}^{(t)}(p,\boldsymbol{\chi})\bigr).
\end{equation}
We can express $\Omega$ as
\begin{equation}
\Omega(r,t) = \mathbb{E}_{\boldsymbol{\chi}} \bigl[\gamma(p)^{A_t(p,\boldsymbol{\chi})} z^{(t)}_r(p, \boldsymbol{\chi}) \bigr],\qquad A_t(p,\boldsymbol{\chi})= \sum_{n=0}^{t-1} \sum_{j=-\infty}^{+\infty} z^{(n)}_j(p,\boldsymbol{\chi}).
\end{equation}
The vacuum state $\boldsymbol{0}:\mathbb{Z}\rightarrow \{0,1\}$ is defined by $\boldsymbol{0}(j)=0 $ for all sites $j$. It is an absorbing state: if, along a trajectory, $\boldsymbol{z}^{(t)}=\boldsymbol{0}$, then $\boldsymbol{z}^{(n)}=\boldsymbol{0}$ for all $n>t$. Let $\mathcal{C}_{t,p}$ denote the event 
\begin{equation}
\mathcal{C}_{t,p} =\{\boldsymbol{\chi} \in \mathcal{X}:\boldsymbol{z}^{(t)}(p,\boldsymbol{\chi}) =\boldsymbol{0}\}.
\end{equation}
We start by calculating the probability that a trajectory evaluated at $p=0$ ever reaches the vacuum state, denoted by $\mathbb{P}(\mathcal{C}_{\infty,0})$, where
\begin{equation}
\mathcal{C}_{\infty, 0} = \bigcup_{t=0}^{\infty} \mathcal{C}_{t,0}.
\end{equation}

At $p=0$, $\widehat{d}(0)=1$ and there are no nucleation transitions $(1,1) \rightarrow (0,0)$. For any $\boldsymbol{\chi} \in\mathcal{C}_{t,0}^c$, define
\begin{equation}
\begin{aligned}
u(t,\boldsymbol{\chi}) &= \inf\{j \in\mathbb{Z}: z^{(t)}_j(0,\boldsymbol{\chi}) =1\},\\
v(t,\boldsymbol{\chi}) &= 1+\sup\{ j \in \mathbb{Z}: z^{(t)}_j(0,\boldsymbol{\chi}) =1 \}.
\end{aligned}
\end{equation}
Let $b_0 =\widehat{b}(0)$ and $a_0= 1-\widehat{b}(0)$. Let $\widetilde{u}(t)$, $\widetilde{v}(t) \in \mathbb{Z}$ be the positions of random walkers moving in discrete time steps on the one-dimensional lattice. Initially $\widetilde{u}(0)=0$ and $\widetilde{v}(0)=2$. At each time step the walkers at $\widetilde{u}(t)$ and $\widetilde{v}(t)$ independently move left or right by one lattice site. The walkers at $\widetilde{u}(t)$ and $\widetilde{v}(t)$ move right with probabilities $a_0$ and $b_0$, respectively. We say the walkers collide at time $t$ if $\widetilde{u}(t)=\widetilde{v}(t)$. We can map each $\boldsymbol{\chi}$ to a realization of $\widetilde{u}(t)$ and $\widetilde{v}(t)$ through a recursive rule such as\footnote{\parbox[t]{\linewidth}{
\raggedright The function $\zeta$ was introduced to make $\widetilde{u}$ and $\widetilde{v}$ move independently for the step immediately after a collision. We made use of an adjacent random variable which is not used by the other walker for this purpose.}}
\begin{equation}
\begin{gathered}
\widetilde{u}(t+1,\boldsymbol{\chi}) = \widetilde{u}(t, \boldsymbol{\chi}) - \phi(\zeta(t+1,\boldsymbol{\chi})),\qquad \widetilde{v}(t+1,\boldsymbol{\chi}) =\widetilde{v}(t, \boldsymbol{\chi}) +\phi(\chi^{(t+1)}_{(\widetilde{v}(t,\boldsymbol{\chi})-1,\widetilde{v}(t,\boldsymbol{\chi}))}),\\
\phi:\mathbb{R}\rightarrow \mathbb{Z},\qquad\phi(\chi)= \begin{cases}
1, & \chi < b_0,\\
-1, & \chi \geq b_0
\end{cases},\\
\zeta:\mathbb{N} \times \mathcal{X} \rightarrow\mathbb{R}, \qquad \zeta(t,\boldsymbol{\chi}) = \begin{cases}
\chi^{(t)}_{(\widetilde{u}(t-1,\boldsymbol{\chi})-1, \widetilde{u}(t-1,\boldsymbol{\chi}))},& \widetilde{u}(t-1,\boldsymbol{\chi}) \neq \widetilde{v}(t-1,\boldsymbol{\chi}),\\
\chi^{(t)}_{(\widetilde{u}(t-1,\boldsymbol{\chi})+1, \widetilde{u}(t-1,\boldsymbol{\chi})+2)},& \widetilde{u}(t-1,\boldsymbol{\chi}) = \widetilde{v}(t-1,\boldsymbol{\chi})
\end{cases}.
\end{gathered}
\end{equation}

On $\mathcal{C}^c_{\infty,0}$, $u(t,\boldsymbol{\chi})=\widetilde{u}(t,\boldsymbol{\chi})$ and $v(t,\boldsymbol{\chi}) = \widetilde{v}(t,\boldsymbol{\chi})$ for all time.  On $\mathcal{C}_{\infty,0}$, $u(t,\boldsymbol{\chi})$ and $v(t,\boldsymbol{\chi})$ agree with the random walker positions before the first collision, at which point the vacuum state $\boldsymbol{0}$ is reached. Hence $\mathbb{P}(\mathcal{C}_{\infty,0})$ is the probability that the random walkers ever collide. 

This can be calculated as follows~\cite{GrinsteadSnell1997}. For any integer $l\geq0$, let $R_l(t)$ denote the position of a particle that starts at $l$ and at each time step moves right with probability $b_0^2$, stays put with probability $2 a_0 b_0$ and steps left with probability $a_0^2$. For any integer $L\geq 1$, let $P^{(L)}_l$ be the probability that this particle hits zero before reaching $L$. It follows that $P^{(L)}_l$ satisfies, for $1\leq l \leq L-1$,
\begin{equation}
b_0^2 P^{(L)}_{l+1} - (a_0^2 +b_0^2) P^{(L)}_l +a_0^2 P^{(L)}_{l-1}=0,\qquad P^{(L)}_0=1, \qquad P^{(L)}_L =0.
\end{equation}
The solution is
\begin{equation}
P^{(L)}_l =\frac{(a_0/b_0)^{2l} -(a_0/b_0)^{2L}}{1-(a_0/b_0)^{2L}}.
\end{equation}
We can make the identification $R_1(t) =\frac{1}{2} (\widetilde{v}(t)-\widetilde{u}(t))$. Thus
\begin{equation}
\mathbb{P}(\mathcal{C}_{\infty,0}) = \lim_{L\rightarrow \infty}P^{(L)}_1 = \bigl(\frac{a_0}{b_0}\bigr)^2=q^{-4}.
\label{eq:collision_probability}
\end{equation}
Next, define
\begin{equation}
F(t,p) =\mathbb{P}(\mathcal{C}^c_{t,p}) \gamma(p)^{v_B t^2},\qquad v_B =\frac{q^2-1}{q^2+1}.
\end{equation}
Using Eq.~\eqref{eq:collision_probability}, we obtain, for any $\tau>0$,\footnote{\parbox[t]{\linewidth}{
\raggedright A direct calculation gives $\lim_{p \rightarrow 0^+} \gamma(p)^{v_B (\lfloor\tau p^{-1/2} \rfloor)^2}=e^{-\kappa \tau^2}$. We later define the event $\mathcal{G}_{t,p}$ in Eq.~\eqref{eq:G_event_definition} and show $ \lim_{p\rightarrow 0^+}\mathbb{P}(\mathcal{G}^c_{\lfloor \tau p^{-1/2}\rfloor,p})=0$. Since $0\leq \mathbb{P}(\mathcal{C}^c_{t,0})-\mathbb{P}(\mathcal{C}^c_{t,p}) \leq \mathbb{P}(\mathcal{G}^c_{t,p})$, it follows that $\lim_{p\rightarrow 0^+} \mathbb{P}(\mathcal{C}^c_{\lfloor \tau p^{-1/2}\rfloor,p})=\mathbb{P}(\mathcal{C}^c_{\infty,0}) = 1-q^{-4}$.}}
\begin{equation}
\lim_{p \rightarrow 0^+} F(\lfloor\tau p^{-1/2} \rfloor,p) =\bigl(1-q^{-4} \bigr) e^{-\kappa \tau^2}.
\end{equation}
Thus, to establish Eq.~\eqref{eq:omega_gaussian_limit}, it is sufficient to show that
\begin{equation}
\lim_{p \rightarrow 0^+} \bigl| \Omega_p(r,\lfloor\tau p^{-1/2} \rfloor) -F(\lfloor\tau p^{-1/2} \rfloor,p) \bigr| =0.
\label{eq:omega_F_limit}
\end{equation}
Define the occupied set
\begin{equation}
\Gamma^{(t)}(p,\boldsymbol{\chi}) = \{j \in \mathbb{Z}: z^{(t)}_j(p,\boldsymbol{\chi}) =1\},
\end{equation}
which satisfies $|\Gamma^{(t)}(p,\boldsymbol{\chi})|\leq 2(t+1)$ for any time $t$. We will show that
\begin{equation}
\Gamma^{(t)}(p,\boldsymbol{\chi})\subseteq\Gamma^{(t)}(0,\boldsymbol{\chi}).
\label{eq:gamma_subset}
\end{equation}
This can be shown by induction through layers. Initially,
\begin{equation}
z^{(0)}_j(p,\boldsymbol{\chi}) \leq z^{(0)}_j(0,\boldsymbol{\chi})
\end{equation}
for every site $j$, because the two processes start from the same initial state. Suppose this inequality holds at every site at time $t$. Then, for every local transition forming $\boldsymbol{z}^{(t+1)}$, the Hamming weight of the two-bit input at $p$ is at most the Hamming weight at $p=0$. Since the probability of transitioning to $(0,0)$ is nonincreasing with the Hamming weight of the input (see Eq.~\eqref{eq:G_transition}) and
\begin{equation}
\widehat{b}(p) \leq\widehat{b}(0),\qquad \widehat{d}(p)\leq \widehat{d}(0),
\end{equation}
using the same uniform random number for both updates gives
\begin{equation}
z^{(t+1)}_j (p,\boldsymbol{\chi}) \leq z^{(t+1)}_j(0,\boldsymbol{\chi})
\end{equation}
for every site $j$. The relation in Eq.~\eqref{eq:gamma_subset} follows by induction.

For any realization $\boldsymbol{\chi}$ and time $t$, we call $\mathcal{Z}(t,0,\boldsymbol{\chi})$ the reference trajectory. We use $\mathcal{G}_{t,p}$ to denote the event that the trajectory $\mathcal{Z}(t,p,\boldsymbol{\chi})$ agrees with the reference trajectory on the spacetime boundary. We define $\mathcal{G}_{t,p}$ formally as follows. For any finite set $O \subset \mathbb{Z}$, define
\begin{equation}
\text{Fill}(O) = \begin{cases}
\{ j \in \mathbb{Z}: \min O \leq j \leq \max O \}, & O \neq \emptyset,\\
\emptyset, & O = \emptyset
\end{cases}.
\end{equation}
For any $t \in \mathbb{N}$, define $t_e$ by
\begin{equation}t_e(t,p,\boldsymbol{\chi}) = \inf \Bigl( \{n \in \mathbb{Z}_{t+1}: \text{Fill}(\Gamma^{(n)}(p, \boldsymbol{\chi})) \neq \text{Fill}(\Gamma^{(n)}(0, \boldsymbol{\chi})) \} \cup  \{t+1\} \Bigr).
\end{equation}
If the trajectory $\mathcal{Z}(t,p,\boldsymbol{\chi})$ disagrees with the reference trajectory at the spatial boundary at any time up to $t$, $t_e(t,p,\boldsymbol{\chi})$ is the earliest such time. Otherwise, it is defined to be $t+1$.  Using this, let $\mathcal{G}_{t,p}$ be the event
\begin{equation}
\begin{gathered}
\mathcal{G}_{t,p} = \mathcal{E}_{t,p} \cap \mathcal{F}_{t,p},\\
\mathcal{E}_{t,p }=\{\boldsymbol{\chi} \in \mathcal{X}:t_e(t,p,\boldsymbol{\chi}) > t\},\qquad\mathcal{F}_{t,p}= \{\boldsymbol{\chi} \in \mathcal{X}:\boldsymbol{z}^{(t)}(p,\boldsymbol{\chi}) = \boldsymbol{z}^{(t)}(0,\boldsymbol{\chi}) \}.
\label{eq:G_event_definition}
\end{gathered}
\end{equation}

\begin{figure}[h]
\centering
\includegraphics[width=0.55\linewidth]{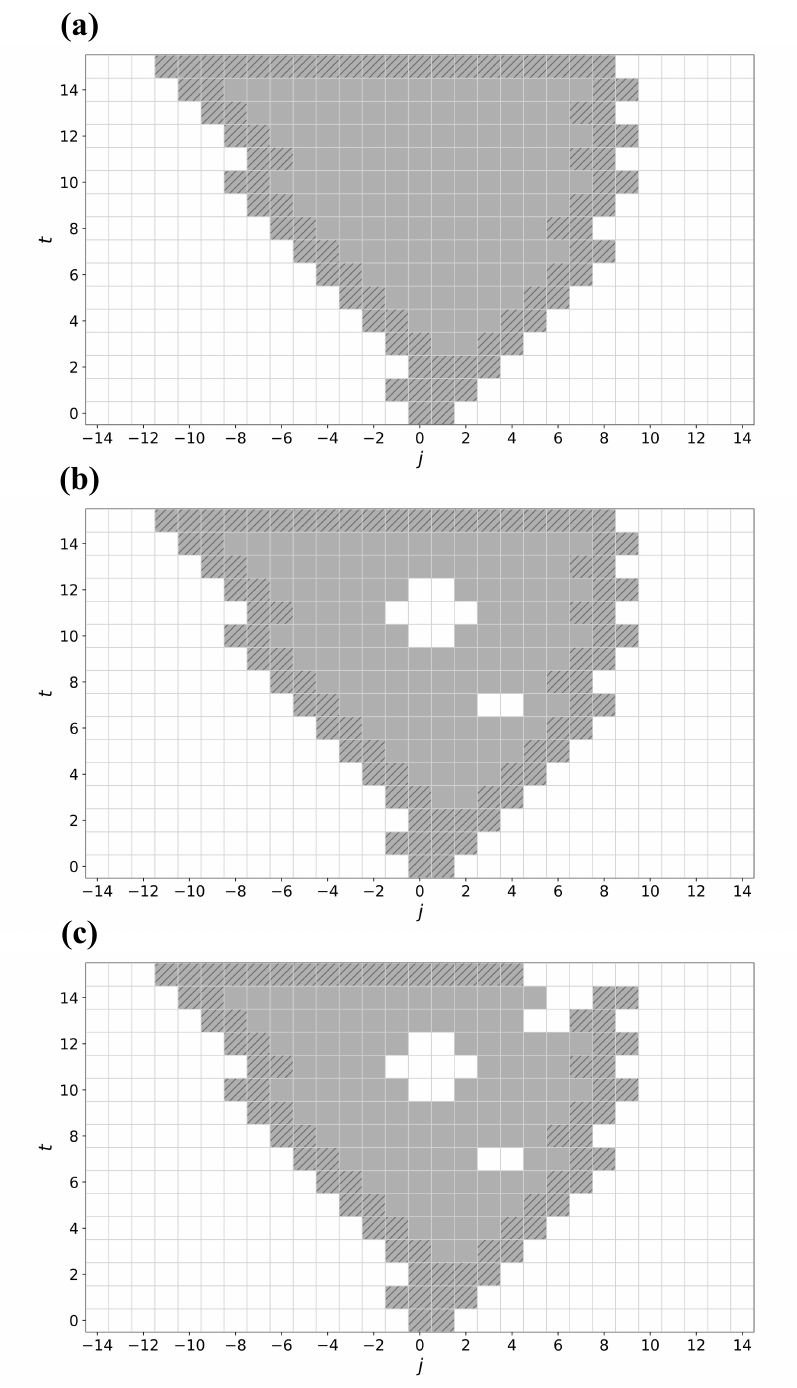}
\caption{Illustration of trajectories up to time $T=15$: (a) an example of a reference trajectory $\mathcal{Z}(T,0,\boldsymbol{\chi})$, (b) a possible corresponding trajectory $\mathcal{Z}(T,p,\boldsymbol{\chi})$ with $\boldsymbol{\chi} \in \mathcal{G}_{T,p}$, (c) a possible corresponding trajectory $\mathcal{Z}(T,p,\boldsymbol{\chi})$ with $\boldsymbol{\chi} \in \mathcal{G}_{T,p}^c$. For visual clarity, zeros and ones are depicted by gray and white squares, respectively. The local transitions on the spacetime boundary are marked by diagonal lines. For $\boldsymbol{\chi} \in \mathcal{G}_{T,p}$, the trajectory $\mathcal{Z}(T,p,\boldsymbol{\chi})$ may have holes in the interior, but must agree with the reference process on the spacetime boundary.}
\label{fig:mc_trajectories}
\end{figure}
Figure~\ref{fig:mc_trajectories}(a) depicts a trajectory of the Monte Carlo process at $p=0$. Examples of possible corresponding trajectories with $\boldsymbol{\chi} \in\mathcal{G}_{t,p}$ and $\boldsymbol{\chi} \in\mathcal{G}^c_{t,p}$ are shown in Fig.~\ref{fig:mc_trajectories}(b) and (c), respectively. For $\boldsymbol{\chi} \in \mathcal{G}_{t,p}$, the trajectory $\mathcal{Z}(t,p,\boldsymbol{\chi})$ must agree with the reference trajectory at the spacetime boundary, but is permitted to have holes (spacetime regions of zeros) in the interior.

For an event $E$ and realization $\boldsymbol{\chi}$, we use the notation $\boldsymbol{1}_{\boldsymbol{\chi}\in E}$ to denote the observable that is $1$ if the realization belongs to $E$ and zero otherwise. We proceed to bound $|\Omega_p(r,t)-F(t,p)|$. For any $p\in [0,\frac{q+1}{q})$:
\begin{equation}
\begin{aligned}
\bigl|\Omega_p(r,t) - F(t,p) \bigr| &= \bigl|\mathbb{E}_{\boldsymbol{\chi}}\bigl[ \gamma^{A_t(p,\boldsymbol{\chi})} z^{(t)}_r(p,\boldsymbol{\chi}) \bigr] - \mathbb{P}(\mathcal{C}_{t,p}^c) \gamma^{v_B t^2} \bigr|\\
&=\mathbb{P}(\mathcal{C}_{t,p}^c) \bigl| \mathbb{E}_{\boldsymbol{\chi}} \bigl[ \gamma^{A_t(p,\boldsymbol{\chi})} z_r^{(t)}(p,\boldsymbol{\chi})\bigr|\mathcal{C}^c_{t,p} \bigr] -\gamma^{v_B t^2} \bigr|\\
&=  \bigl| \mathbb{E}_{\boldsymbol{\chi}} \bigl[ \boldsymbol{1}_{\boldsymbol{\chi} \in \mathcal{C}_{t,p}^c} \bigl( \gamma^{A_t(p,\boldsymbol{\chi})} z^{(t)}_r(p,\boldsymbol{\chi}) -\gamma^{v_B t^2} \bigr) \bigr]|\\
&\leq  \mathbb{E}_{\boldsymbol{\chi}} \bigl[ \boldsymbol{1}_{\boldsymbol{\chi} \in \mathcal{C}_{t,p}^c} \bigl| \gamma^{A_t(p,\boldsymbol{\chi})} z^{(t)}_r(p,\boldsymbol{\chi}) -\gamma^{v_B t^2 } \bigr| \bigr]\\
&=  \mathbb{E}_{\boldsymbol{\chi}} \bigl[ \boldsymbol{1}_{\boldsymbol{\chi} \in \mathcal{C}_{t,p}^c \cap \mathcal{G}_{t,p}} \bigl| \gamma^{A_t(p,\boldsymbol{\chi})} z^{(t)}_r(p,\boldsymbol{\chi}) -\gamma^{v_B t^2 } \bigr| \bigr] \\ 
&\quad+ \mathbb{E}_{\boldsymbol{\chi}} \bigl[ \boldsymbol{1}_{\boldsymbol{\chi} \in \mathcal{C}_{t,p}^c \cap \mathcal{G}_{t,p}^c} \bigl| \gamma^{A_t(p,\boldsymbol{\chi})} z^{(t)}_r(p,\boldsymbol{\chi}) -\gamma^{v_B t^2 } \bigr| \bigr]\\
&\leq  \mathbb{E}_{\boldsymbol{\chi}} \bigl[ \boldsymbol{1}_{\boldsymbol{\chi} \in \mathcal{C}_{t,p}^c \cap \mathcal{G}_{t,p}} \bigl| \gamma^{A_t(p,\boldsymbol{\chi})} z^{(t)}_r(p,\boldsymbol{\chi}) -\gamma^{v_B t^2 } \bigr| \bigr] + \mathbb{P}(\mathcal{C}^c_{t,p} \cap \mathcal{G}_{t,p}^c)\\
&\leq  \mathbb{E}_{\boldsymbol{\chi}} \bigl[ \boldsymbol{1}_{\boldsymbol{\chi} \in \mathcal{C}_{t,p}^c \cap \mathcal{G}_{t,p}} \bigl| \gamma^{A_t(p,\boldsymbol{\chi})} z^{(t)}_r(p,\boldsymbol{\chi}) -\gamma^{v_B t^2 } \bigr| \bigr] + \mathbb{P}(\mathcal{G}_{t,p}^c)\\
&= \mathbb{E}_{\boldsymbol{\chi}} \bigl[ \boldsymbol{1}_{\boldsymbol{\chi} \in \mathcal{C}_{t,p}^c \cap \mathcal{G}_{t,p}} \bigl| (\gamma^{A_t(p,\boldsymbol{\chi})} - \gamma^{v_B t^2})z^{(t)}_r(p,\boldsymbol{\chi}) + \gamma^{v_B t^2} (z^{(t)}_r(p,\boldsymbol{\chi})-1) \bigr| \bigr] + \mathbb{P}(\mathcal{G}_{t,p}^c)\\
&\leq  \mathbb{E}_{\boldsymbol{\chi}} \bigl[\boldsymbol{1}_{\boldsymbol{\chi} \in \mathcal{C}_{t,p}^c \cap \mathcal{G}_{t,p}} z^{(t)}_r(p,\boldsymbol{\chi})\bigl| \gamma^{A_t(p,\boldsymbol{\chi})} -\gamma^{v_B t^2} \bigr|\bigr] \\ &\quad +  \mathbb{E}_{\boldsymbol{\chi}} \bigl[\boldsymbol{1}_{\boldsymbol{\chi} \in \mathcal{C}_{t,p}^c \cap \mathcal{G}_{t,p}} \gamma^{v_B t^2} \bigl| z^{(t)}_r(p,\boldsymbol{\chi})-1 \bigr| \bigr]+\mathbb{P}(\mathcal{G}_{t,p}^c).
\end{aligned}
\end{equation}
Since $\gamma^{v_B t^2} \leq 1$ and $z^{(t)}_r\leq 1$, we have
\begin{equation}
\begin{aligned}
\bigl|\Omega_p(r,t) - F(t,p) \bigr| & \leq \mathbb{E}_{\boldsymbol{\chi}} \bigl[\boldsymbol{1}_{\boldsymbol{\chi} \in \mathcal{C}_{t,p}^c \cap \mathcal{G}_{t,p}} \bigl| \gamma^{A_t(p,\boldsymbol{\chi})} -\gamma^{v_B t^2} \bigr|\bigr] \\ &\quad + \mathbb{E}_{\boldsymbol{\chi}} \bigl[\boldsymbol{1}_{\boldsymbol{\chi} \in \mathcal{C}_{t,p}^c \cap \mathcal{G}_{t,p}}  \bigl| z^{(t)}_r(p,\boldsymbol{\chi})-1 \bigr| \bigr]+\mathbb{P}(\mathcal{G}_{t,p}^c)\\
& =\mathbb{E}_{\boldsymbol{\chi}} \bigl[\boldsymbol{1}_{\boldsymbol{\chi} \in \mathcal{C}_{t,p}^c \cap \mathcal{G}_{t,p}} \bigl| \gamma^{A_t(p,\boldsymbol{\chi})} -\gamma^{v_B t^2} \bigr|\bigr]  +\mathbb{P}( \{\boldsymbol{\chi} \in \mathcal{C}^c_{t,p} \cap \mathcal{G}_{t,p}: z^{(t)}_r(p,\boldsymbol{\chi})=0 \} )+  \mathbb{P}(\mathcal{G}_{t,p}^c)\\
& \leq \mathbb{E}_{\boldsymbol{\chi}} \bigl[\boldsymbol{1}_{\boldsymbol{\chi} \in \mathcal{C}_{t,p}^c \cap \mathcal{G}_{t,p}} \bigl| \gamma^{A_t(p,\boldsymbol{\chi})} -\gamma^{v_B t^2} \bigr|\bigr]  +\mathbb{P}( \{\boldsymbol{\chi} \in \mathcal{C}^c_{t,0}: z^{(t)}_r(0,\boldsymbol{\chi})=0 \} )+  \mathbb{P}(\mathcal{G}_{t,p}^c).
\end{aligned}
\end{equation}
The elementary inequality\footnote{\parbox[t]{\linewidth}{
\raggedright
For a given $\boldsymbol{\chi}$, let $y=\min(A_t,v_B t^2)$, $z = \max(A_t, v_B t^2)$, and $f(x)=\gamma^x$. If $y=z$, the bound is trivial. Otherwise, by the mean value theorem, there exists $ \xi \in [y,z]$ such that $f(z)-f(y)=(z-y)\ln \gamma \gamma^{\xi} \geq (z-y)\ln \gamma$.
}} 
\begin{equation}
|\gamma^{A_t} -\gamma^{v_B t^2}| \le (-\ln \gamma) |A_t -v_B t^2|
\end{equation}
gives
\begin{equation}
\bigl|\Omega_p(r,t) - F(t,p) \bigr| \leq  (-\ln\gamma) \mathbb{E}_{\boldsymbol{\chi}} \bigl[\boldsymbol{1}_{\boldsymbol{\chi} \in \mathcal{C}_{t,p}^c \cap \mathcal{G}_{t,p}} \bigl| {A_t(p,\boldsymbol{\chi})} -v_B t^2 \bigr|\bigr]  +\mathbb{P}( \{\boldsymbol{\chi} \in \mathcal{C}^c_{t,0}: z^{(t)}_r(0,\boldsymbol{\chi})=0 \} )+  \mathbb{P}(\mathcal{G}_{t,p}^c).
\end{equation}
First, we derive an upper bound on the area term. For nonnegative integers $n$, define 
\begin{equation}
\begin{gathered}
H^{(n)}(p,\boldsymbol{\chi}) = \boldsymbol{1}_{\boldsymbol{\chi} \in \mathcal{E}_{n,p}}Q^{(n)}(p,\boldsymbol{\chi}),\\
Q^{(n)}(p,\boldsymbol{\chi}) = \frac{1}{2} (|\Gamma^{(n)}(0,\boldsymbol{\chi})| -|\Gamma^{(n)}(p,\boldsymbol{\chi})|).
\end{gathered}
\label{eq:H_definition}
\end{equation}
The expectation value of $H^{(n+1)}(p,\boldsymbol{\chi})$ conditional on the history $\mathcal{M}_n(\boldsymbol{\chi})=(\boldsymbol{\chi}^{(1)},\ldots ,\boldsymbol{\chi}^{(n)})$ satisfies
\begin{equation}
\begin{aligned}
\mathbb{E}_{\boldsymbol{\chi}^{(n+1)}}[H^{(n+1)}(p,\boldsymbol{\chi})|\mathcal{M}_n(\boldsymbol{\chi})] &=\mathbb{E}_{\boldsymbol{\chi}^{(n+1)}} [\boldsymbol{1}_{\boldsymbol{\chi} \in\mathcal{E}_{n+1,p}} Q^{(n+1)}(p,\boldsymbol{\chi})|\mathcal{M}_n(\boldsymbol{\chi})]\\
&\leq \mathbb{E}_{\boldsymbol{\chi}^{(n+1)}} [\boldsymbol{1}_{\boldsymbol{\chi} \in\mathcal{E}_{n,p}} Q^{(n+1)}(p,\boldsymbol{\chi})|\mathcal{M}_n(\boldsymbol{\chi})]\\
&=\boldsymbol{1}_{\boldsymbol{\chi} \in \mathcal{E}_{n,p}} \mathbb{E}_{\boldsymbol{\chi}^{(n+1)}} [ Q^{(n+1)}(p,\boldsymbol{\chi})|\mathcal{M}_n(\boldsymbol{\chi})].
\end{aligned}
\end{equation}
Define the tuple of two-bit strings at time $n$ by
\begin{equation}
\mathcal{B}^{(n)}(p,\boldsymbol{\chi}) =\bigl(z^{(n)}_{e_0}(p,\boldsymbol{\chi}) ,z^{(n)}_{e_1}(p,\boldsymbol{\chi}) \bigr)_{e=(e_0,e_1) \in \mathbb{S}_{n+1} },
\end{equation}
where bonds are ordered from left to right. For $x \in \{(0,1),(1,0),(1,1)\}$, let $N^{(n)}_{x}(p,\boldsymbol{\chi})$ denote the number of occurrences of the bit pattern $x$ in $\mathcal{B}^{(n)}(p,\boldsymbol{\chi})$. The conditional expectation of $Q^{(n+1)}(p,\boldsymbol{\chi})$ is
\begin{equation}
\begin{aligned}
\mathbb{E}_{\boldsymbol{\chi}^{(n+1)}} \bigl[Q^{(n+1)}(p,\boldsymbol{\chi}) \bigr|\mathcal{M}_n(\boldsymbol{\chi})\bigr] 
 &= Q^{(n)}(p,\boldsymbol{\chi})+ v_B \boldsymbol{1}_{\boldsymbol{\chi}\in\mathcal{C}^c_{n,0}}-\frac{1}{2}(2\widehat{b}(p)-1)(N^{(n)}_{(0,1)}(p,\boldsymbol{\chi}) +N^{(n)}_{(1,0)}(p,\boldsymbol{\chi}))  \\ &\quad +(1-\widehat{d}(p)) N^{(n)}_{(1,1)}(p,\boldsymbol{\chi}).
\label{eq:Q_recursion}
\end{aligned}
\end{equation}
If $\boldsymbol{\chi} \in \mathcal{C}_{n,0}$, the reference state is vacuum. Since $\Gamma^{(n)}(p,\boldsymbol{\chi}) \subseteq \Gamma^{(n)}(0,\boldsymbol{\chi})$, both processes are vacuum and remain so. Consequently, $Q^{(n)}(p,\boldsymbol{\chi})=Q^{(n+1)}(p,\boldsymbol{\chi})=0$. On $\mathcal{E}_{n,p} \cap \mathcal{C}^c_{n,0}$, we have
\begin{equation}
N^{(n)}_{(0,1)}(p,\boldsymbol{\chi}) + N^{(n)}_{(1,0)}(p,\boldsymbol{\chi}) \geq 2+2 \boldsymbol1_{\boldsymbol{\chi} \in \mathcal{J}_{n,p}},\qquad \mathcal{J}_{n,p}=\{\boldsymbol{\chi} \in \mathcal{X}:H^{(n)}(p,\boldsymbol{\chi})>0\}.
\end{equation}
This follows because the state at time $n$ must have two exterior mixed input pairs, and if $H^{(n)}(p,\boldsymbol{\chi})>0$, it also has at least one interior hole, contributing two additional mixed input pairs. Finite propagation also gives $N^{(n)}_{(1,1)}(p,\boldsymbol{\chi}) \leq n$. Hence, for sufficiently small $p$ and $\boldsymbol{\chi} \in \mathcal{C}^c_{n,0} $, the conditional expectation satisfies\footnote{\parbox[t]{\linewidth}{
\raggedright
Here sufficiently small means $p \leq \frac{q(q+1)(q^2-3)}{q^4+3}$, which is satisfied at any qudit dimension $q\geq2$ when $p \leq \frac{6}{19} $. It follows that $2\widehat{b}(p)-1 \geq \frac{1}{2}$.}}
\begin{equation}
\begin{aligned}
\mathbb{E}_{\boldsymbol{\chi}^{(n+1)}}[H^{(n+1)}(p,\boldsymbol{\chi})|\mathcal{M}_n(\boldsymbol{\chi})] &\leq \boldsymbol{1}_{\boldsymbol{\chi} \in \mathcal{E}_{n,p}} \mathbb{E}_{\boldsymbol{\chi}^{(n+1)}} [ Q^{(n+1)}(p,\boldsymbol{\chi})|\mathcal{M}_n(\boldsymbol{\chi})]\\
&= H^{(n)}(p,\boldsymbol{\chi}) + \boldsymbol{1}_{\boldsymbol{\chi}\in \mathcal{E}_{n,p}} \bigl[ v_B -\frac{1}{2}(2\widehat{b}(p)-1)(N^{(n)}_{(0,1)}(p,\boldsymbol{\chi}) +N^{(n)}_{(1,0)}(p,\boldsymbol{\chi}))  \\ &\quad +(1-\widehat{d}(p)) N^{(n)}_{(1,1)}(p,\boldsymbol{\chi})\bigr]\\
&\leq H^{(n)}(p,\boldsymbol{\chi}) +v_B -(2 \widehat{b}(p)-1) (1+\boldsymbol{1}_{\boldsymbol{\chi} \in \mathcal{J}_{n,p}})+ (1-\widehat{d}(p))n\\
&= H^{(n)}(p,\boldsymbol{\chi}) -(2 \widehat{b}(p)-1) \boldsymbol{1}_{\boldsymbol{\chi}\in \mathcal{J}_{n,p} } +(1-\widehat{d}(p))n + 2(\widehat{b}(0)-\widehat{b}(p))\\
&\leq H^{(n)}(p,\boldsymbol{\chi}) - \frac{1}{2} \boldsymbol{1}_{\boldsymbol{\chi} \in \mathcal{J}_{n,p}} +2 \theta p (n+1).
\end{aligned}
\label{eq:conditional_H_inequality}
\end{equation}
For convenience, we define $\theta$ as
\begin{equation}
\theta = \frac{q^3}{(q+1)(q^2+1)}.
\end{equation}
Since $ \mathcal{J}_{n,p} \subseteq \mathcal{C}^c_{n,0}$, Eq.~\eqref{eq:conditional_H_inequality} also holds for $\boldsymbol{\chi} \in \mathcal{C}_{n,0}$, where $H^{(n)}(p,\boldsymbol{\chi}) =H^{(n+1)}(p,\boldsymbol{\chi})=0$. Thus, averaging the conditional inequality over the history gives
\begin{equation}
\mathbb{E}_{\boldsymbol{\chi}}[H^{(n+1)}(p,\boldsymbol{\chi})] +\frac{1}{2} \mathbb{P}(\mathcal{J}_{n,p}) \leq \mathbb{E}_{\boldsymbol{\chi}}[H^{(n)}(p,\boldsymbol{\chi})]+ 2 \theta p(n+1).
\label{eq:H_inequality_update}
\end{equation}
From the initial condition $\mathbb{E}_{\boldsymbol{\chi}}[H^{(0)}(p,\boldsymbol{\chi})]=0$ and induction using Eq.~\eqref{eq:H_inequality_update}, it follows that, for every $n \in\mathbb{N}$,
\begin{equation}
\mathbb{E}_{\boldsymbol{\chi}}[H^{(n)}(p,\boldsymbol{\chi})] +\frac{1}{2} \sum_{m=0}^{n-1}\mathbb{P}(\mathcal{J}_{m,p})\leq \theta p n(n+1) \leq 2 \theta p n^2.
\label{eq:H_inequality}
\end{equation}

For all $\boldsymbol{\chi}\in \mathcal{G}_{t,p}$,
\begin{equation}
A_t(0,\boldsymbol{\chi})-A_t(p,\boldsymbol{\chi}) = 2 \sum_{n=0}^{t-1} H^{(n)}(p,\boldsymbol{\chi}).
\end{equation}
Using Eq.~\eqref{eq:H_inequality}, we get
\begin{equation}
\begin{aligned}
\mathbb{E}_{\boldsymbol{\chi}} \bigl[\boldsymbol{1}_{\boldsymbol{\chi}\in\mathcal{C}_{t,p}^c \cap \mathcal{G}_{t,p}} \bigl| {A_t(p,\boldsymbol{\chi})} -v_B t^2 \bigr|\bigr] &=\mathbb{E}_{\boldsymbol{\chi}} \bigl[\boldsymbol{1}_{\mathcal{C}_{t,p}^c \cap \mathcal{G}_{t,p}} \bigl| {A_t(0,\boldsymbol{\chi})} - 2 \sum_{n=0}^{t-1}H^{(n)}(p,\boldsymbol{\chi}) -v_B t^2 \bigr|\bigr]\\
&\leq \mathbb{E}_{\boldsymbol{\chi}} \bigl[\boldsymbol{1}_{\boldsymbol{\chi} \in \mathcal{C}_{t,p}^c \cap \mathcal{G}_{t,p}} \bigl| {A_t(0,\boldsymbol{\chi})} -v_B t^2 \bigr|\bigr]+2 \sum_{n=0}^{t-1} \mathbb{E}_{\boldsymbol{\chi}}[\boldsymbol{1}_{\boldsymbol{\chi} \in \mathcal{C}_{t,p}^c \cap \mathcal{G}_{t,p}}H^{(n)}(p,\boldsymbol{\chi})]\\
&\leq \mathbb{E}_{\boldsymbol{\chi}}[\boldsymbol{1}_{\boldsymbol{\chi} \in \mathcal{C}_{t,p}^c}\bigl| {A_t(0,\boldsymbol{\chi})} -v_B t^2 \bigr|\bigr] +2 \sum_{n=0}^{t-1} \mathbb{E}_{\boldsymbol{\chi}}[H^{(n)}(p,\boldsymbol{\chi})]\\
& \leq \mathbb{E}_{\boldsymbol{\chi}}[\boldsymbol{1}_{\boldsymbol{\chi} \in \mathcal{C}_{t,0}^c}\bigl| {A_t(0,\boldsymbol{\chi})} -v_B t^2 \bigr|\bigr] +4 \theta p\sum_{n=0}^{t-1}n^2 \\
& \leq \mathbb{E}_{\boldsymbol{\chi}}[\boldsymbol{1}_{\boldsymbol{\chi} \in \mathcal{C}_{t,0}^c}\bigl| {A_t(0,\boldsymbol{\chi})} -v_B t^2 \bigr|\bigr] +\frac{4}{3} \theta p t^3.
\end{aligned}
\label{eq:area_inequality}
\end{equation}
Next, we derive an upper bound on $\mathbb{P}(\mathcal{G}_{t,p}^c)$. Consider a realization $\boldsymbol{\chi} \in \mathcal{E}^c_{t,p}$. For any time $n \in \{1,2,\ldots, t_e(t,p,\boldsymbol{\chi})-1\}$, the state $\boldsymbol{z}^{(n)}(0,\boldsymbol{\chi})$ consists of a contiguous region of ones of length $l \in \{2,4,\ldots\}$ with infinite regions of zeros to either side
\footnote{\parbox[t]{\linewidth}{
\raggedright
If either process reached vacuum before $t_e(t,p,\boldsymbol{\chi})$, agreement of the filled configurations would force both processes to be vacuum. Absorption would then prevent any later disagreement, contradicting $\boldsymbol{\chi} \in \mathcal{E}_{t,p}^c$.}}. Denote the leftmost and rightmost occupied positions at time $n$ by
\begin{equation}
k_L^{(n)}(p,\boldsymbol{\chi}) = \inf\{j \in \mathbb{Z}: z^{(n)}_j(p,\boldsymbol{\chi})=1 \},\qquad k_R^{(n)}(p,\boldsymbol{\chi}) = \sup\{j \in \mathbb{Z}: z^{(n)}_j(p,\boldsymbol{\chi})=1 \}.
\end{equation}
Denote the edge bond sites for the reference process at time $n$ by
\begin{equation}
e^{(n)}_L(\boldsymbol{\chi})=(k_L^{(n-1)}(0,\boldsymbol{\chi})-1, k_L^{(n-1)}(0,\boldsymbol{\chi})),\qquad e^{(n)}_R(\boldsymbol{\chi})=(k_R^{(n-1)}(0,\boldsymbol{\chi}), k_R^{(n-1)}(0,\boldsymbol{\chi})+1).
\end{equation}
If the reference island at time $n-1$ has length $l\geq 4$, we can additionally define the interior bond sites nearest the edge by
\begin{equation}
f^{(n)}_L(\boldsymbol{\chi})=(k_L^{(n-1)}(0,\boldsymbol{\chi})+1, k_L^{(n-1)}(0,\boldsymbol{\chi})+2),\qquad f^{(n)}_R(\boldsymbol{\chi})=(k_R^{(n-1)}(0,\boldsymbol{\chi})-2, k_R^{(n-1)}(0,\boldsymbol{\chi})-1).
\end{equation}
Since $\boldsymbol{\chi} \in \mathcal{E}_{t,p}^c$, either the process at $p$ reaches vacuum at $\sigma =t_e(t,p,\boldsymbol{\chi})$, while the reference process does not,
\begin{equation}
\boldsymbol{z}^{(\sigma)}(p,\boldsymbol{\chi} )=\boldsymbol{0},\qquad \boldsymbol{z}^{(\sigma)}(0,\boldsymbol{\chi} ) \neq \boldsymbol{0},
\end{equation}
or both states are nonvacuum and their edge positions disagree,
\begin{equation}
(k^{(\sigma)}_L(p,\boldsymbol{\chi}),k^{(\sigma)}_R(p,\boldsymbol{\chi})) \neq (k^{(\sigma)}_L(0,\boldsymbol{\chi}),k^{(\sigma)}_R(0,\boldsymbol{\chi})).
\end{equation}
Thus, in the transition from $\boldsymbol{z}^{(\sigma-1)}$ to $\boldsymbol{z}^{(\sigma)}$, either at least one local edge transition disagreed
\begin{equation}
\begin{gathered}
\exists e =(e_0,e_1) \in \{e^{(\sigma)}_L(\boldsymbol{\chi}), e_R^{(\sigma)}(\boldsymbol{\chi}) \}:\\(z^{(\sigma)}_{e_0}(p,\boldsymbol{\chi}),z^{(\sigma)}_{e_1}(p,\boldsymbol{\chi})) \neq (z^{(\sigma)}_{e_0}(0,\boldsymbol{\chi}),z^{(\sigma)}_{e_1}(0,\boldsymbol{\chi})),
\end{gathered}
\label{eq:direct_edge_disagreement}
\end{equation}
or an edge and its nearest interior neighbor both transitioned to $(0,0)$:\footnote{\parbox[t]{\linewidth}{
\raggedright
For a given edge, these conditions come from directly considering the possible combinations of the edge and adjacent interior transition outcomes for the process at $p$ and the reference process. If $l=2$, only Eq.~\eqref{eq:direct_edge_disagreement} is possible.
}}
\begin{equation}
\begin{gathered}
\exists (e,f) =((e_0,e_1),(f_0,f_1)) \in \{(e^{(\sigma)}_L(\boldsymbol{\chi}),f^{(\sigma)}_L(\boldsymbol{\chi})),(e^{(\sigma)}_R(\boldsymbol{\chi}),f^{(\sigma)}_R(\boldsymbol{\chi}))\}:\\ 
(z^{(\sigma)}_{e_0}(p,\boldsymbol{\chi}),z^{(\sigma)}_{e_1}(p,\boldsymbol{\chi})) =(z^{(\sigma)}_{f_0}(p,\boldsymbol{\chi}),z^{(\sigma)}_{f_1}(p,\boldsymbol{\chi})) =(0,0).
\end{gathered}
\end{equation}
For nonnegative integers $n$ and $l$, define
\begin{equation}
\mathcal{L}_l^{(n)} = \{ \boldsymbol{\chi} \in \mathcal{X}: |\Gamma^{(n)}(0,\boldsymbol{\chi})|\geq l\}.
\end{equation}
Using this, define the events
\begin{equation}
\begin{gathered}
\mathcal{P}_{t,p} = \bigcup_{n=1}^{t} \{ \boldsymbol{\chi} \in \mathcal{L}^{(n-1)}_2 \cap \mathcal{E}_{n-1,p} | \exists e=(e_0,e_1) \in \{e_L^{(n)}(\boldsymbol{\chi}),e_R^{( n)}(\boldsymbol{\chi}) \}: (z^{(n)}_{e_0}(p,\boldsymbol{\chi}),z^{(n)}_{e_1}(p,\boldsymbol{\chi})) \neq (z^{(n)}_{e_0}(0,\boldsymbol{\chi}),z^{(n)}_{e_1}(0,\boldsymbol{\chi}))\},\\
\mathcal{Q}_{t,p} = \bigcup_{n=1}^{t} \{ \boldsymbol{\chi} \in\mathcal{L}_4^{(n-1)} |  \exists f=(f_0,f_1)  \in\{f_L^{(n)}(\boldsymbol{\chi}),f_R^{( n)}(\boldsymbol{\chi}) \} :(z^{(n)}_{f_0}(p,\boldsymbol{\chi}),z^{(n)}_{f_1}(p,\boldsymbol{\chi}))=(0,0) \}\cap \mathcal{P}^c_{t,p},\\
\mathcal{R}_{t,p} =\mathcal{E}_{t,p}\cap \mathcal{F}_{t,p}^c.
\end{gathered}
\end{equation}
Since $\mathcal{G}_{t,p}^c \subseteq \mathcal{P}_{t,p} \cup \mathcal{Q}_{t,p} \cup \mathcal{R}_{t,p}$, we have 
\begin{equation}
\mathbb{P}(\mathcal{G}_{t,p}^c) \leq \mathbb{P}(\mathcal{P}_{t,p})+\mathbb{P}(\mathcal{Q}_{t,p})+\mathbb{P}(\mathcal{R}_{t,p}).
\end{equation}

Next, we derive upper bounds for each of these probabilities. First, $\mathbb{P}(\mathcal{P}_{t,p})$ is the probability that at least one of at most $2t$ local edge Bernoulli trials results in different outcomes for the process at $p$ and the reference process: 
\begin{equation}
\begin{aligned}
\mathbb{P}(\mathcal{P}_{t,p}) &\leq 1-\bigl[1 -  \bigl(\widehat{b}(0)-\widehat{b}(p)\bigr)\bigr]^{2t}\\
&= (\widehat{b}(0)-\widehat{b}(p)) \sum_{n=0}^{2t-1} \bigl(1-(\widehat{b}(0)-\widehat{b}(p))\bigr)^n\\
&\leq 2 t \bigl(\widehat{b}(0)-\widehat{b}(p) \bigr) \\
&\leq \theta p t.
\end{aligned}
\end{equation}
The last step is valid for sufficiently small $p$. Next, we bound $\mathbb{P}(\mathcal{Q}_{t,p})$. For every $D\in \mathbb{N}$ satisfying $D<t$, define the event 
\begin{equation}
\begin{gathered}
\mathcal{D}_{t,p}= \bigcup_{n=1}^{t} \{\boldsymbol{\chi}\in \mathcal{L}^{(n-1)}_4 \cap \mathcal{Q}_{t,p}| \exists(e_0,e_1) \in \mathbb{S}_n:[(k^{(n-1)}_R(0,\boldsymbol{\chi})-e_1 \leq D) \lor (e_0-k^{(n-1)}_L(0,\boldsymbol{\chi}) \leq D)]\\,(z^{(n-1)}_{e_0}(p,\boldsymbol{\chi}),z^{(n-1)}_{e_1}(p,\boldsymbol{\chi}))=(1,1),(z^{(n)}_{e_0}(p,\boldsymbol{\chi}),z^{(n)}_{e_1}(p,\boldsymbol{\chi}))=(0,0) \}.
\end{gathered}
\end{equation}
$\mathcal{D}_{t,p}$ is the subevent of $\mathcal{Q}_{t,p}$ in which at least one nucleation transition $(1,1)\rightarrow (0,0)$ occurs within a distance $D$ of the domain walls of the reference process. At each time there are at most $ 2\lceil D/2 \rceil \leq D+1$ candidate bonds for such transitions. Thus
\begin{equation}
\begin{aligned}
\mathbb{P}(\mathcal{D}_{t,p}) &\leq 1- \bigl[ \widehat{d}(p)\bigr]^{2t(D+1)}\\
&\leq 1- \bigl[ d(p)\bigr]^{2t(D+1)}\\
&=\Bigl( \frac{2q^3 p -q^2(q-1)p^2}{(q+1)(q^2+1)} \Bigr) \sum_{n=0}^{2t(D+1)-1}\Bigl[1-\Bigl( \frac{2q^3 p -q^2(q-1)p^2}{(q+1)(q^2+1)} \Bigr) \Bigr]^{n}\\
&\leq 4 \theta(D+1) p t.
\end{aligned}
\end{equation}
The remaining realizations in $\mathcal{D}_{t,p}^c \cap\mathcal{Q}_{t,p}$ correspond to trajectories where holes are formed through nucleation transitions further in the interior that later reach an interior bond nearest an edge of the reference process. For a given realization  $\boldsymbol{\chi} \in\mathcal{D}_{t,p}^c \cap\mathcal{Q}_{t,p}$, let $t_d\in \{1,\ldots,t\}$ be the first time such that 
\begin{equation}
\exists f=(f_0,f_1) \in\{f_L^{(t_d)}(\boldsymbol{\chi}),f_R^{( t_d)}(\boldsymbol{\chi}) \}: (z^{(t_d)}_{f_0}(p,\boldsymbol{\chi}),z^{(t_d)}_{f_1}(p,\boldsymbol{\chi}))=(0,0).
\end{equation}
By reflection symmetry, we can without loss of generality consider the case where this occurs at $f_R^{(t_d)}(\boldsymbol{\chi})$. For every $n \in \{1,2,\ldots,t_d\}$ such that $\Gamma^{(n)}(p,\boldsymbol{\chi}) \neq \Gamma^{(n)}(0,\boldsymbol{\chi})$, define the position of the rightmost missing site by
\begin{equation}
s^{(n)}(p,\boldsymbol{\chi}) = \sup\bigl(\Gamma^{(n)}(0,\boldsymbol{\chi}) \setminus \Gamma^{(n)}(p,\boldsymbol{\chi}) \bigr).
\end{equation}
Trace backwards in time from $n=t_d$. The local input to the bond containing the rightmost missing site is
\begin{equation}
(z^{(n-1)}_{s^{(n)}-1},z^{(n-1)}_{s^{(n)}} ).
\end{equation}
If this input is $(1,1)$, we have reached a nucleation transition; denote its time by $m$. Otherwise, at least one input site was already missing at time $n-1$, so 
\begin{equation}
s^{(n)}\leq s^{(n-1)}+1.
\end{equation}
Continue backwards until a nucleation is reached. This must occur because the initial configurations coincide. Since $\boldsymbol{\chi} \notin \mathcal{D}_{t,p}$, this nucleation occurred more than $D$ sites from the reference right edge. At time $t_d$,
\begin{equation}
s^{(t_d)} = k_R^{(t_d-1)}-1.
\end{equation}
The rightmost missing site moves right by at most one per nonnucleation step, while the reference right edge moves left by at most one per step. Consequently,
\begin{equation}
D < k^{(m-1)}_R(0,\boldsymbol{\chi})-s^{(m)} \leq 2(t_d-m)+1,
\end{equation}
which implies $t_d-m\geq \lceil D/2 \rceil$. At every time $n=m,\ldots,t_d-1$, missing sites are present and the filled configurations still agree, so $H^{(n)}>0$. Thus, for every $\boldsymbol{\chi}\in \mathcal{X}$,
\begin{equation}
\lceil D/2 \rceil \boldsymbol{1}_{\boldsymbol{\chi} \in\mathcal{D}_{t,p}^c \cap\mathcal{Q}_{t,p}} \leq \sum_{n=0}^{t-1} \boldsymbol{1}_{\mathcal{J}_{n,p}} .
\end{equation}
Taking the expectation value and using the inequality in Eq.~\eqref{eq:H_inequality} gives
\begin{equation}
\begin{aligned}
\mathbb{P}(\mathcal{D}_{t,p}^c \cap\mathcal{Q}_{t,p}) &\leq \frac{1}{ \lceil D/2 \rceil} \sum_{n=0}^{t-1} \mathbb{P} (\mathcal{J}_{n,p})\\
&\leq \frac{4 \theta p t^2}{ \lceil D/2 \rceil} \\
&\leq \frac{8 \theta p t^2}{D}.
\end{aligned}
\end{equation}
Next, we bound $\mathbb{P}(\mathcal{R}_{t,p})$ using a similar method. For any $\boldsymbol{\chi} \in \mathcal{R}_{t,p}$, either there is at least one nucleation event at a time in $\{t-D,\ldots,t\}$  or at least one hole formed by a prior nucleation must survive for at least $D$ time steps. Define the event
\begin{equation}
\mathcal{T}_{t,p} = \bigcup_{n=t-D}^{t} \{ \boldsymbol{\chi} \in \mathcal{L}^{(n-1)}_4 \cap \mathcal{R}_{t,p}| \exists (e_0,e_1) \in \mathbb{S}_n: (z^{(n-1)}_{e_0}(p,\boldsymbol{\chi}),z^{(n-1)}_{e_1}(p,\boldsymbol{\chi})) =(1,1), (z^{(n)}_{e_0}(p,\boldsymbol{\chi}),z^{(n)}_{e_1}(p,\boldsymbol{\chi})) =(0,0) \}.
\end{equation}
At any time $n$, the number of candidate pairs of sites for a nucleation is at most $n$. Hence
\begin{equation}
\begin{aligned}
\mathbb{P}(\mathcal{T}_{t,p}) &\leq 1- [\widehat{d}(p)]^{(\sum_{n=t-D}^tn )}\\
&\leq 1- [\widehat{d}(p)]^{(D+1)t}\\
&\leq 1- [d(p)]^{(D+1)t}\\
&=\Bigl( \frac{2q^3 p -q^2(q-1)p^2}{(q+1)(q^2+1)} \Bigr) \sum_{n=0}^{(D+1)t-1}\Bigl[1-\Bigl( \frac{2q^3 p -q^2(q-1)p^2}{(q+1)(q^2+1)} \Bigr) \Bigr]^{n}\\
&\leq 2 \theta (D+1)pt.
\end{aligned}
\end{equation}
For $\boldsymbol{\chi} \in \mathcal{T}_{t,p}^c \cap \mathcal{R}_{t,p}$, the same backward-tracing argument reaches a nucleation at a time $m\leq t-D-1$. Since the filled configurations agree throughout, $H^{(n)}(p,\boldsymbol{\chi})>0$ for every $n=m,\ldots,t-1$, giving at least $D+1$ distinct times. Taking expectations and using Eq.~\eqref{eq:H_inequality} gives
\begin{equation}
\begin{aligned}
\mathbb{P}(\mathcal{T}_{t,p}^c \cap \mathcal{R}_{t,p}) &\leq \frac{1}{D+1}\sum_{n=0}^{t-1} \mathbb{P} (\mathcal{J}_{n,p})\\
&\leq \frac{4 \theta p t^2}{D+1}.
\end{aligned}
\end{equation}
Thus, for sufficiently small $p$, we have shown
\begin{equation}
\mathbb{P}(\mathcal{G}_{t,p}^c) \leq \bigl(6D+7\bigr) \theta p t+\frac{12 \theta p t^2}{D}.
\end{equation}
Together with Eq.~\eqref{eq:area_inequality}, we have
\begin{equation}
\begin{aligned}
\bigl|\Omega_p(r,t) - F(t,p) \bigr|  & \leq (-\ln \gamma(p)) \mathbb{E}_{\boldsymbol{\chi}}[\boldsymbol{1}_{\boldsymbol{\chi} \in \mathcal{C}_{t,0}^c}\bigl| {A_t(0,\boldsymbol{\chi})} -v_B t^2 \bigr|\bigr]+ \mathbb{P}( \{\boldsymbol{\chi} \in \mathcal{C}^c_{t,0}: z^{(t)}_r(0,\boldsymbol{\chi})=0 \} )\\
&\quad + \bigl(6D+7\bigr) \theta p t+\frac{12 \theta p t^2}{D} +\frac{4}{3} \theta(-\ln \gamma(p)) pt^3.
\end{aligned}
\end{equation}
Setting $t= \lfloor\tau p^{-1/2} \rfloor$ and $D=\lfloor\sqrt{t}\rfloor$, we have $D \sim \sqrt{\tau} p^{-1/4}$ and $-\ln \gamma(p)= O(p)$. The three terms in the second line are therefore $O(p^{1/4})$, $O(p^{1/4})$, and $O(p^{1/2})$, respectively, and vanish as $p\rightarrow 0^+$. 

The remaining terms involve only the reference trajectories. Furthermore, since only trajectories which do not reach vacuum through time $t$ contribute to these terms, they can be expressed using the random walkers at $\widetilde{u}(t,\boldsymbol{\chi})$ and $\widetilde{v}(t,\boldsymbol{\chi})$ discussed previously. On $\mathcal{C}^c_{t,0}$, we can express $A_t(0,\boldsymbol{\chi})$ as
\begin{equation}
A_t(0,\boldsymbol{\chi}) = \sum_{n=0}^{t-1} (\widetilde{v}(n,\boldsymbol{\chi})-\widetilde{u}(n,\boldsymbol{\chi})).
\end{equation}
Thus
\begin{equation}
\begin{aligned}
\mathbb{E}_{\boldsymbol{\chi}} \bigl[ \boldsymbol{1}_{\boldsymbol{\chi} \in \mathcal{C}^c_{t,0}} |A_t(0,\boldsymbol{\chi}) - v_B t^2| \bigr] &=\mathbb{E}_{\boldsymbol{\chi}} \bigl[ \boldsymbol{1}_{\boldsymbol{\chi} \in \mathcal{C}^c_{t,0}} | \sum_{n=0}^{t-1} \bigl(\widetilde{v}(n,\boldsymbol{\chi})-\widetilde{u}(n,\boldsymbol{\chi})\bigr) - v_B t^2 | \bigr]\\
& \leq \mathbb{E}_{\boldsymbol{\chi}} \bigl[  | \sum_{n=0}^{t-1} \bigl(\widetilde{v}(n,\boldsymbol{\chi})-\widetilde{u}(n,\boldsymbol{\chi}) \bigr) - v_B t^2 | \bigr].
\end{aligned}
\end{equation}
Adding and subtracting a sum over $\mathbb{E}_{\boldsymbol{\chi}} [\widetilde{v}(n, \boldsymbol{\chi}) -\widetilde{u}(n, \boldsymbol{\chi})]=2(1+n v_B)$ inside the absolute value gives
\begin{equation}
\begin{aligned}
\mathbb{E}_{\boldsymbol{\chi}} \bigl[ \boldsymbol{1}_{\boldsymbol{\chi} \in \mathcal{C}^c_{t,0}} |A_t(0,\boldsymbol{\chi}) - v_B t^2| \bigr] &\leq \mathbb{E}_{\boldsymbol{\chi}} \bigl[| \sum_{n=0}^{t-1} \bigl( \widetilde{v}(n,\boldsymbol{\chi}) -\widetilde{u}(n,\boldsymbol{\chi}) -\mathbb{E}_{\boldsymbol{\chi}}[\widetilde{v}(n,\boldsymbol{\chi}) -\widetilde{u}(n,\boldsymbol{\chi})]\bigr) +(2-v_B)t |\bigr]\\
&\leq (2-v_B)t + \mathbb{E}_{\boldsymbol{\chi}} \bigl[| \sum_{n=0}^{t-1} \bigl( \widetilde{v}(n,\boldsymbol{\chi}) -\widetilde{u}(n,\boldsymbol{\chi}) -\mathbb{E}_{\boldsymbol{\chi}}[\widetilde{v}(n,\boldsymbol{\chi}) -\widetilde{u}(n,\boldsymbol{\chi})]\bigr) |\bigr] \\
&\leq (2-v_B)t +\sum_{n=0}^{t-1} \sqrt{\text{Var}(\widetilde{v}(n,\boldsymbol{\chi})-\widetilde{u}(n,\boldsymbol{\chi}))}\\
&=(2-v_B)t + \sqrt{2(1-v_B^2)}\sum_{n=0}^{t-1}\sqrt{n}\\
&\leq(2-v_B)t + \sqrt{2(1-v_B^2)}t^{3/2}.
\end{aligned}
\end{equation}
Setting $t=\lfloor \tau p^{-1/2} \rfloor$ and multiplying by $-\ln \gamma(p)$ gives an $O(p^{1/4})$ term, which vanishes as $p\rightarrow 0^+$. It remains to show
\begin{equation}
\lim_{t\rightarrow \infty}\mathbb{P}( \{\boldsymbol{\chi} \in \mathcal{C}^c_{t,0}: z^{(t)}_r(0,\boldsymbol{\chi})=0 \} )=0.
\end{equation}
For sufficiently large $t$, we get \footnote{\parbox[t]{\linewidth}{
\raggedright Chebyshev's inequality states that, for an observable $X$ with mean $\mu=\mathbb{E}[X]$ and finite variance, and any $\xi>0$, $\mathbb{P}(|X-\mu|\geq \xi) \leq \text{Var}(X)/ \xi^2$. Here the thresholds are $r+v_B t$ and $2+v_B t-r$, both positive when $t> (|r|+2)/v_B$.}}
\begin{equation}
\begin{aligned}
\mathbb{P}( \{\boldsymbol{\chi} \in \mathcal{C}^c_{t,0}: z^{(t)}_r(0,\boldsymbol{\chi})=0\}) &= \mathbb{P}(\{\boldsymbol{\chi} \in \mathcal{C}^c_{t,0} : \widetilde{u}(t,\boldsymbol{\chi}) >r \lor \widetilde{v}(t,\boldsymbol{\chi}) \leq r\}  )\\
&\leq \mathbb{P}(\{\boldsymbol{\chi} \in \mathcal{X} : \widetilde{u}(t,\boldsymbol{\chi}) >r \lor \widetilde{v}(t,\boldsymbol{\chi}) \leq r\}  )\\
&\leq \mathbb{P}(\{\boldsymbol{\chi} \in \mathcal{X} : \widetilde{u}(t,\boldsymbol{\chi}) >r \}) + \mathbb{P}(\{\boldsymbol{\chi} \in \mathcal{X} : \widetilde{v}(t,\boldsymbol{\chi}) \leq r \})\\
&\leq \mathbb{P}(\{\boldsymbol{\chi}\in \mathcal{X}: |\widetilde{u}(t,\boldsymbol{\chi}) +v_B t|\geq r+v_B t\}) \\  &\quad+\mathbb{P}(\{\boldsymbol{\chi} \in \mathcal{X}: |\widetilde{v}(t,\boldsymbol{\chi}) -2 -v_B t| \geq 2 +v_B t-r\})\\
&\leq \frac{(1-v_B^2)t}{(r+v_Bt)^2}+\frac{(1-v_B^2)t}{(2+v_Bt -r)^2}.
\end{aligned}
\end{equation}
The last line made use of Chebyshev's inequality for the random walkers. This vanishes in the limit $t\rightarrow \infty$, at fixed $r$. Thus, we have established Eq.~\eqref{eq:omega_F_limit} and hence the Gaussian limit in Eq.~\eqref{eq:omega_gaussian_limit}.
%\section{Comparing OTOC to other Scrambling Metrics}

%\begin{center}

 %   {\large\bfseries Supplemental Material}

  %  \vspace{0.5em}

   % {\bfseries Non-unitary Quantum Butterfly Effect}

%\end{center}

%\vspace{1em}

% Optional: label supplemental equations/figures/tables S1, S2, ...

%\setcounter{equation}{0}

%\renewcommand{\theequation}{S\arabic{equation}}

%\setcounter{figure}{0}

%\renewcommand{\thefigure}{S\arabic{figure}}

%\setcounter{table}{0}

%\renewcommand{\thetable}{S\arabic{table}}

% ===========================================================

\section{Reproducibility and Seeding Protocol}
\label{app:reproducibility-seeding}
Each realization defined above is assigned a set of pseudorandom-number-generator (PRNG) seeds. For $n$ realizations, each seed is determined by a randomness stream $A\in \{0,1,2,3\}$, a user-specified base seed $b$, the circuit depth $L$, and the realization index $r$. The four values of $A$ label independent randomness streams: $A=0$ for angle sampling, $A=1$ for initial-state sampling, $A=2$ for forward-measurement sampling, and $A=3$ for backward-measurement sampling. For each stream, the seed is defined as
\begin{equation}
s=(b+A)\times 10^9+Ln+r.
\end{equation}

The offset $((b+A)\times10^9)$ places the four randomness streams in non-overlapping integer ranges, while the term $Ln+r$ assigns a distinct seed to each realization at each circuit depth. Provided that $Ln+r<10^9$, seeds belonging to different streams cannot overlap. This construction assigns distinct PRNG seeds to different realizations, while the perturbed and unperturbed circuits within each pair share the corresponding seeds, retaining exact reproducibility of the classical random sampling.

However, subsequent circuit execution is intrinsically stochastic. Consequently, even when the same circuit and PRNG seeds are reproduced exactly, measurement outcomes need not coincide between runs, whether obtained from quantum hardware or simulation. 

\section{Cartan Sampling with TK1 and TK2 Gates}
\label{app:cartan-sampling}
Generic two-qubit unitaries are generated using the Cartan, or (KAK), decomposition of (SU(4)). Up to an irrelevant global phase, a two-qubit unitary may be written as
\begin{equation}
U=
\left(k_1^{\mathrm{post}}\otimes k_2^{\mathrm{post}}\right)
A(c_x,c_y,c_z)
\left(k_1^{\mathrm{pre}}\otimes k_2^{\mathrm{pre}}\right),
\end{equation}

where the four local factors $k_i^{\mathrm{pre/post}}$ are in $SU(2)$, while

\begin{equation}
A(c_x,c_y,c_z)
=
\exp\left[
-\frac{i}{2}
\left(
c_x X\otimes X+
c_y Y\otimes Y+
c_z Z\otimes Z
\right)
\right]
\end{equation}

contains the non-local part of the two-qubit transformation. Redundancies under local rotations permit the Cartan coordinates to be restricted to the Weyl chamber
\begin{equation}
0\leq c_y\leq c_x\leq\frac{\pi}{2},
\qquad
|c_z|\leq c_y.
\end{equation}

The Haar measure on $SU(4)$, when expressed in these $KAK$ coordinates, factorizes into Haar measures for the four local $SU(2)$ rotations and a Haar-induced measure over the three non-local Cartan coordinates. Consequently, the four local transformations may be sampled independently according to the Haar measure on $SU(2)$.

For each local transformation we use the Euler-angle representation
\begin{equation}
k=
R_z(\phi)R_y(\theta)R_z(\lambda),
\end{equation}
with
\begin{equation}
\phi,\lambda\sim\mathrm{Unif}(0,2\pi),
\qquad
\theta=\arccos(1-2u),
\qquad
u\sim\mathrm{Unif}(0,1).
\end{equation}

The transformation $u\mapsto\theta$ generates the Haar density
\begin{equation}
p(\theta)=\frac{1}{2}\sin\theta,
\end{equation}
such that
\begin{equation}
d\mu_{SU(2)}
=
\frac{1}{8\pi^2}
\sin\theta\, d\phi\,d\theta\,d\lambda. 
\end{equation}

To express these rotations in the native TK1 parameterization, we use
\begin{equation}
R_z(\phi)R_y(\theta)R_z(\lambda)= 
R_z\left(\phi+\frac{\pi}{2}\right)
R_x(\theta)
R_z\left(\lambda-\frac{\pi}{2}\right),
\end{equation}
giving
\begin{equation}
\mathrm{TK1}\left(
\frac{\phi}{\pi}+\frac12,
\frac{\theta}{\pi},
\frac{\lambda}{\pi}-\frac12
\right).  
\end{equation}

The non-local Cartan coordinates are sampled separately. The Haar measure on $SU(4)$ induces the following density over the Weyl chamber:
\begin{equation}
d\mu_{\mathrm{C}}
\propto
J(c_x,c_y,c_z)\,
dc_x\,dc_y\,dc_z,
\end{equation}
where
\begin{equation}
J(c_x,c_y,c_z)
={}
\left|
\sin(c_x+c_y)\sin(c_x-c_y)
\sin(c_x+c_z)\sin(c_x-c_z)
\right.\left.
\sin(c_y+c_z)\sin(c_y-c_z)
\right|.
\end{equation}

The six sine factors constitute the Jacobian associated with the change from the invariant $SU(4)$ Haar measure to the local and Cartan coordinates of the $KAK$ decomposition.

We sample this distribution by rejection sampling. First, four independent random numbers
\begin{equation}
u_x,u_y,u_z,u_a\sim\mathrm{Unif}(0,1)   
\end{equation}

are generated, and a candidate Cartan point is constructed according to
\begin{equation}
c_x=\frac{\pi}{2}u_x^{1/3},
\qquad
c_y=c_x\sqrt{u_y},
\qquad
c_z=c_y(2u_z-1).
\end{equation}

Although these transformations are non-linear, their joint probability density is constant over the Weyl chamber. Indeed,
\begin{equation}
p(c_x)
=
\frac{3c_x^2}{(\pi/2)^3},
\qquad
p(c_y|c_x)=\frac{2c_y}{c_x^2},
\qquad
p(c_z|c_y)=\frac{1}{2c_y},
\end{equation}
and therefore
\begin{equation}
p(c_x,c_y,c_z)=
\frac{3}{(\pi/2)^3},
\end{equation}

independent of the Cartan coordinates. The proposal consequently samples the Weyl chamber uniformly without requiring additional rejection to enforce its boundaries.

The uniform candidate is subsequently accepted according to the Haar-induced Cartan weight. Since
\begin{equation}
J(c_x,c_y,c_z)\leq\frac14
\end{equation}

within the Weyl chamber, the candidate is accepted whenever
\begin{equation}
u_a<4J(c_x,c_y,c_z).
\end{equation}

Conditioning the uniform proposal distribution on this acceptance criterion produces the desired Haar-induced Cartan distribution.

Finally, the accepted coordinates are mapped directly to the native TK2 parameters,
\begin{equation}
\mathrm{TK2}(t_1,t_2,t_3)
=
\exp\left[
-\frac{i\pi}{2}
\left(
t_1X\otimes X+
t_2Y\otimes Y+
t_3Z\otimes Z
\right)
\right], 
\end{equation}

through
\begin{equation}
t_1=\frac{c_x}{\pi},
\qquad
t_2=\frac{c_y}{\pi},
\qquad
t_3=\frac{c_z}{\pi}.
\end{equation}

The resulting Haar-random two-qubit gate is therefore implemented natively as
\begin{equation}
U=
\left(
\mathrm{TK1}^{\mathrm{post}}_1
\otimes
\mathrm{TK1}^{\mathrm{post}}_2
\right)
\mathrm{TK2}(t_1,t_2,t_3)
\left(
\mathrm{TK1}^{\mathrm{pre}}_1
\otimes
\mathrm{TK1}^{\mathrm{pre}}_2
\right). 
\end{equation}

For the inverse transformation required on the backward branch of the circuit, the gate order is reversed and each constituent operation is inverted. In particular,
\begin{equation}
\mathrm{TK1}(a,b,c)^\dagger
=
\mathrm{TK1}(-c,-b,-a),
\qquad
\mathrm{TK2}(t_1,t_2,t_3)^\dagger=
\mathrm{TK2}(-t_1,-t_2,-t_3).
\end{equation}

Thus, the same sampled parameters completely specify both the forward transformation $U$ and its inverse $U^\dagger$.
\end{document}